\documentclass[10pt,journal]{IEEEtran}
\pdfoutput=1
\ifCLASSOPTIONcompsoc
  \usepackage[nocompress]{cite}
\else
  \usepackage{cite}
\fi
 
\usepackage{ragged2e}
\usepackage{array,enumerate}
\usepackage{multirow}
\usepackage{graphicx}
\usepackage{algorithm}
\usepackage{algpseudocode}
\usepackage{amssymb}
\usepackage{amsmath}
\usepackage{balance}
\usepackage{listings}
\newcommand{\subcaption}[1]{\centerline{{\footnotesize  #1}}\vspace{0.1in}}
\newlength{\minipagewidth}
\newlength{\figurewidthFour}

\begin{document}

\title{MetaFlow: a Scalable Metadata Lookup Service for Distributed File Systems in Data Centers}

\author{Peng~Sun,
        Yonggang~Wen, ~\IEEEmembership{Senior Member,~IEEE,}
        Ta~Nguyen~Binh~Duong, 
        and Haiyong~Xie \\

\IEEEcompsocitemizethanks{
\IEEEcompsocthanksitem
Peng Sun, Yonggang Wen and Ta Nguyen Binh Duong are with School of Computer Science and Engineering,  Nanyang Technological University, Singapore. Email: \{sunp0003, ygwen, donta\}@ntu.edu.sg.
\IEEEcompsocthanksitem
Haiyong Xie is with the China Academy of Electronics and Information Technology, Beijing 100041, China. Email: haiyong.xie@gmail.com.}
}

\thispagestyle{empty}

\IEEEtitleabstractindextext{
\begin{abstract}
\justifying
In large-scale distributed file systems, efficient metadata operations are critical since most file operations have to interact with metadata servers first. In existing distributed hash table (DHT) based metadata management systems, the lookup service could be a performance  bottleneck due to its significant CPU overhead. Our investigations showed that the lookup service could reduce system throughput by up to $70\%$, and increase system latency by a factor of up to $8$ compared to ideal scenarios. In this paper, we present MetaFlow, a scalable metadata lookup service utilizing  software-defined networking (SDN) techniques to distribute lookup workload over network components. MetaFlow tackles the lookup bottleneck problem by leveraging B-tree, which is constructed over the physical topology, to manage flow tables for SDN-enabled switches. Therefore, metadata requests can be forwarded to appropriate servers using only switches. Extensive performance evaluations in both simulations and testbed showed that MetaFlow increases system throughput by a factor of up to $3.2$, and reduce system latency by a factor of up to $5$ compared to DHT-based systems. We also deployed MetaFlow in a distributed file system, and demonstrated  significant performance improvement.
\end{abstract}

}

\maketitle
\IEEEdisplaynontitleabstractindextext
\IEEEpeerreviewmaketitle

\section{Introduction}\label{sec:introduction}

\IEEEPARstart{M}{etadata}  for file systems is ``data about data'' \cite{ross2000pvfs} and plays a crucial role in file system management. Specifically, metadata summarizes the basic information regarding files and directories in file systems. A metadata object is commonly represented as a \emph{key-value} pair, where the \emph{key} denotes the file name, and the \emph{value} consists of a set of attributes (e.g., file size, permission, access time, disk block, etc.) for the file or directory. In file systems, before users perform any file-related operations such as open, read, write, delete, etc., they have to acquire the files' metadata first. Therefore, metadata management strategies play an important role in determining the system performance.

Distributed file systems like HDFS \cite{shvachko2010hadoop}, GFS \cite{ghemawat2003google}, Lustre \cite{schwan2003lustre}, Ceph \cite{weil2006ceph}, and PVFS \cite{ ross2000pvfs} usually separate the management of metadata from storage servers. Such separation could make it easier to scale the storage capacity and bandwidth of the file system, since new storage servers can be added to the cluster when needed \cite{abadmetadata}. In these systems, clients should interact with the metadata server first to fetch files'  addresses and other attributes, after that they could perform operations on the desired files. It has been shown in previous work e.g., \cite{carns2009small,roselli2000comparison} that more than  80\% of file operations need to interact with metadata servers. Therefore, efficient metadata operations are critical for distributed file systems' performance. In \cite{meshram2011can}, the authors showed that an optimized metadata management system could improve the performance of directory operations in Lustre  and PVFS2 significantly (more than $20$ times).

The traditional single metadata server scheme cannot cope with the increasing workload in large scale storage systems \cite{shvachko2010hdfs, shvachko2010hadoop}. Therefore, most modern distributed file systems like GFS, Ceph, and Lustre deploy a cluster to share the metadata workload. In these systems, a very large metadata table is partitioned into smaller parts located on separate servers. It might be straightforward to achieve large-scale metadata storage by just adding more servers; however the same cannot be said for improving the system performance. Many novel distributed metadata management systems have been proposed to provide high performance metadata services, for instance \cite{weil2006ceph, aguilera2008practical, li2006dynamic}.

Existing approaches, e.g., \cite{weil2006ceph, aguilera2008practical, li2006dynamic}, focused on building overlay-based metadata management systems, which could be centralized or decentralized. These systems usually provides two main operations: lookup and I/O. In particular, \emph{the lookup operation aims to locate the desired metadata; and the I/O operation retrieves the metadata itself from the storage server using the address returned by the lookup operation.} However, the throughput and latency of these overlay-based systems could be significantly degraded due to the bottleneck created by a large number of lookup operations. Such operations actually compete for CPU cycles with I/O operations, which might lead to  reduced system throughput. Meanwhile, it could take a long time to locate a metadata object in overlay-based systems, which might increase the system latency.

In this paper, we propose MetaFlow, a new, efficient and fast distributed lookup service for metadata management. Rather than setting up a separate lookup operation, MetaFlow utilizes network components to locate the desired metadata with two techniques: software-defined-networking (SDN) and a B-tree based overlay network. SDN provides the ability for network switches to forward packets based on metadata identifiers. A B-tree based overlay is constructed over all physical switches and servers in the data center to generate and maintain forwarding tables for SDN-enabled switches. In this way, a metadata request can reach its target server directly, without the need for a separate lookup operation to query the destination. As a result, the latency in metadata operations could be reduced; and more available CPU cycles would be dedicated to I/O operations to improve the overall system throughput.

The primary contributions of this paper are as follows:
\begin{enumerate}
  \item We propose MetaFlow, a new distributed lookup service for metadata management, which transfers the lookup workload from  servers to network components with minimal overhead. 
  \item We design and develop a working implementation of MetaFlow using SDN and B-tree.
  \item We conduct extensive experiments using both large-scale simulations and a real, working testbed. The results show that MetaFlow could improve the metadata system's performance significantly. 
  \item We deploy MetaFlow in a real-word distributed file system. The results show that MetaFlow is able to demonstrate significant performance gain.
  \end{enumerate}

The rest of the paper is structured as follows. Section 2 presents existing lookup services for distributed metadata management. Section 3 identifies the performance bottlenecks through a measurement study on two DHT-based metadata management systems. Section 4 introduces MetaFlow-based system design, which harnesses the capability of SDN, to solve the lookup performance issue. Section 5 describes the algorithm to generate flow tables for SDN-enabled switches using B-tree. In Section 6, we introduce the flow table update algorithm when new node joins or leaves the system. The evaluation results are detailed in Section 7. In Section 8, we compare MetaFlow to information centric networking (ICN). Section 9 concludes the paper.

\section{Existing Approaches}

Lookup services for metadata management, which map a metadata object to its location, i.e., a metadata storage server, have been receiving much attention. The following summarizes existing lookup services for distributed  metadata management, including subtree partitioning, hash-based mapping and distributed hash table.

\subsection{Subtree Partitioning}

Static subtree partitioning \cite{pawlowski1994nfs} is a simple way of locating metadata objects used in many file systems such as NFS \cite{pawlowski1994nfs}. This approach requires an administrator to manually divide the directory tree, assign subtrees to different metadata servers, and store the partition information at some well-known locations. Clients can use the static partition information to locate metadata objects easily. This scheme works well when the file access pattern is uniform in the file system. However, in real file systems, the file access pattern is highly skewed \cite{zhu2008hba}. Thus, the static subtree partition scheme may have significant unbalance workloads. Dynamic subtree partitioning \cite{weil2004dynamic}, which dynamically maps subtrees to metadata servers based on their workload, is proposed to solve this problem. However, when a metadata server joins or leaves the system, the dynamic subtree partitioning scheme needs to refresh all subtree's information, generating high overhead in large-scale distributed file systems \cite{zhu2008hba}. 

\subsection{Hash-based Mapping}

Hash-based mapping \cite{brandt2003efficient} is another distributed metadata lookup service used in 
Lustre \cite{schwan2003lustre}, etc. This approach hashes a file name to an integer $k$, and assigns its metadata to a server according to the remainder value when dividing $k$ by the total number of metadata servers. There is no lookup overhead on storage servers with this approach, since such lookup is done on the client side using the hash function. However, the hash-based mapping approach might not be practical due to two reasons. First, since all metadata objects must be re-allocated when a metadata server joins or leaves the system. Second, hash-based mapping may have high POSIX directory access overhead \cite{zhu2008hba} \cite{weil2004dynamic}. Specifically, since the hash-based mapping scheme eliminates all hierarchical locality, metadata objects within the same directory may be allocated to different servers. Therefore, when trying to satisfy POSIX directory access semantics, such a scheme needs to interact with multiple metadata servers, generating high overhead.

Three approaches have been proposed to tackle the  POSIX directory access's performance issue for hash-based mapping. First, GFS shows that there is no need to satisfy POSIX semantics strictly for most cloud computing applications \cite{ghemawat2003google} \cite{caesar2010efficient}. Second, an efficient metadata caching system  also helps  improve the directory access performance \cite{caesar2010efficient}. Finally, Lazy Hybrid (LH) \cite{brandt2003efficient} has been proposed to combine the best of both subtree partitioning and hash-based mapping to address the  directory access problem.

\setlength{\minipagewidth}{0.48\textwidth}
\setlength{\figurewidthFour}{\minipagewidth}
\begin{figure} 
    \centering
    \begin{minipage}[t]{\minipagewidth}
    \begin{center}
    \includegraphics[width=\figurewidthFour]{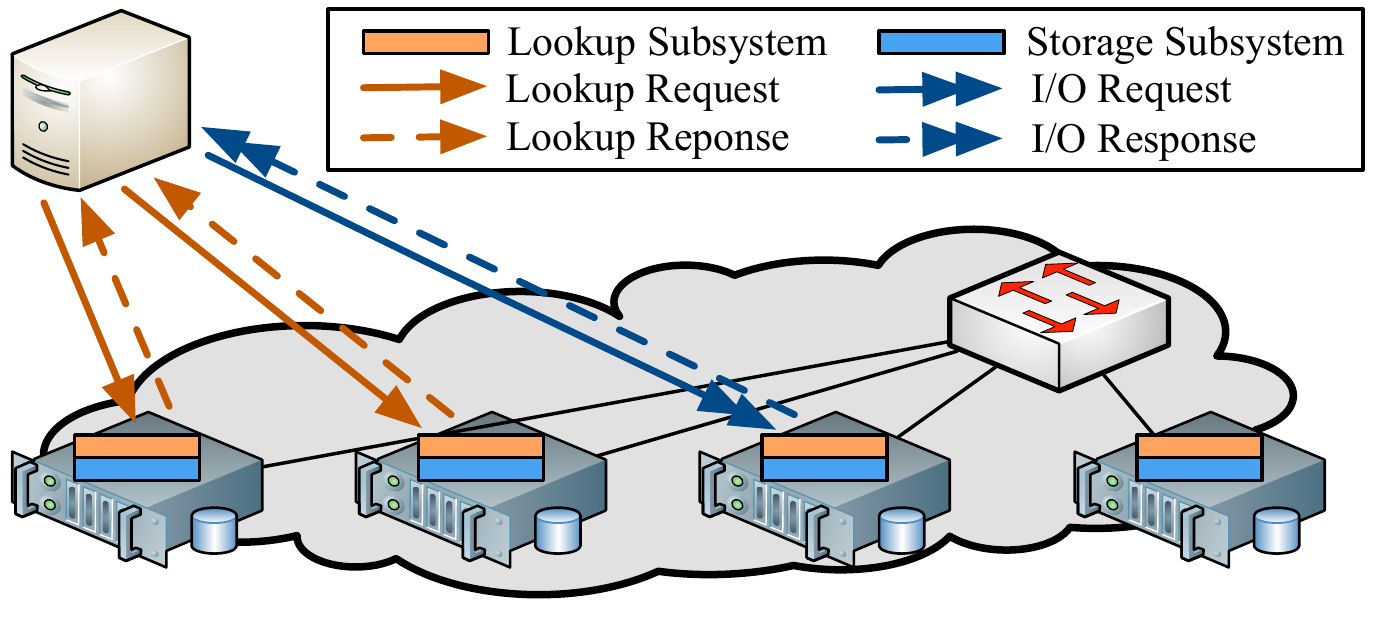}
    \end{center}
    \end{minipage}
    \caption{DHT-based distributed metadata management system architecture. In such systems, each server consists of two subsystems, namely lookup and storage subsystem, to deal with lookup and I/O operations separately. }
\label{Fig: dht_storage_system}
\end{figure}

\subsection{Distributed Hash Table}

Recent work have been focusing on using the Distributed Hash Table (DHT) \cite{ratnasamy2001scalable} model for the distributed metadata lookup service, because it allows nodes to join or leave the system dynamically with minimal overhead. In these systems, a metadata operation consist of two sequential operations: lookup and I/O. As shown in Figure \ref{Fig: dht_storage_system}, an I/O operation must wait for the completion of its associated lookup operation, since it needs to know the  destination to send the I/O request. Generally, each node has two subsystems to process lookup requests and I/O requests separately. The lookup subsystem maintains some storage information to locate metadata objects. The storage subsystem, which is usually a high performance in-memory storage system \cite{shvachko2010hdfs}, deals with the I/O requests on metadata objects. Two DHT-based approaches are widely used: (i) \emph{{Chord}} \cite{stoica2001chord}. In a $N$-node system, each Chord node maintains  $O(\log N)$ other nodes' storage information. On average, each lookup request needs to interact with $O(\log N)$ nodes to locate a metadata object. For example, Arpeggio \cite{clements2005arpeggio}, a peer-to-peer file-sharing network, uses Chord to support distributed metadata lookup. (ii) \emph{{One-Hop}} \cite{gupta2003one}. One-Hop allows each node to maintain all other nodes'  information. Therefore, any lookup requests will be processed by only one node. As shown in \cite{caesar2010efficient}, One-Hop could achieve high throughput and low latency for metadata operations.

\section{Problem Identification}

We argue that overlay-based systems might not be the best solutions for metadata management, due to bottlenecks caused by the lookup service. We construct a testbed to carry out a series of experiments to identify potential performance issues.

\subsection {Experiment Configurations}

We conduct experiments using a testbed with up to $200$ Linux containers \cite{merkel2014docker} and $3$ physical switches. In the testbed, we implement two DHT-based models (i.e., Chord and One-Hop), and one centralized model (i.e., Central Coordinator \cite{lua2005survey}, which uses a central server to locate metadata objects) to provide the lookup service. In the experiment, each file and directory's metadata object is a $250$ and $290$ bytes of key-value pair, respectively \cite{shvachko2010hdfs}. We set up a set of clients to generate a mix \emph{get} and \emph{put} metadata operations with a ratio of $20\%$ and $80\%$, respectively, to simulate the real metadata workload \cite{dandong2012decentralized}. In the \emph{get} operation, a client retrieves a metadata object using the given file name. In the \emph{put} operation, a client writes new data into a metadata object. We measure the throughput and latency respectively to identify the performance issues:
\begin{enumerate}
\item[\textbullet] \emph{\textbf{Throughput}}: We compare the DHT-based systems to an ideal system, which provides linear throughput performance without any performance reduction in theory.

\item[\textbullet] \emph{\textbf{Latency}}: We use the hash-based mapping approach, which has no additional lookup latency, as the baseline.
\end{enumerate}

Since the traditional hardware virtualization techniques (e.g., Kernel Virtual Machine (KVM), Xen \cite{barham2003xen}, etc.) may reduce the system performance significantly \cite{kivity2007kvm}, we use the container-based virtualization \cite{merkel2014docker} to run our experiments, reducing the additional overhead caused by the hypervisors \cite{felter2014updated}.  Specifically,  we create up to $200$ Linux containers using Docker \cite{merkel2014docker}, a lightweight Linux container management system. Each Linux container is allocated with a $2$ GHz CPU core and $4$ GB memory. In the experiments, we use four different storage subsystems:
\begin{enumerate}
\item[\textbullet] \emph{\textbf{Redis}} \cite{Redis}: Redis is a pure in-memory key-value storage system. Metadata management systems in HDFS, GFS, PVFS, etc. use only the memory to store the metadata.

\item[\textbullet] \emph{\textbf{LevelDB (HDD)}} \cite{LevelDB}: LevelDB (HDD) is a fast key-value storage system using both the memory and the hard disk drive (HDD). Some file systems like Tablefs \cite{ren2013tablefs} store  the metadata using LevelDB.

\item[\textbullet] \emph{\textbf{LevelDB (SSD)}}: LevelDB (SDD) uses both the memory and the solid state disk (SSD) to manage key-value data.

\item[\textbullet] \emph{\textbf{MySQL (HDD)}}: MySQL is a conventional relational database management system (RDBMS). In the experiments, we deploy MySQL on the HDD. To measure HDD's impact, we disable the  query cache  function in MySQL. This serves as a lower bound for storage subsystems' performance.
\end{enumerate}

\subsection {DHT: Throughput}

\setlength{\minipagewidth}{0.24\textwidth}
\setlength{\figurewidthFour}{\minipagewidth}
\begin{figure}
    \centering
    \begin{minipage}[t]{\minipagewidth}
    \begin{center}
    \includegraphics[width=\figurewidthFour]{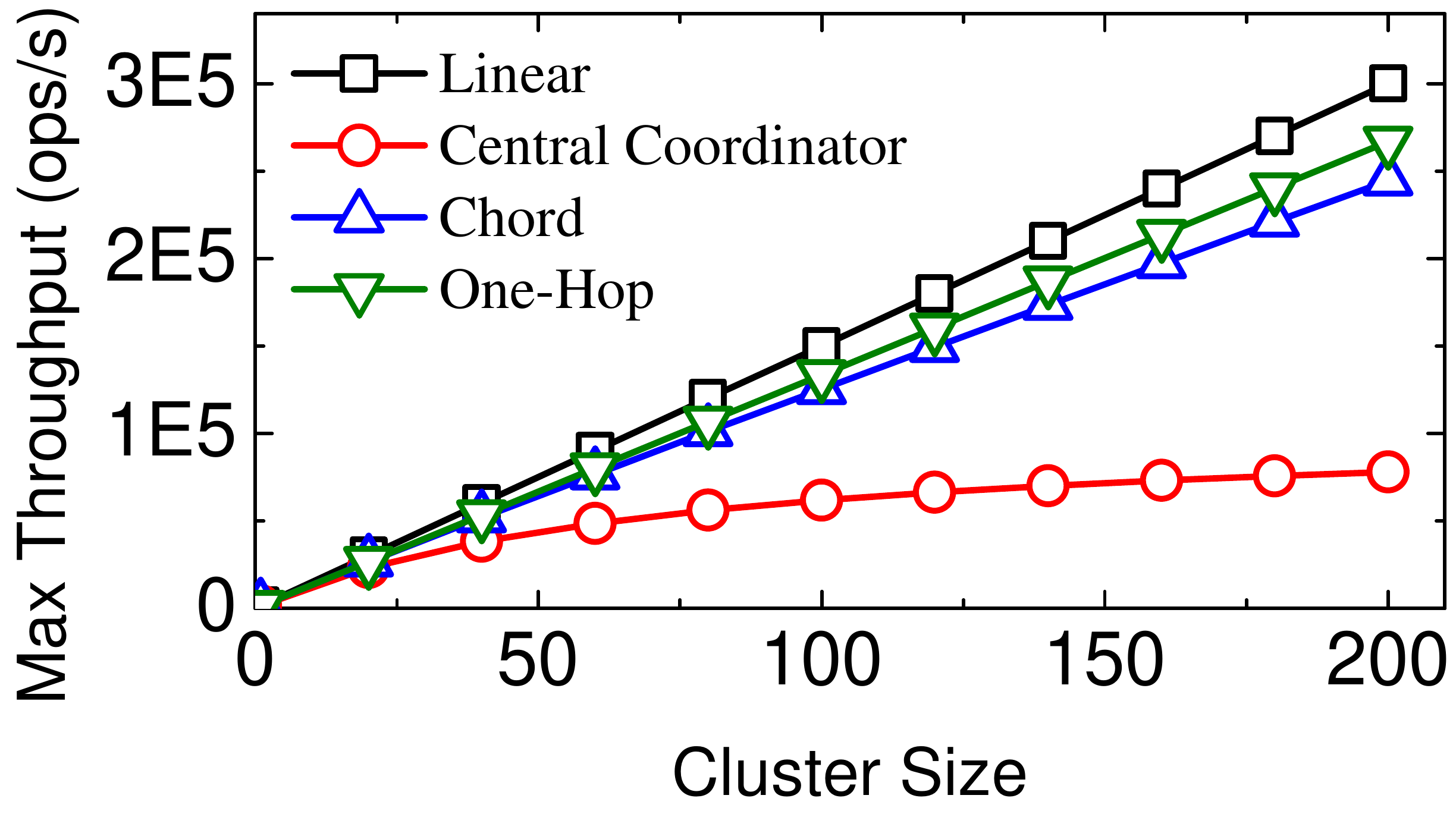}
    \subcaption{(a) MySQL (HDD)}
    \end{center}
    \end{minipage}
    \centering
    \begin{minipage}[t]{\minipagewidth}
    \begin{center}
    \includegraphics[width=\figurewidthFour]{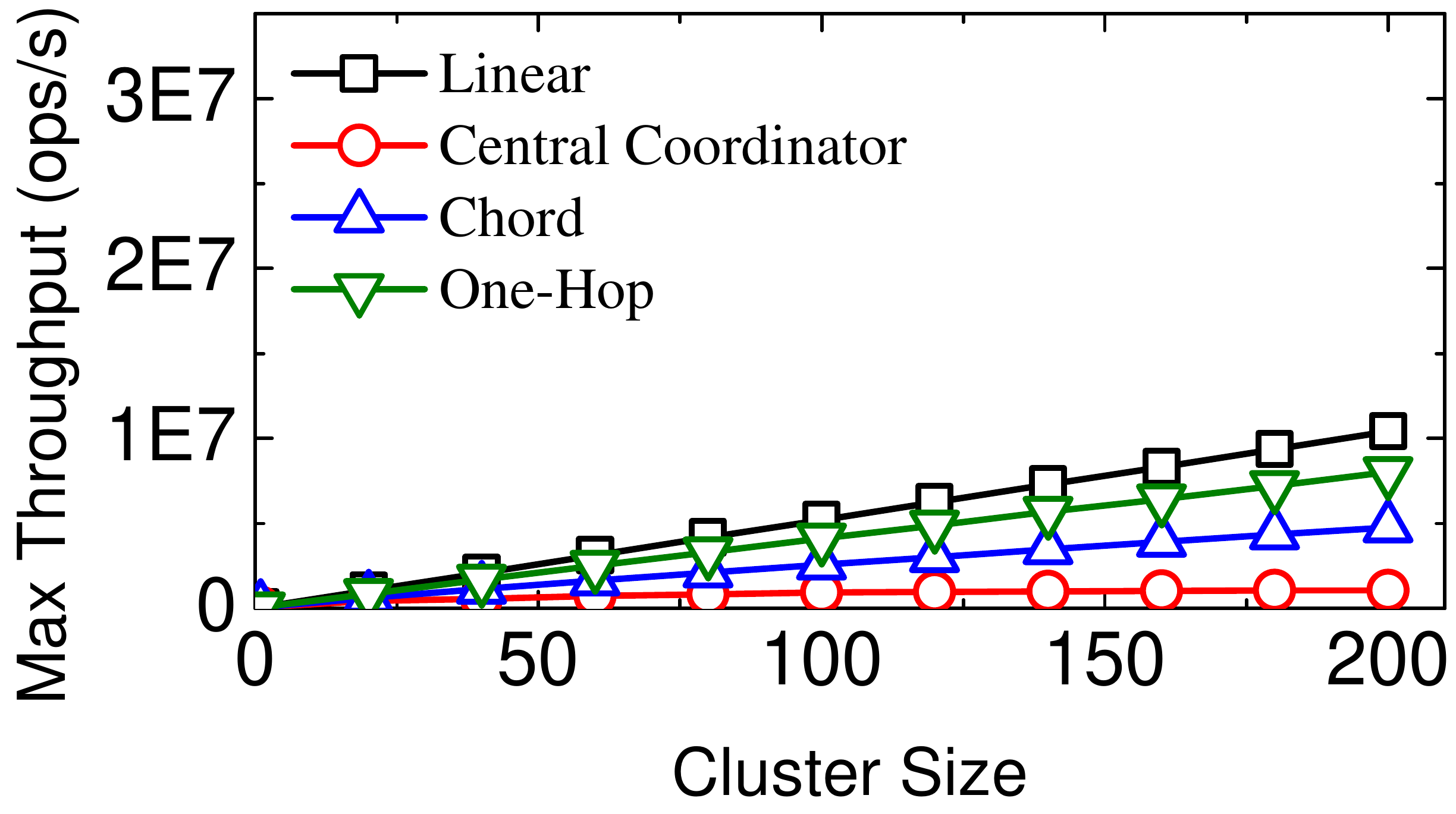}
    \subcaption{(b) LevelDB (HDD)}
    \end{center}
    \end{minipage}
    \centering
    \begin{minipage}[t]{\minipagewidth}
    \begin{center}
    \includegraphics[width=\figurewidthFour]{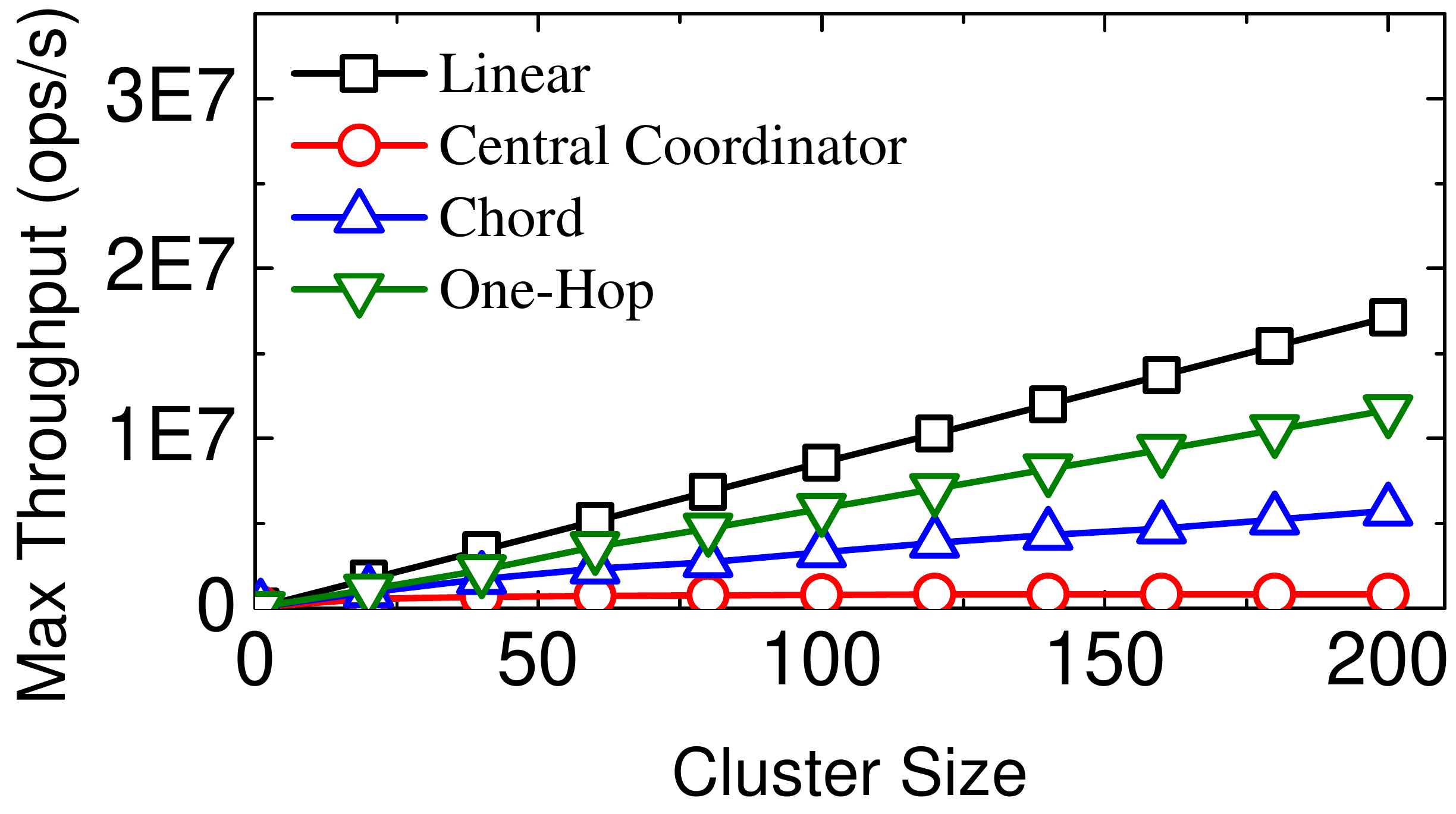}
    \subcaption{(c) LevelDB (SSD)}
    \end{center}
    \end{minipage}
    \centering
    \begin{minipage}[t]{\minipagewidth}
    \begin{center}
    \includegraphics[width=\figurewidthFour]{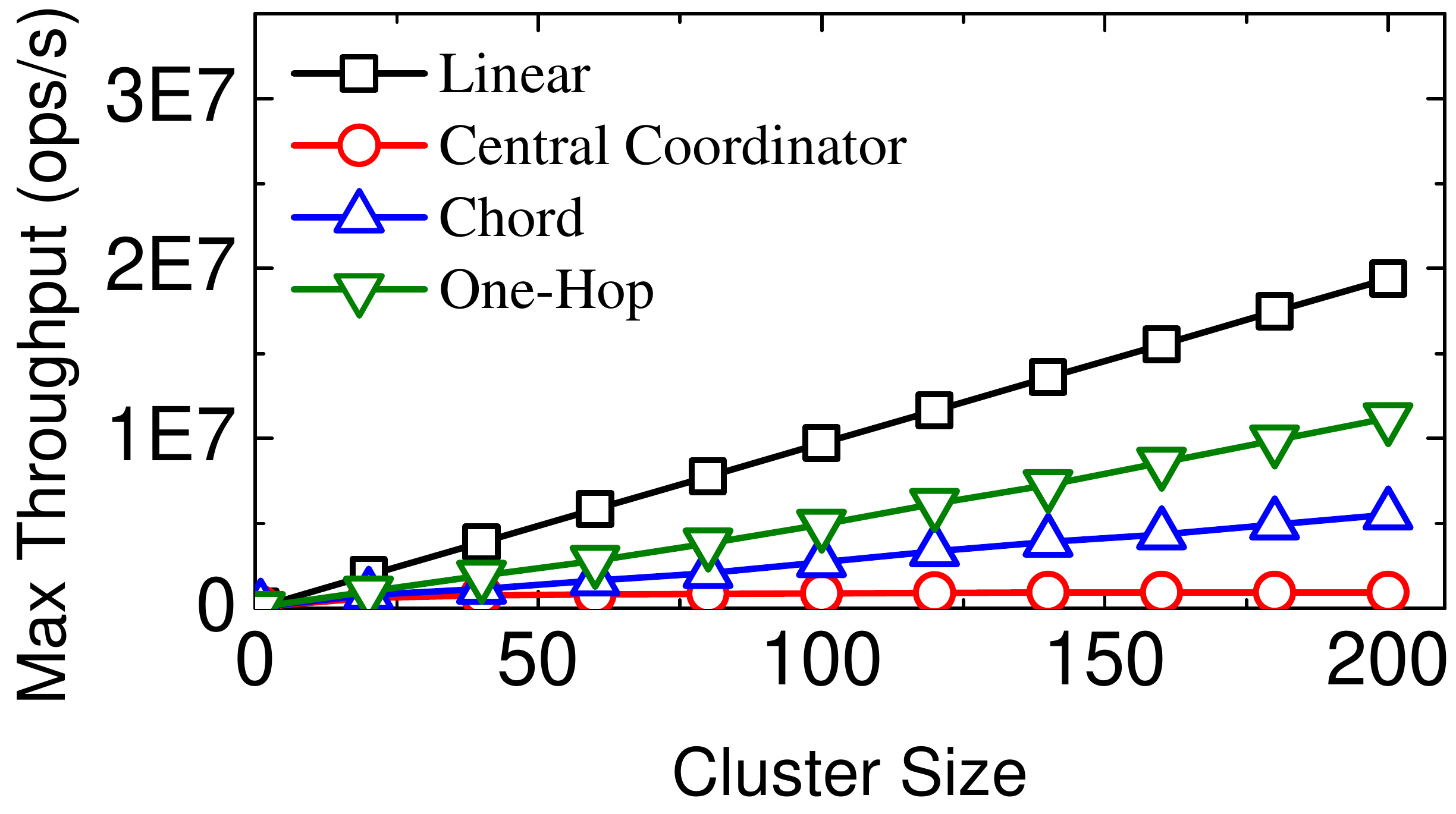}
    \subcaption{(d) Redis}
    \end{center}
    \end{minipage}
    \centering
    \vspace{-0.1 in}
    \caption{DHT-based metadata management systems' throughput comparison in a testbed with four types of storage subsystems.}
\label{Fig: throughput_dht}
\end{figure}

Figure \ref{Fig: throughput_dht} shows that DHT could cause large throughput reductions, where the throughput is defined as the maximum number of metadata operations that a metadata cluster can deal with.  DHT solves the single node performance bottleneck problem in Central Coordinator, and provides higher throughput. However, compared to the ideal system, Chord has roughly $70\%$ throughput reduction with $200$ Redis servers as shown in Figure \ref{Fig: throughput_dht} (a). The corresponding measure for One-Hop is $50\%$. When using LevelDB (HDD) and LevelDB (SSD), DHT-based systems still have no less than $20\%$ throughput reduction as shown in Figure \ref{Fig: throughput_dht} (b) (c). Even when we use a storage subsystem with low I/O throughput like MySQL (Figure 2 (a)), there is still roughly $10\%$ performance reduction for DHT-based approaches with $200$ servers.

\setlength{\minipagewidth}{0.235\textwidth}
\setlength{\figurewidthFour}{\minipagewidth}
\begin{figure}
    \centering
    \begin{minipage}[t]{\minipagewidth}
    \begin{center}
    \includegraphics[width=\figurewidthFour]{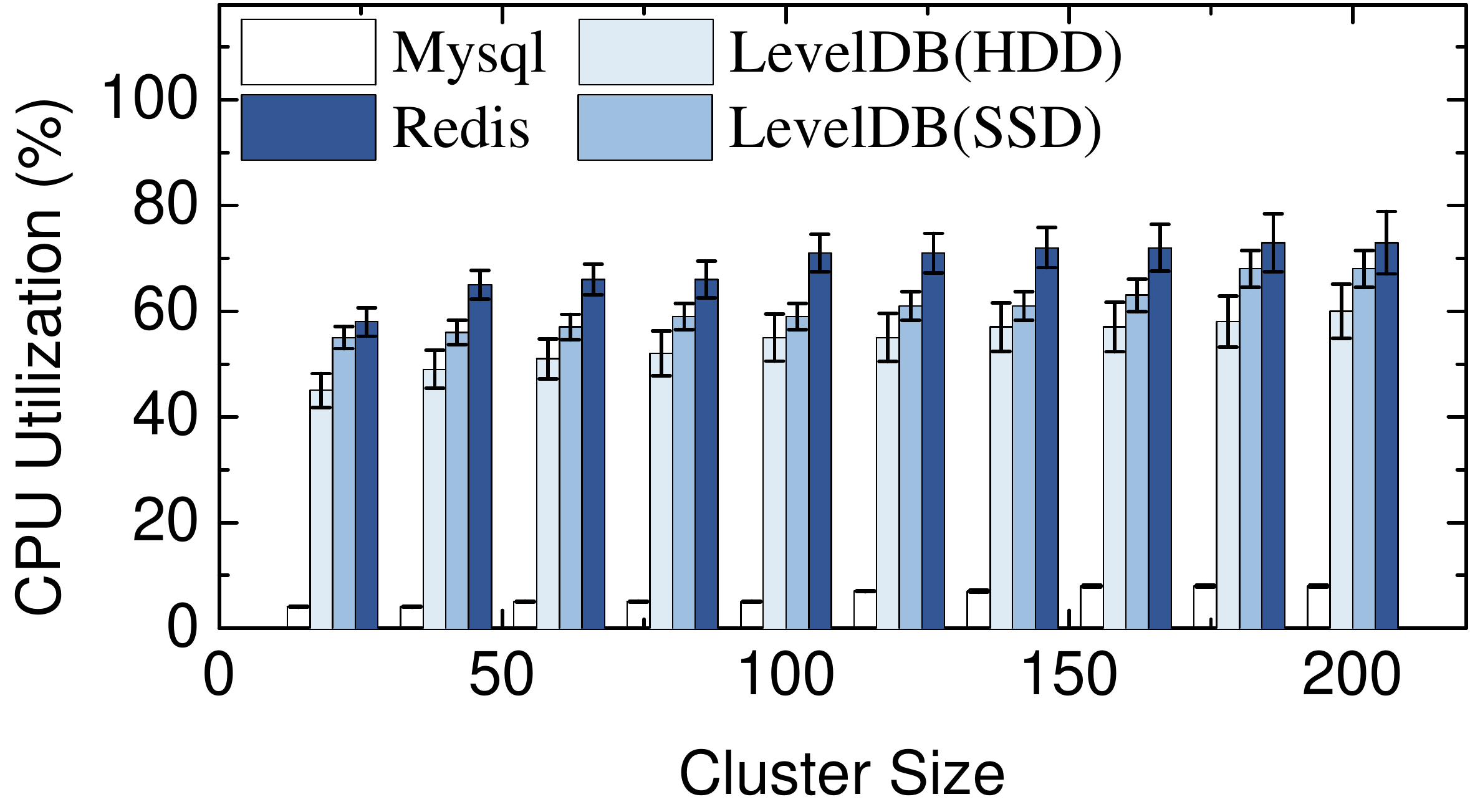}
    \subcaption{(a) Chord}
    \end{center}
    \end{minipage}
    \centering
    \begin{minipage}[t]{\minipagewidth}
    \begin{center}
    \includegraphics[width=\figurewidthFour]{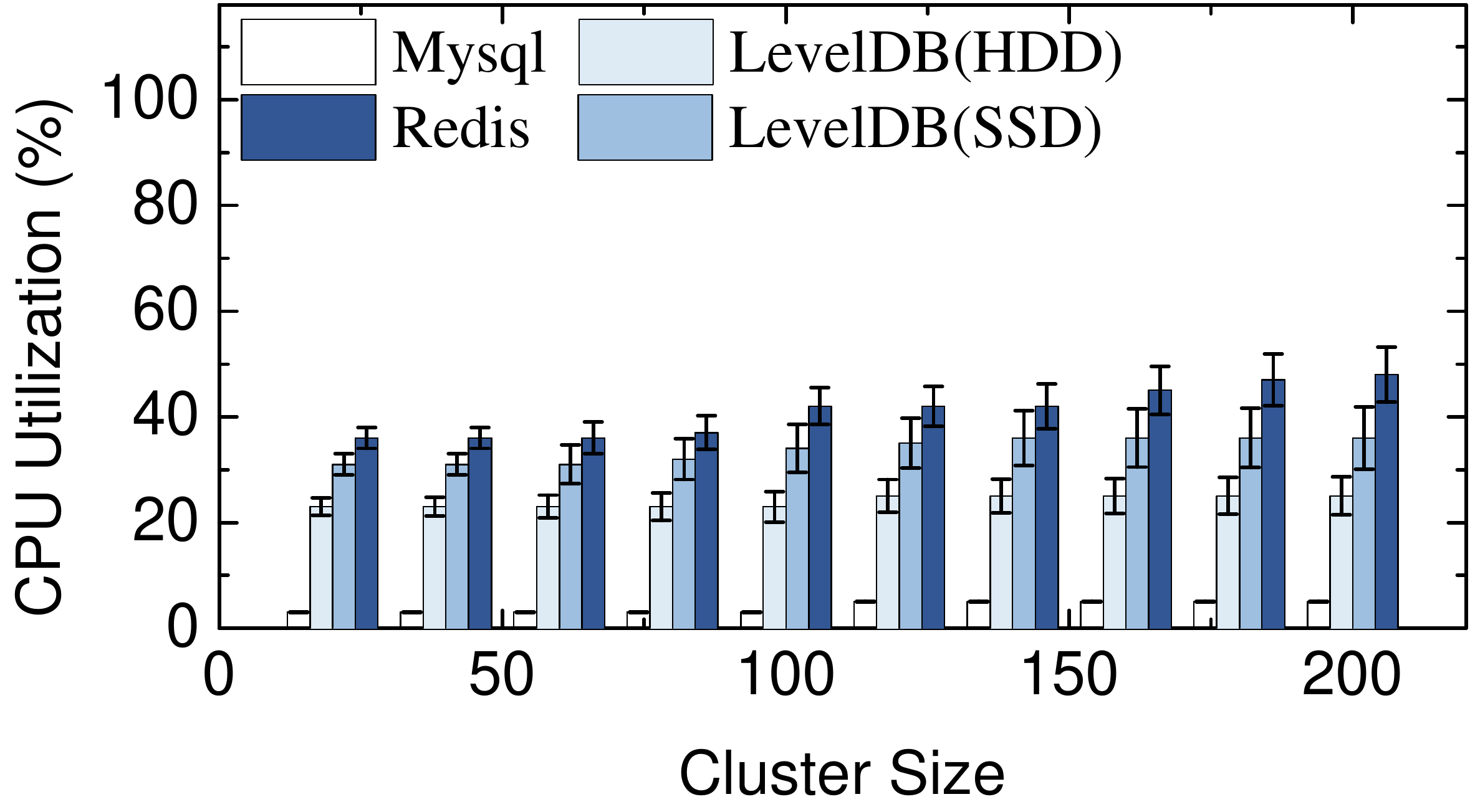}
    \subcaption{(b) One Hop}
    \end{center}
    \end{minipage}
    \centering
    \centering
    \vspace{-0.1 in}
    \caption{The lookup subsystem's CPU overhead in DHT-based metadata management systems using four types of storage subsystems.}
\label{Fig: lookup_dht_CPU}
\end{figure}

We carry out system profiling with \emph{Valgrind} \cite{valgrind} to identify the source of significant throughput reduction in DHT-based systems with LevelDB and Redis. Figure \ref{Fig: lookup_dht_CPU} shows that such reductions are mainly caused by high CPU cycle consumption of the lookup service. In the Chord-based system, the lookup subsystem consumes roughly $70\%$ of CPU cycles in our experiments with more than $100$ Redis nodes. The corresponding measures for LevelDB (HDD) and LevelDB (SSD) are $55\%$ and $60\%$, respectively. As a result, the storage subsystem might not have enough CPU resources to deal with I/O operations. Although the One-Hop based system has better performance, its lookup service still consumes roughly $40\%$, $35\%$, and $25\%$ of CPU cycles when using Redis, LevelDB (SSD), and LevelDB (HDD) as the storage subsystems with more than $100$ servers, respectively.

In summary, the lookup service could reduce the system throughput significantly in DHT-based systems, especially when used with memory intensive storage subsystems like Redis, because of the CPU resource competition.

\setlength{\minipagewidth}{0.235\textwidth}
\setlength{\figurewidthFour}{\minipagewidth}
\begin{figure} 
    \centering
    \begin{minipage}[t]{\minipagewidth}
    \begin{center}
    \includegraphics[width=\figurewidthFour]{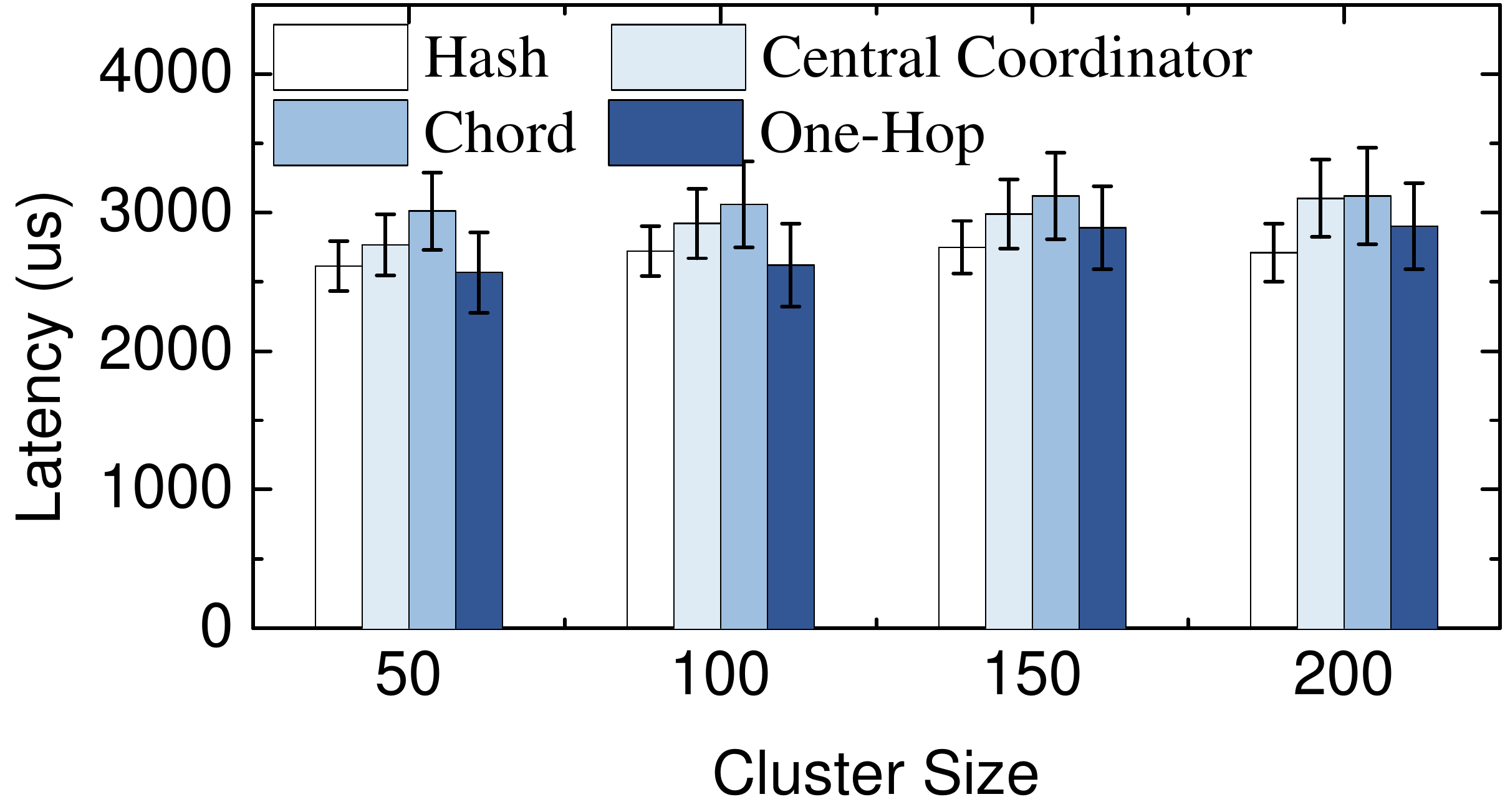}
    \subcaption{(a) MySQL (HDD)}
    \end{center}
    \end{minipage}
    \centering
    \begin{minipage}[t]{\minipagewidth}
    \begin{center}
    \includegraphics[width=\figurewidthFour]{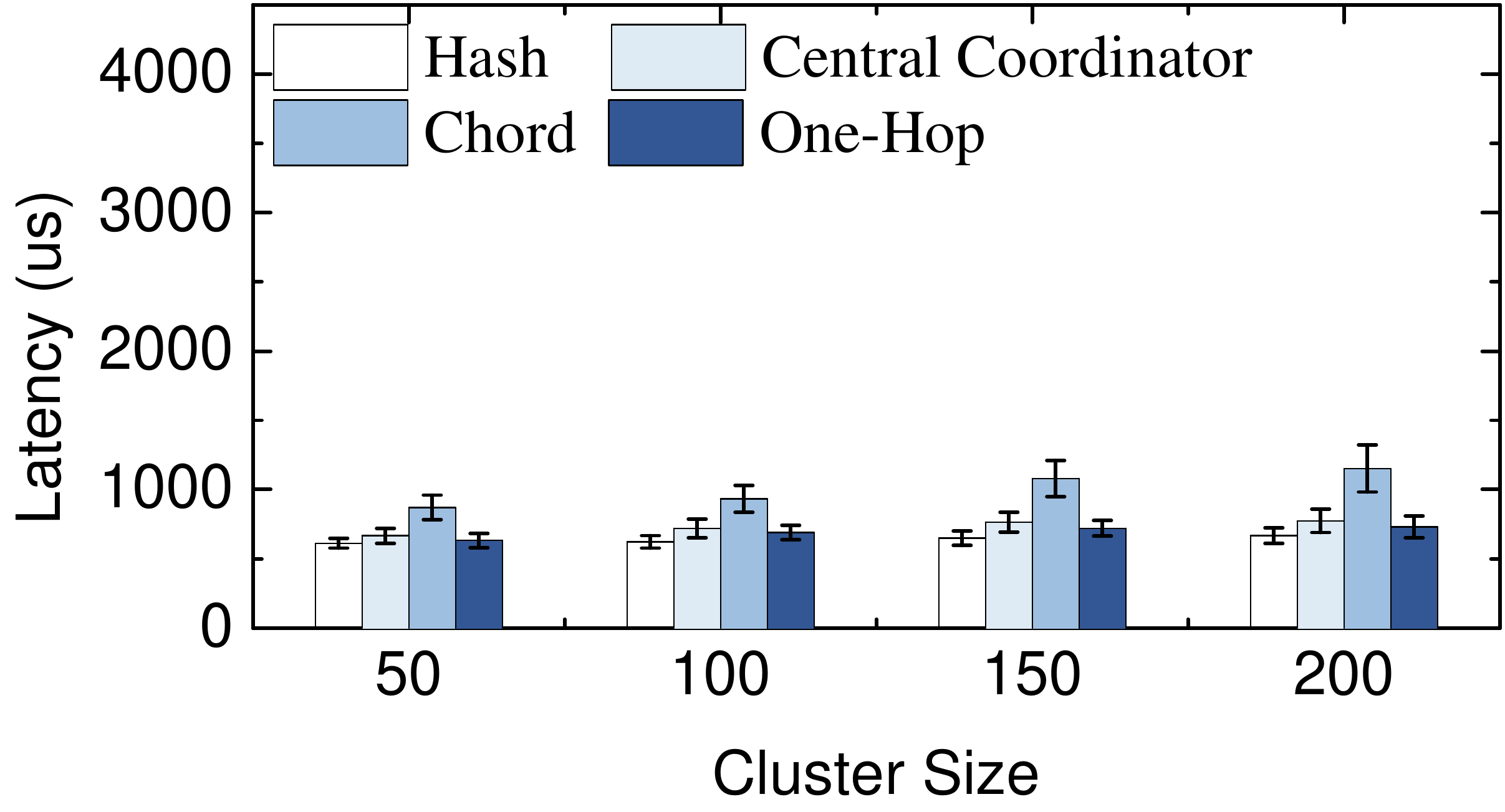}
    \subcaption{(b) LevelDB (HDD)}
    \end{center}
    \end{minipage}
    \centering
    \begin{minipage}[t]{\minipagewidth}
    \begin{center}
    \includegraphics[width=\figurewidthFour]{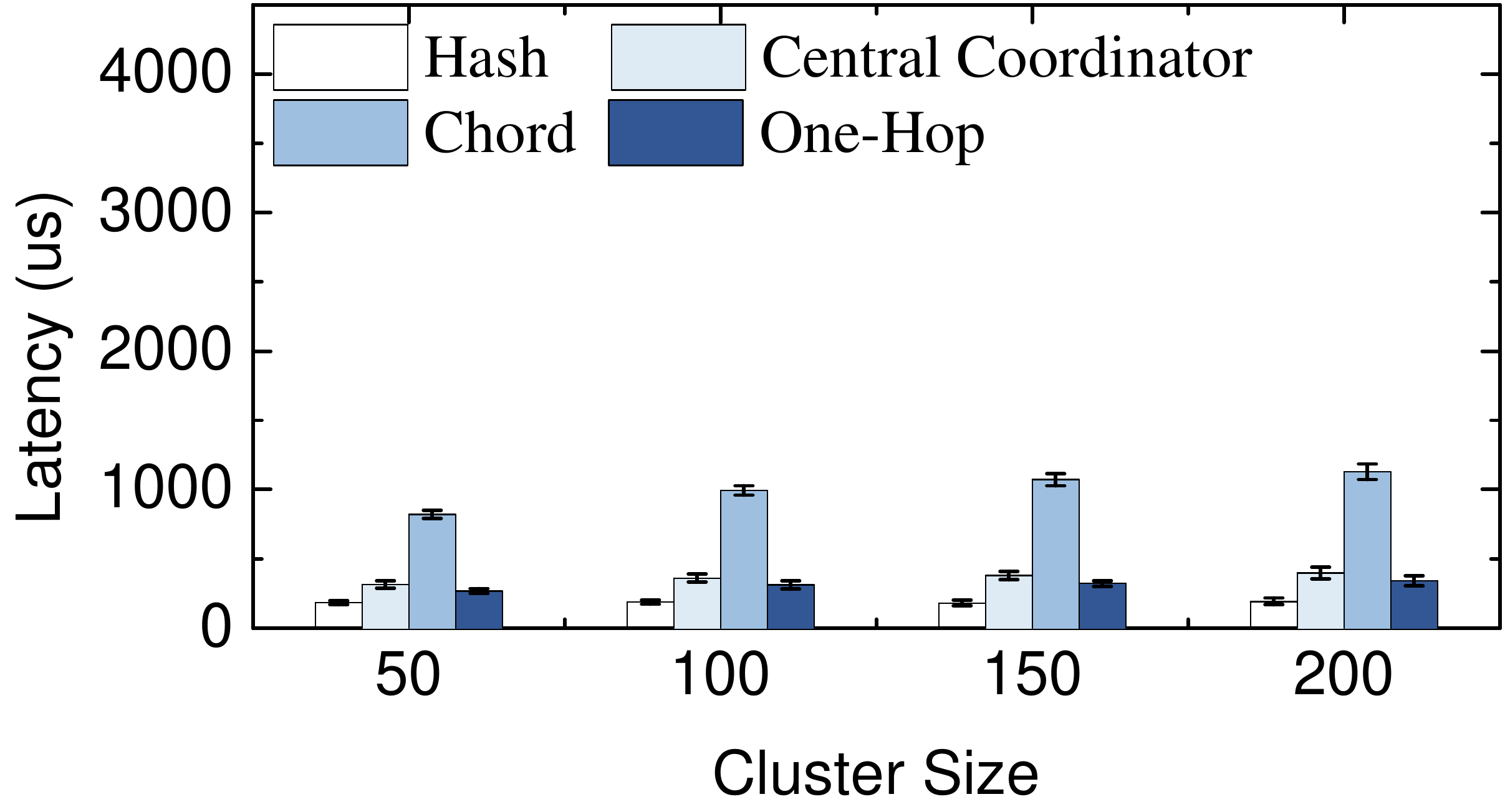}
    \subcaption{(c) LevelDB (SSD)}
    \end{center}
    \end{minipage}
    \centering
    \begin{minipage}[t]{\minipagewidth}
    \begin{center}
    \includegraphics[width=\figurewidthFour]{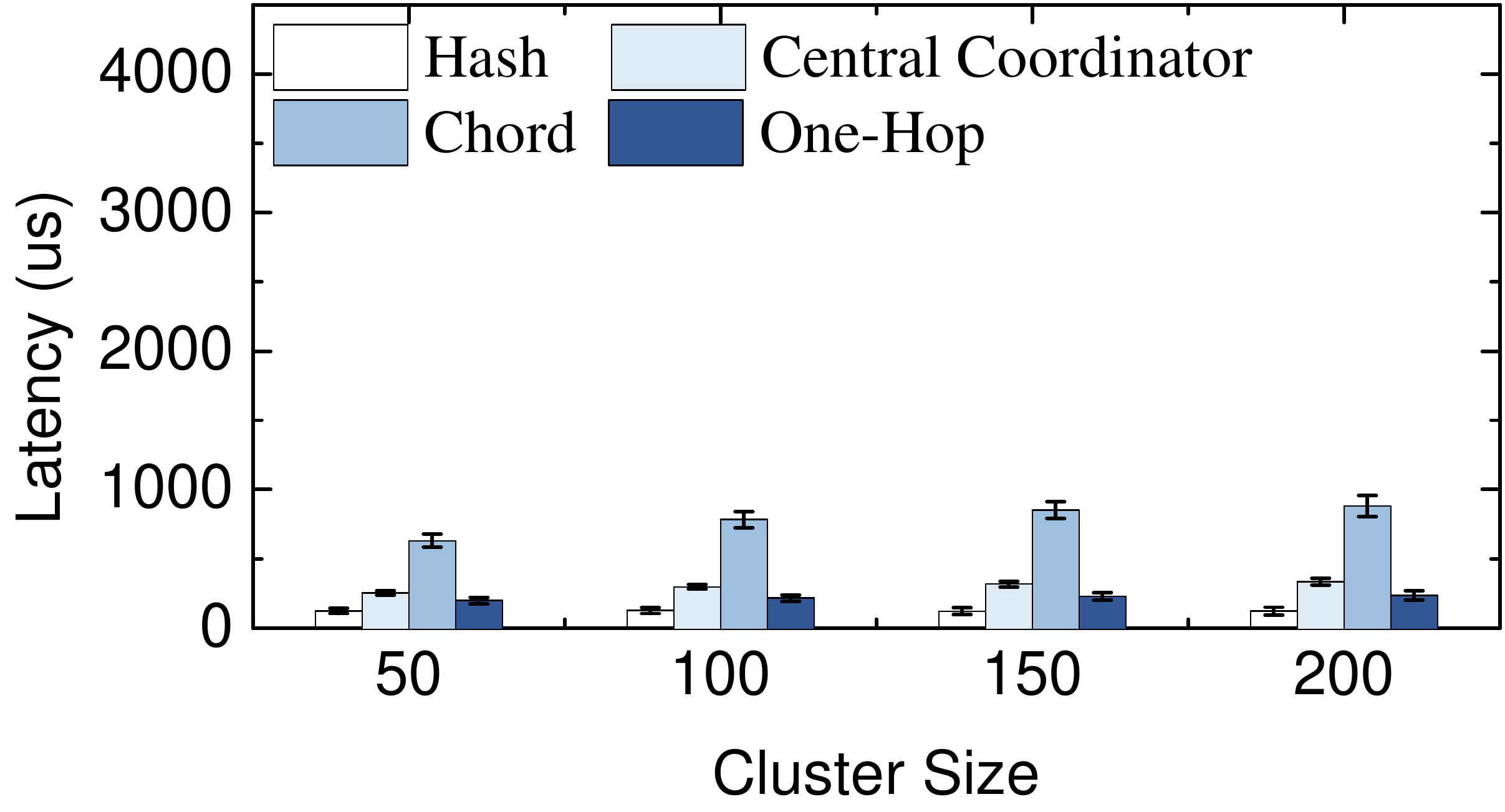}
    \subcaption{(d) Redis}
    \end{center}
    \end{minipage}
    \centering
    \vspace{-0.1 in}
    \caption{DHT-based metadata management systems' latency comparison in a testbed  with four types of storage subsystems.}
\label{Fig: latency_dht}
\end{figure}

\subsection {DHT: Latency}

Figure \ref{Fig: latency_dht} shows that DHT-based systems have high system latency, which is defined as the time used to complete a \emph{get} or \emph{put} metadata operation. In particular, the Chord-based system is about $8$ times slower than the hash-based system when using Redis as the storage subsystem as shown in Figure \ref{Fig: latency_dht} (d). The One-Hop-based system is much faster than the Chord-based system, but it is still $2$ times slower than the hash-based system. When using LevelDB (HDD) and LevelDB (SSD) as the storage subsystems, Chord and One-Hop also have obviously higher system latency than the hash-based system by a factor of at least $1.8$ and $1.3$, respectively, as shown in Figure \ref{Fig: latency_dht} (b) and (c). Figure \ref{Fig: latency_dht} (a) shows that DHT-based systems perform acceptably when used with MySQL. However, compared to other systems, storage systems using MySQL have much higher latency.

\setlength{\minipagewidth}{0.24\textwidth}
\setlength{\figurewidthFour}{\minipagewidth}
\begin{figure}
    \centering
    \begin{minipage}[t]{\minipagewidth}
    \begin{center}
    \includegraphics[width=\figurewidthFour]{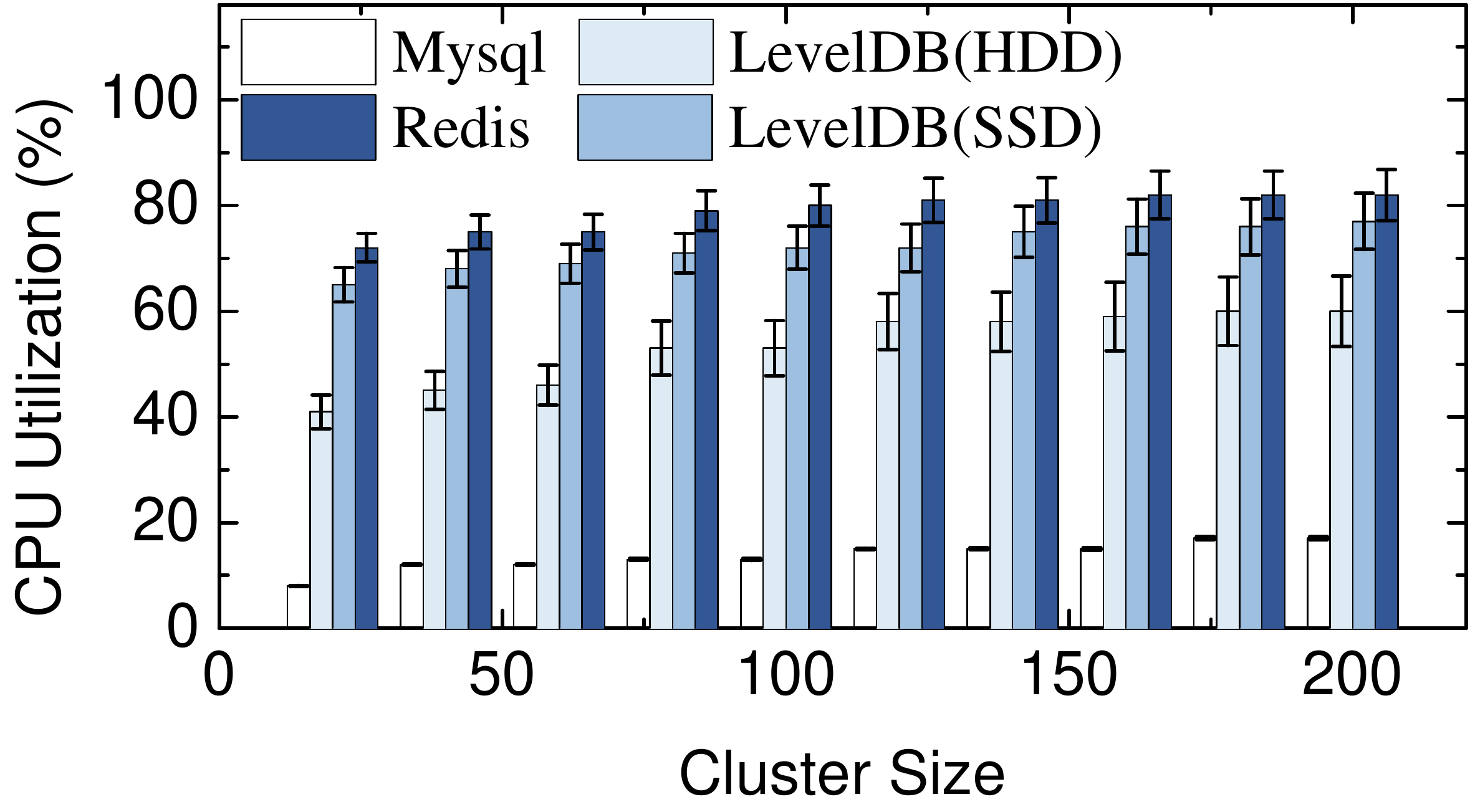}
    \subcaption{(a) Chord}
    \end{center}
    \end{minipage}
    \centering
    \begin{minipage}[t]{\minipagewidth}
    \begin{center}
    \includegraphics[width=\figurewidthFour]{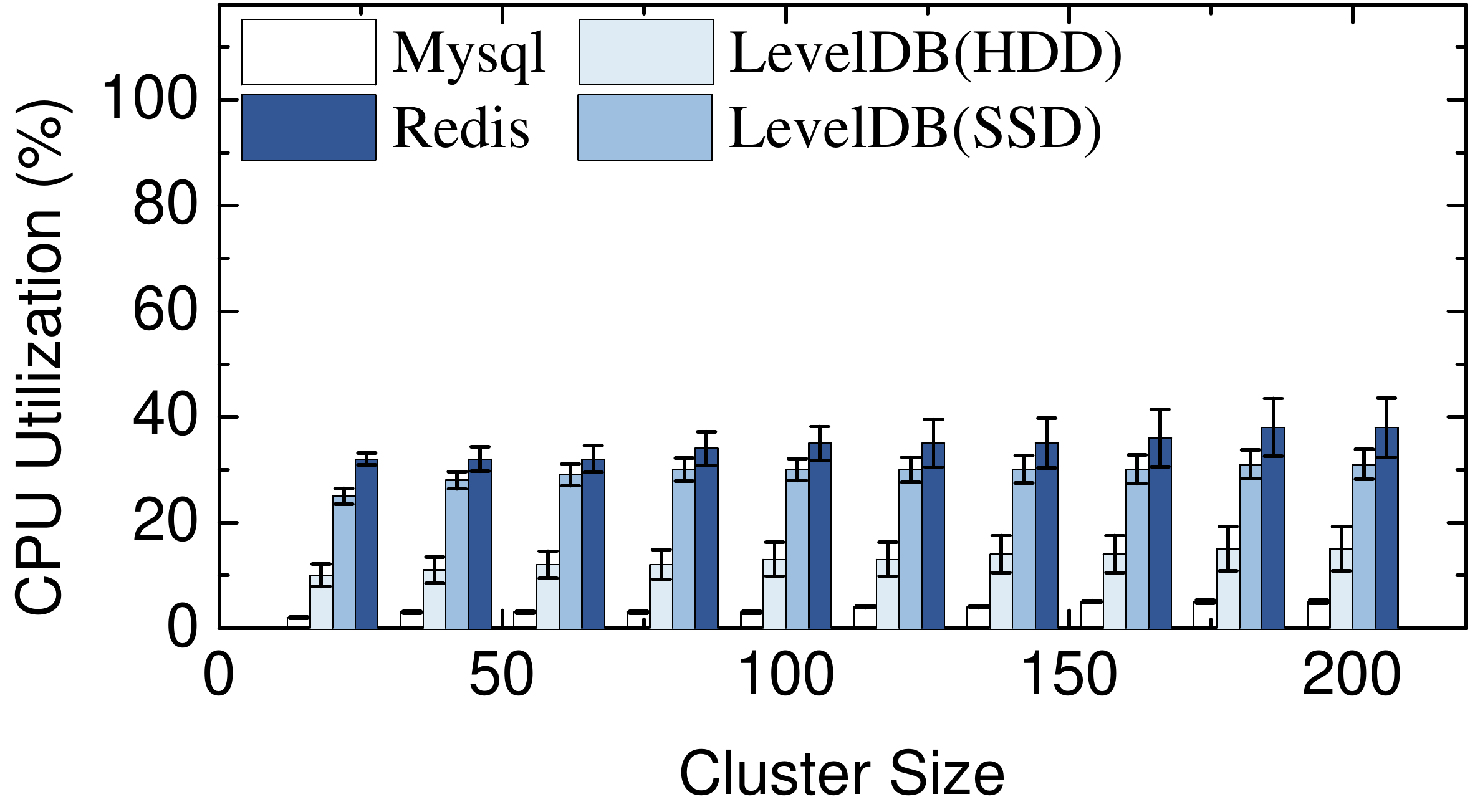}
    \subcaption{(b) One Hop}
    \end{center}
    \end{minipage}
    \centering
    \centering
    \vspace{-0.1 in}
    \caption{The lookup subsystem's latency overhead in DHT-based metadata management systems using four types of storage subsystems.}
\label{Fig: lookup_dht_time}
\end{figure}

\setlength{\minipagewidth}{1\textwidth}
\setlength{\figurewidthFour}{\minipagewidth}
\begin{figure*}[htb]
    \centering
    \includegraphics[width=\figurewidthFour]{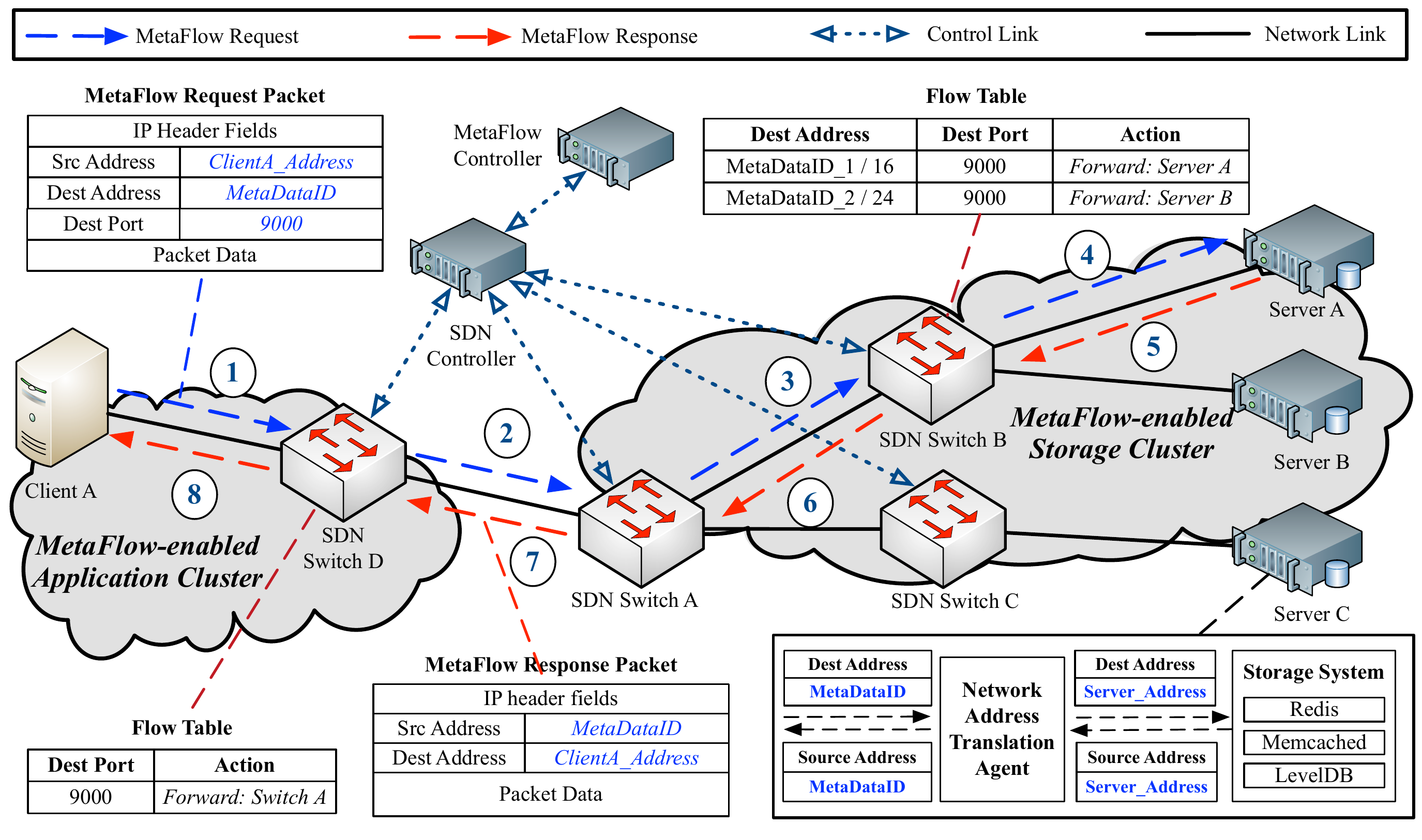}
    \caption{A MetaFlow-based distributed metadata management system architecture. There are two clusters in the system: the storage cluster hosts a set of storage servers to provide metadata service with a tree topology; and the application cluster manages a set of clients to query metadata objects. MetaFlow harnesses the capacity of SDN to forward \emph{MetaFlow Requests} to their associated storage servers using \emph{MetaDataIDs} as the destination IP addresses. Thus, there is no separate lookup operation for a metadata operation in MetaFlow.}
\label{Fig: metaflow_system}
\end{figure*}

Profiling results from  \emph{Valgrind}  in Figure \ref{Fig: lookup_dht_time} show that the high system latency in DHT-based systems is mainly caused by the lookup operation. In our experiments, the lookup operation in the Chord-based system could account for $72\%$ to $84\%$ of the total system latency when using Redis as the storage subsystem. Although the One-Hop-based system has better performance, its lookup operation still takes roughly $35\%$ of the total system latency. When using LevelDB (HDD) as the storage subsystem, the lookup operation takes at least $40\%$ and $10\%$ of the total system latency for the Chord-based system and the One-Hop-based system, respectively.

In summary, the lookup service adds considerable latency into metadata operations in DHT-based systems. The main reason is that a lookup operation needs to invoke at least one remote procedure call (RPC) on storage servers. Based on Chord's properties, the Chord-based system invokes $\log_{2}{M}$ RPCs on average to locate a metadata object in a $M$-node cluster. The  One-Hop-based system needs to one RPC per metadata operation.

\section{MetaFlow: Objective and Design}

We propose MetaFlow to solve the performance bottleneck caused by lookup operations in existing DHT-based approaches for the metadata management. This section describes the design objective, system architecture, and the lookup workflow in MetaFlow.

\subsection{Objective}

We design MetaFlow to provide a fast lookup service with minimal overhead incurred on metadata storage servers. Essentially,  MetaFlow maps a \emph{MetaDataID}, which is the hash value of a metadata object's file name, to the location of a server storing this metadata object. Compared to existing overlay-based approaches,
MetaFlow has three key features:
\begin{enumerate}
\item[\textbullet] \emph{\textbf{In-Network Lookup}}: MetaFlow places the lookup workload on network components instead of metadata storage servers. More specifically, it takes advantage of SDN-enabled switches to send metadata requests directly to storage servers using just \emph{MetaDataIDs} instead of \emph{IP} or \emph{MAC} addresses. As a result, MetaFlow could avoid the CPU resource competition problem between the lookup subsystem and the storage subsystem in conventional DHT-based approaches.

\item[\textbullet] \emph{\textbf{Compatibility}}: MetaFlow is compatible with existing lookup services and network infrastructures. MetaFlow can detect its packets according to the destination TCP port, and process them using specific rules. Other network packets will be forwarded normally using existing layer 2/3 switching techniques. 

\item[\textbullet] \emph{\textbf{Zero-Hop}}: MetaFlow does not have  a separate step to fetch the location of the desired metadata object. MetaFlow allows the client to establish a network connection directly to a metadata storage server to perform I/O operations such as \emph{get}, \emph{put}, \emph{update}, \emph{delete}, etc.  with \emph{MetaDataIDs}. In the following example, we illustrate how this could be done using a TCP connection for four basic metadata operations:
\end{enumerate}

\begin{lstlisting}
MetaDataOperation(FileName, Method){
    MetaDataID=Hash(FileName)
    Connection=TCPConnect(MetaDataID, Port)
    Connection.request(Method, ...)
    Connection.getresponse()
    Connection.close()
}
\end{lstlisting}

It should be noted that we focus on the lookup service in this paper. MetaFlow could also support more features for distributed metadata management such as fault tolerance, load balancing, and POSIX directory access optimization. In particular, MetaFlow could leverage  SDN-based approaches \cite{jarschel2014interfaces}, \cite{xia2015survey} to achieve fault tolerance and load balancing. Other approaches such as  GFS\cite{ghemawat2003google}, metadata caching system \cite{caesar2010efficient}, and LH \cite{brandt2003efficient} could be used in MetaFlow to improve the directory access performance.

\subsection{System Design}

MetaFlow contains four key components: \emph{Storage Cluster}, \emph{Application Cluster}, \emph{SDN-based Networking}, and \emph{MetaFlow Controller}.

\subsubsection{\textbf{Storage Cluster}}

The storage cluster consists of a set of storage servers and their associated switches. In this work, we manage all storage servers using a tree topology such as the tier tree \cite{bari2013data} or the fat tree \cite{aguilera2008practical}. Each storage server has two subsystems: (i) a  high performance key-value storage system such as Redis or LevelDB to maintain metadata objects; and (ii) a network address translation (NAT) agent to manage source and destination IP addresses for MetaFlow requests and responses.

\subsubsection{\textbf{Application Cluster}}

The application cluster manages a set of clients. For example, it can be a MapReduce cluster, which queries metadata objects for HDFS related operations. It should be noted that the storage and application cluster could either be located in the same physical cluster or two separated physical clusters. 

\subsubsection{\textbf{SDN-based Networking}}

MetaFlow employs SDN to realize its design objective. SDN is one of the recent approaches to programmable networks.  Based on the fact that the basic function of a switch is to forward packets according to a set of rules \cite{lara2014network}, SDN decouples the control and data planes of a network. A centralized software-based \emph{SDN controller} manages the rules for the switch to forward packets.

\noindent
{\textbf{OpenFlow. }} MetaFlow uses OpenFlow \cite{mckeown2008openflow} as the standard for SDN. There are three components in an OpenFlow architecture: an OpenFlow-enabled switch, which uses flow entries to forward packets; an OpenFlow controller, which manages flow tables; and a secure channel, which connects the controller to all switches. A packet is examined with regard to the flow entries by using one or more its header fields. If there is a match, the packet is processed according to the instruction in the flow entry. If not, the packet is sent to the OpenFlow controller for further processing. OpenFlow 1.0.0 is one of the most widely used specifications \cite{lara2014network}. It  supports $12$ header fields, which include Source/Destination IP Address, Source/Destination TCP/UDP Port, etc. MetaFlow uses destination IP address and destination TCP port to forward packets.

\noindent
\textbf{{Network Packet Format. }} MetaFlow  packets are normal IP packets. \emph{MetaFlow Request} is the packet sent by clients to query a metadata object with the \emph{MetaDataID}, and \emph{MetaFlow Response} is the packet sent from storage servers with the desired metadata object as the packet's content. MetaFlow packets differ from common TCP packets in two aspects: destination IP address and destination TCP port:

\begin{enumerate}
\item[\textbullet]  {\emph{Destination IP Address}}: MetaFlow uses the \emph{MetaDataID}, which is the hash value of a metadata object's file name, as the \emph{MetaFlow Request's} destination IP address. Depending on the IP protocol in use, \emph{MetaDataID} could have a different length. \emph{MetaDataID} is a $32$-bit integer when using \emph{IPv4}. In \emph{IPv6}, \emph{MetaDataID} is a $128$-bit integer.
\item[\textbullet] {\emph{Destination TCP Port}}: MetaFlow uses the destination TCP port to distinguish MetaFlow packets from other packets in the cluster. As shown in Figure \ref{Fig: metaflow_system}, the \emph{MetaFlow Request} uses $9000$ as its destination TCP port. SDN-enabled switches can detect \emph{MetaFlow Requests} according to the destination TCP port, and process them using the appropriate flow tables. Other packets will be relayed normally using existing layer 2/3 switching techniques. It should be noted that normal network packets except MetaFlow packets should not use $9000$ as their destination TCP ports in the example in Figure \ref{Fig: metaflow_system}. Otherwise, they will not be forwarded properly.
\end{enumerate}


\subsubsection{\textbf{MetaFlow Controller}}

The \emph{MetaFlow Controller} is in charge of generating and maintaining flow tables for SDN-enabled switches to forward metadata packets.

\subsection{MetaFlow Packets Forwarding}

MetaFlow leverages SDN-enabled switches to forward MetaFlow packets properly. Specifically, it relays \emph{MetaFlow Requests} to corresponding storage servers based  on \emph{MetaDataIDs}. It also forwards  \emph{MetaFlow Responses} back to clients.

\subsubsection{\textbf{Forward MetaFlow Requests}}

As shown in Figure \ref{Fig: metaflow_system}, MetaFlow forwards a \emph{MetaFlow Request} packet via following three steps:

\begin{enumerate}
\item[\textbullet] \emph{From the application cluster to the storage cluster.} The SDN-enabled switch in the application cluster forwards the \emph{MetaFlow Request} based on the destination TCP port. As shown in Figure \ref{Fig: metaflow_system}, \emph{SwitchD} is configured to forward the \emph{MetaFlow Request}, whose destination TCP port is $9000$, to \emph{SwitchA}, which is located in the storage cluster.

\item[\textbullet] \emph{From the storage cluster to the storage server.}
The SDN-enabled switch in the storage cluster forwards \emph{MetaFlow Requests} based on both the destination TCP port and the destination IP address. As shown in Figure \ref{Fig: metaflow_system}, when receiving  packets from \emph{SwitchA}, \emph{SwitchB} recognizes the \emph{MetaFlow Request} based on the destination TCP port, and compares its \emph{MetaDataID} against its flow table based on a longest prefix match algorithm. According to the instruction from the matched flow entry, \emph{SwitchB} forwards the \emph{MetaFlow Request} to \emph{ServerA}.

\item[\textbullet] \emph{From the network layer to the application layer in the metadata storage server.}
The server in the storage cluster forwards the \emph{MetaFlow Request} to the application layer from the network layer. Normally, the storage server will drop the received \emph{MetaFlow Request}, since its destination address is the \emph{MetaDataID} rather than the server's IP address. To solve this problem, MetaFlow deploys a network address translation (NAT) agent on each storage server to replace the \emph{MetaFlow Request's} destination address with its physical IP address. Therefore, the application layer in the storage server could receive and process \emph{MetaFlow Requests}.
\end{enumerate}

\subsubsection{\textbf{Forward MetaFlow Responses}}

\emph{MetaFlow Responses} are relayed back to clients based on clients' physical IP addresses. Normally, the storage server will put its physical IP address in the \emph{MetaFlow Response's} source address field. However,  the client will drop these \emph{MetaFlow Responses}, since the requests' destination address differ from the responses' source address. For example, a client sends out a request using the \emph{MetaDataID} 155.69.146.43 as the destination IP address. The corresponding storage server sends back the response using its  physical IP address 192.168.0.1 as the source IP address. In this case, the client will drop the response, since it expects a response from 155.69.146.43 rather than 192.168.0.1. To solve this problem, MetaFlow uses the storage server's NAT agent to replace the \emph{MetaFlow Response's} source address field with the original \emph{MetaDataID} before sending out the response.

\section{MetaFlow: Flow Table Generation}

The central problem in implementing MetaFlow is how to generate flow tables for the SDN-enabled switches in the storage cluster. Since MetaFlow places the lookup workload on network components, we have to generate flow tables for the SDN-enabled switches in both the application cluster and the storage cluster. Flow tables for the application cluster's switches could be generated easily, since these switches forward \emph{MetaFlow Requests} to a pre-determined destination, which is the storage cluster. However, in the storage cluster, the \emph{MetaFlow Requests'} destination storage servers are not known in advance. Therefore, existing IP-based routing protocols cannot work any more. In this section, we describe the flow table generation  algorithm in MetaFlow. 

MetaFlow generates appropriate flow tables based on a logical B-tree data structure, which is mapped from the physical tree network topology. More specifically, the \emph{MetaFlow Controller} maps the storage cluster's  physical network topology, which could be a tier tree \cite{bari2013data} or fat tree \cite{aguilera2008practical}, to a logical B-tree data structure. Using the B-tree's property, MetaFlow then distributes the metadata objects across storage servers, and generates appropriate flow tables for SDN-enabled switches. 

\subsection{Physical Tree Topology}

MetaFlow has to be able to work with different tree topologies in data centers. There are two widely used tree topologies: tier tree  and fat tree.

A tier tree network consists of two or three layers of network switches. A three-tier tree network contains an edge layer, connecting servers via top of rack (ToR) switches; an aggregation layer, using end of rack (EoR) switches to connect ToR switches; and a core layer at the root of the tree. There is no aggregation layer in a two-tier tree network.

A fat tree network is an extended version of the three-tier tree network. \emph{Pod} is the basic cell of a fat tree network. Assume that each switch has $n$ switch ports in a fat tree network, a \emph{Pod} consists of $n/2$ aggregation layer switches, $n/2$ edge layer switches, and their connected servers. Therefore, each edge layer switch connects $n/2$ aggregation layer switches, and each core layer switch connects $n/2$ core switches in the fat tree network.

\subsection{Logical B-tree}

\setlength{\minipagewidth}{0.48\textwidth}
\setlength{\figurewidthFour}{\minipagewidth}
\begin{figure}
    \centering
    \begin{minipage}[t]{\minipagewidth}
    \begin{center}
    \includegraphics[width=\figurewidthFour]{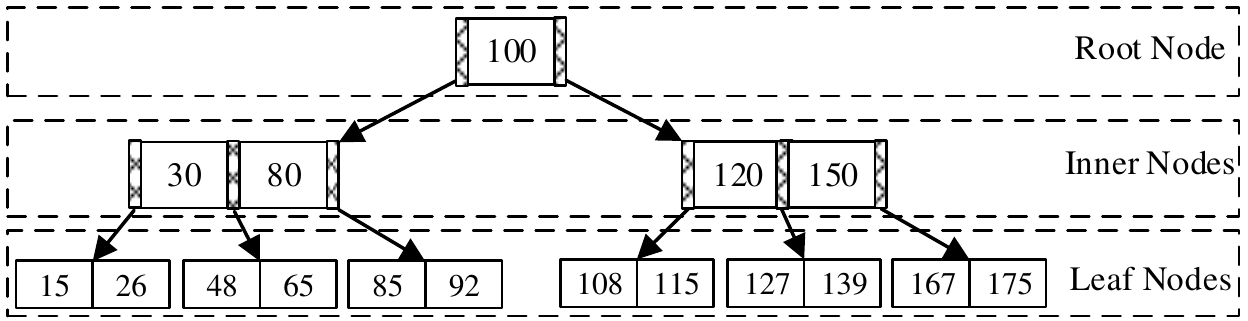}
    \end{center}
    \end{minipage}
    \centering
    \caption{A B-tree example.}
\label{Fig: B-tree}
\end{figure}

B-tree is a self-balancing tree data structure, which has two key features: 1) it distributes the key-value data across its nodes in a balanced manner, 2) it allows lookup operations in logarithmic time. A B-tree is made up of three types of nodes: leaf node, inner node and root node. We adopt the widely-accepted definition by Comer~\cite{comer1979ubiquitous}, where these nodes construct a sample B-tree as shown in Figure \ref{Fig: B-tree}.

A B-tree stores key-value pairs in its nodes in non-decreasing order of the keys' values. The key-value pair stored in the non-leaf node also acts as the partition value to separate the subtree. 
To search a key, the B-tree is recursively traversed from the top to the bottom starting at the root node. At each level, the search algorithm chooses the subtree according to the comparison result between the desired key and stored partition values.  For example, in Figure \ref{Fig: B-tree}, if a client queries the key \emph{65}, it will choose the left subtree at the root node. Then it chooses the middle subtree at the inner node. Finally, the client can fetch the value of key \emph{65} at the leaf node.

\subsection{Mapping Physical Tree Topology to Logical B-tree}

MetaFlow uses a logical B-tree to manage the storage cluster. In order to do that, MetaFlow first discovers the physical storage cluster's topology through OpenFlow, and then carries out a mapping operation from the physical topology, which includes storage servers, SDN-enabled switches, and network links, to a logical B-tree. This is done in the \emph{MetaFlow Controller} via the following steps:

\begin{enumerate}
\item[\textbullet] Storage servers are mapped to the B-tree's leaf nodes.
\item[\textbullet] The core switch is mapped to the B-tree's root node.
\item[\textbullet] The aggregation and edge switches are mapped to the inner nodes in the B-tree based on the layers that they are in.
\item[\textbullet] Network links are mapped to the logical connections between parent nodes and child nodes in the B-tree.
\end{enumerate}

This mapping strategy works for both the tier tree network and the fat tree network. It is straightforward to map a tier tree network to a B-tree as shown in Figure \ref{Fig: map_tier_tree_topology}, since they are quite similar in terms of structure. To map a fat tree network to a B-tree, MetaFlow might need to map multiple switches to one B-tree node. For example,  Figure \ref{Fig: map_fat_tree_topology} shows a fat-tree, in which all switches have $4$ switch ports. In this fat tree, a \emph{pod} contains $2$ aggregation layer switches, $2$ edge layer switches and $4$ storage servers. There are $4$ core switches. To map this fat tree to a B-tree, the $4$ core layer switches are mapped to one B-tree root node. The $2$ aggregation layer switches in the same \emph{Pod} are mapped to one inner node. Edge layer switches and storage servers are mapped to inner nodes and leaf nodes, respectively.

MetaFlow uses two states, which are \emph{idle} and \emph{busy} states, to simulate the B-tree node creation operation. The classical B-tree can create new nodes dynamically for node split operations. MetaFlow uses \emph{idle} and \emph{busy} states to simulate this operation. In the \emph{idle} state, the physical node, which can be a storage server or a SDN-enabled switch, contains no data. The \emph{busy} state means that the physical node manages some keys. For example, in Figure \ref{Fig: map_fat_tree_topology}, there is no data stored in the storage cluster initially. In this case, all the nodes are in the \emph{idle} state. When some key-value pairs are inserted, some nodes' states are transformed into \emph{busy}. When a node is full, the mapped B-tree activates an \emph{idle} node to store roughly half of the full node's data.

\setlength{\minipagewidth}{0.48\textwidth}
\setlength{\figurewidthFour}{\minipagewidth}
\begin{figure}
    \centering
    \begin{minipage}[t]{\minipagewidth}
    \begin{center}
    \includegraphics[width=\figurewidthFour]{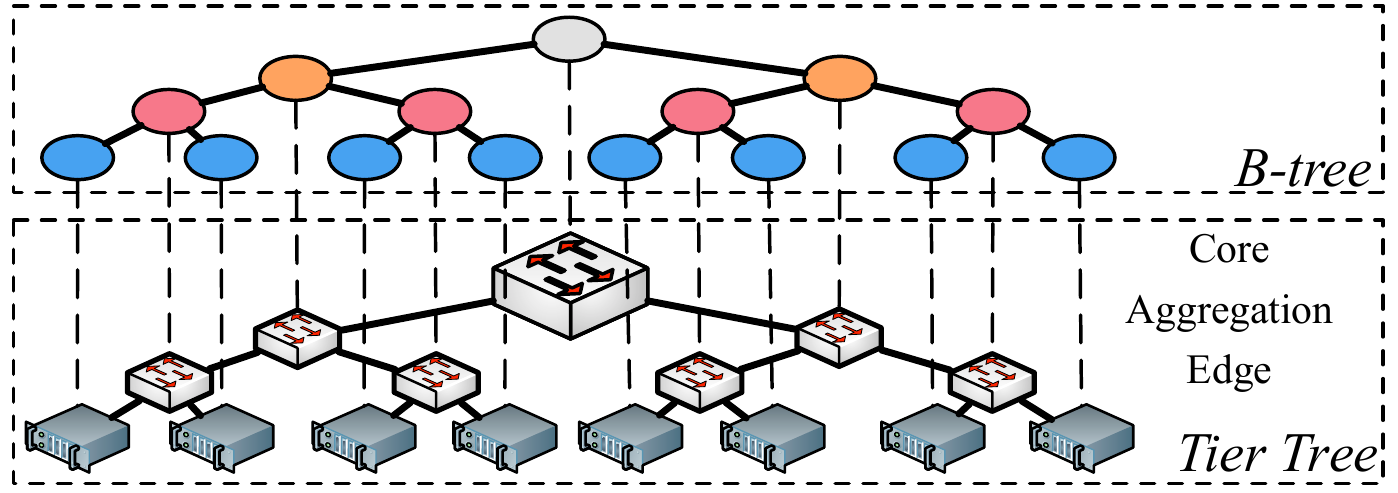}
    \end{center}
    \end{minipage}
    \centering
    \caption{Mapping a three-tier tree network to a logical B-tree. }
\label{Fig: map_tier_tree_topology}
\end{figure}

\setlength{\minipagewidth}{0.48\textwidth}
\setlength{\figurewidthFour}{\minipagewidth}
\begin{figure}
    \centering
    \begin{minipage}[t]{\minipagewidth}
    \begin{center}
    \includegraphics[width=\figurewidthFour]{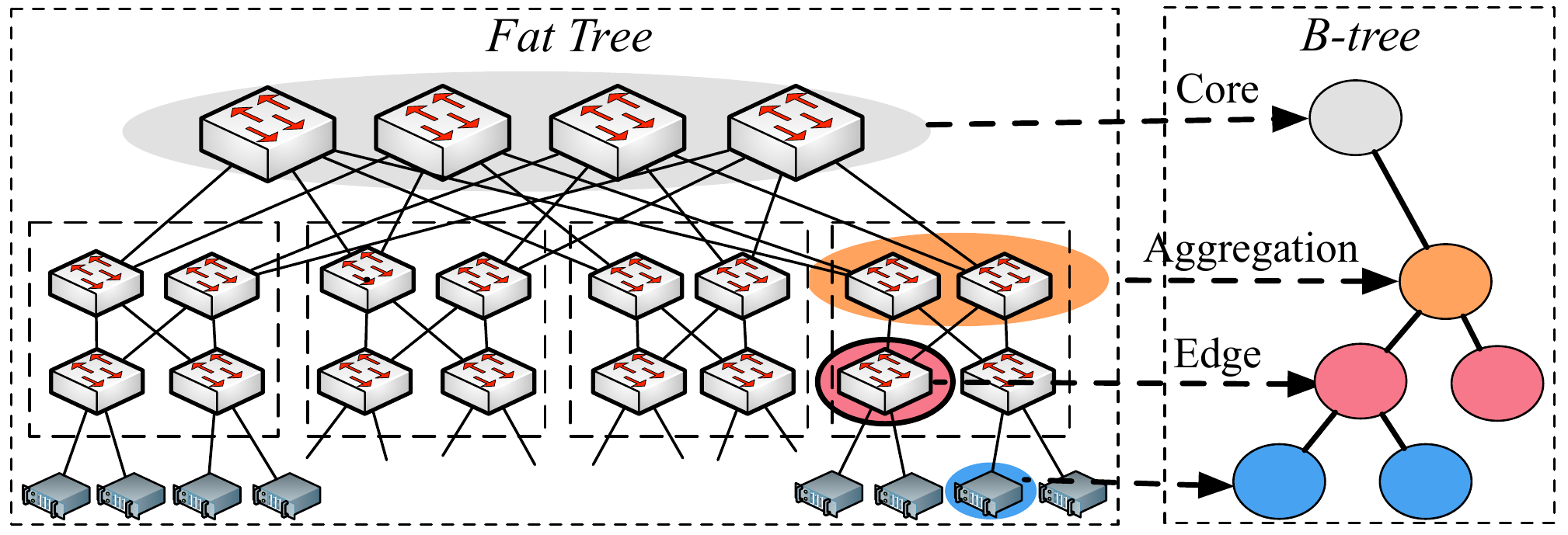}
    \end{center}
    \end{minipage}
    \centering
    \caption{Mapping a fat tree network to a logical B-tree.}
\label{Fig: map_fat_tree_topology}
\end{figure}

Because of the limitations from the physical components, the mapped B-tree has the following properties:
\begin{enumerate}
\item[\textbullet]  { All the key-value pairs are stored in the  leaf nodes.} The root node and inner nodes only store keys to partition subtrees without associated values. The reason is that most switches do not have key-value data storage capacity.

\item[\textbullet]  { Leaf nodes have much higher storage capacity than the non-leaf nodes.} It is common for a storage server to manage millions of key-value pairs. However, most SDN-enabled switches can only support a few thousands of flow entries, which are used to store partition values.

\item[\textbullet]  { The mapped B-tree's depth is a fixed value.} In a two-tier tree, the mapped B-tree's depth is $3$. In a fat tree or a three-tier tree, the mapped B-tree's depth is $4$.
\end{enumerate}

\subsection{Generating Flow Table}

MetaFlow transforms the  B-tree's partition values to SDN-enabled switches' flow tables. The key challenge of this transformation is how to represent the partition values in a format that can be recognized by  SDN-enabled switches. Since state-of-the-art OpenFlow-enabled switches only support longest prefix matching algorithm to deal with the destination IP address field \cite{lara2014network}, MetaFlow uses Classless Inter-Domain Routing (CIDR) \cite{fuller1993classless} blocks to represent partition values in the B-tree.

A CIDR block is a group of IP addresses with the same routing prefix. For example, the IPv4 CIDR block $155.69.146.0/24$ represents $256$ IPv4 addresses from $155.69.146.0$ to  $155.69.146.255$. IPv6 also works under CIDR. The IPv6 CIDR block 2001:db8::/48 represents the block of IPv6 addresses from 2001:db8:0:0:0:0:0:0 to 2001:db8:0:ffff:ffff:ffff:ffff:ffff. An OpenFlow-enabled switch can use a CIDR block as its forwarding table entry. For example, the entry ``$155.69.146.0/24 \to 192.168.0.1$'' means that a packet will be forwarded to $192.168.0.1$ if its destination IP address ranges from $155.69.146.0$ to $155.69.146.255$.

MetaFlow uses CIDR blocks to map the B-tree's partition values to SDN-enabled switches' flow tables. As shown in Figure \ref{Fig: CIDR_partition}, \emph{SwitchC} splits the set of all metadata objects into two partitions, which are $0.0.0.0/1$ and $128.0.0.0/1$, using $128.0.0.0$ as the partition value in IPv4 format. In this case, \emph{SwitchA} is responsible for metadata objects with \emph{MetaDataIDs} less than $128.0.0.0$. \emph{SwitchB} is responsible for metadata objects with \emph{MetaDataIDs} no less than $128.0.0.0$. 

\setlength{\minipagewidth}{0.48\textwidth}
\setlength{\figurewidthFour}{\minipagewidth}
\begin{figure}
    \centering
    \begin{minipage}[t]{\minipagewidth}
    \begin{center}
    \includegraphics[width=\figurewidthFour]{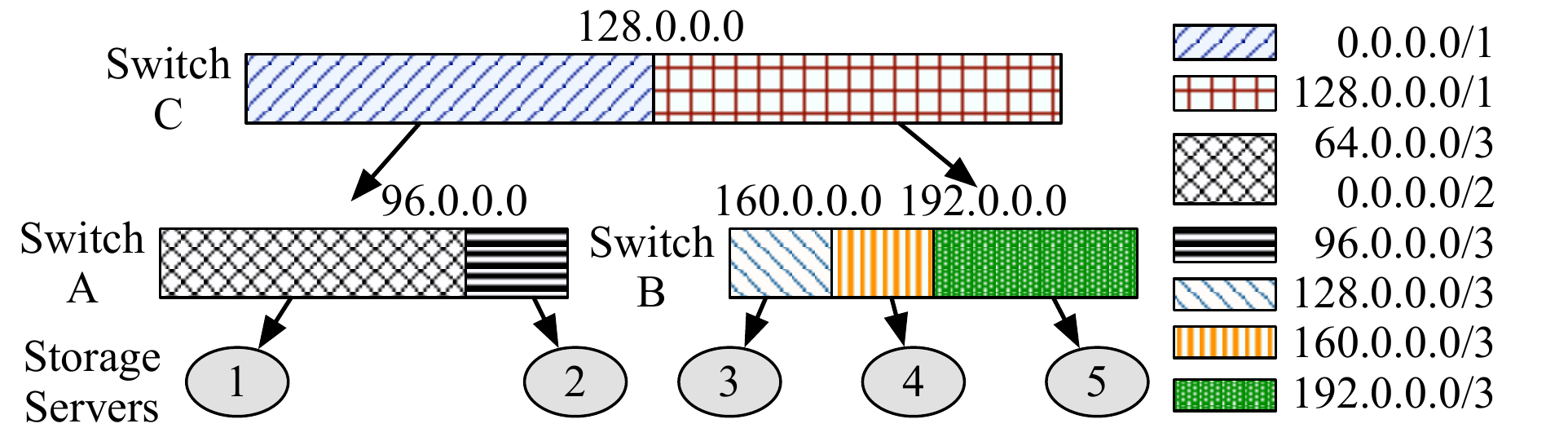}
    \end{center}
    \end{minipage}
    \centering
    \caption{Using CIDR blocks to generate flow tables in a logical B-tree.}
\label{Fig: CIDR_partition}
\end{figure}

Normally, a B-tree partition value could generate a list of flow entries. As shown in  Figure \ref{Fig: CIDR_partition}, \emph{SwitchA} splits the allocated metadata objects using $96.0.0.0$ as the partition value. In this case, MetaFlow uses block $0.0.0.0/2$ and block $64.0.0.0/3$ to represent the left partition of \emph{SwitchA}. Block $96.0.0.0/3$ represents the right partition of \emph{SwitchA}. MetaFlow then generates three flow entries for \emph{SwitchA} to represent the partition value $96.0.0.0$ in the B-tree:

\begin{lstlisting}
 Dest Addr  Dest Port       Action
0.0.0.0 /2     9000   Forward to Server1
64.0.0.0/3     9000   Forward to Server1
96.0.0.0/3     9000   Forward to Server2
\end{lstlisting}

\section{MetaFlow: Flow Table Maintenance}

In a dynamic network, storage servers and switches can join and leave the system any time. MetaFlow needs to update appropriate flow tables in the SDN-enabled switches to maintain proper lookup operations.

\subsection{Node Joins \& Leaves}
When a new storage server or a switch is added to the storage cluster, MetaFlow creates a new node in the existing B-tree at the appropriate location, and sets its state to \emph{idle}. Initially, the new node will not be allocated with metadata objects immediately when it joins the system. Therefore, there is no change to the existing flow tables.

If a storage server fails, the corresponding B-tree node will be deleted. In this case, MetaFlow activates an \emph{idle} node having the same parent node to replace the failed node. MetaFlow identifies the parent switch of the failed node and the newly activated node. Then, it updates appropriate flow entries in the parent switch, using the newly activated node to replace the failed node. If there is no available \emph{idle} node to handle the failed node, it means that more storage servers should be added to the cluster to meet the storage requirement.

\subsection{Node Splits}

A  B-tree node will be split into two nodes when it is full. The mapped B-tree in MetaFlow activates an \emph{idle} node into the \emph{busy} state, and transfers part of the metadata objects from the full node to the newly activated node. Finally, MetaFlow updates appropriate flow entries to maintain proper lookup operations. For example, in Figure \ref{Fig: split_b_tree}, \emph{ServerA} is split into two nodes using $80.0.0.0$ as the partition value.  Before the split operation, \emph{SwitchD} has the following flow table for the left subtree:
\vspace{-1 mm}
\begin{lstlisting}
 Dest Addr   Dest Port      Action
0.0.0.0 /2     9000   Forward to ServerA
64.0.0.0/3     9000   Forward to ServerA
\end{lstlisting}
After the split, one flow entry is added, and another one is modified in \emph{SwitchD's} flow table:
\begin{lstlisting}
 Dest Addr   Dest Port       Action
0.0.0.0 /2     9000    Forward to ServerA
64.0.0.0/4     9000    Forward to ServerA
80.0.0.0/4     9000    Forward to ServerC
\end{lstlisting}

\setlength{\minipagewidth}{0.48\textwidth}
\setlength{\figurewidthFour}{\minipagewidth}
\begin{figure}
    \centering
    \begin{minipage}[t]{\minipagewidth}
    \begin{center}
    \includegraphics[width=\figurewidthFour]{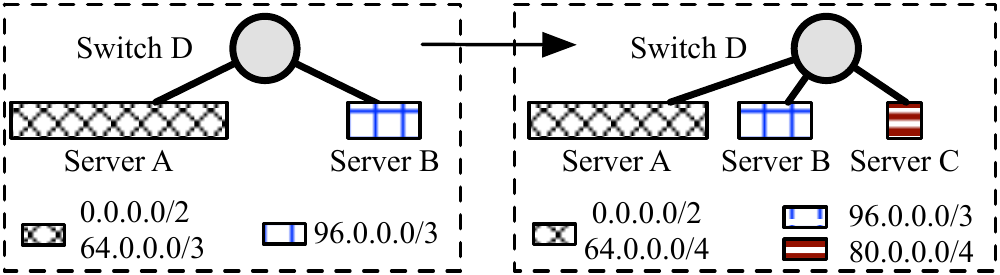}
    \end{center}
    \end{minipage}
    \centering
    \caption{Splitting a  B-tree node. In this example, when \emph{ServerA} is full, the MetaFlow controller activates \emph{ServerC}, uses $80.0.0.0$ to split \emph{ServerA}, and transfers the CIDR block $80.0.0.0/4$ from \emph{ServerA} to \emph{ServerC}.}
\label{Fig: split_b_tree}
\end{figure}


MetaFlow uses a traversal algorithm to select a partition value to  partition the full B-tree node. Given a full B-tree node, which is allocated with several ordered CIDR blocks of metadata objects, MetaFlow splits it to two sets: a \emph{left set} and a \emph{right set}. The \emph{left set} will be left in the existing node; and the \emph{right set} will be transfered to the newly activated node. MetaFlow splits a full node via following steps:
\emph{Step1.} MetaFlow traverses the ordered CIDR blocks in the full node, and checks the number of metadata objects for each block.
\emph{Step2.}  During the traversal operation, MetaFlow puts the incoming CIDR block into the \emph{left set}, until the \emph{left set's} number of metadata objects  exceeds $40\%$ of the full node's number of metadata objects. There are two cases:
\begin{enumerate}
\item[\textbullet]  {The number of metadata objects in the \emph{left set} is smaller than $60\%$ of that in the full node.} In this case, MetaFlow puts the rest of CIDR blocks into the \emph{right set}. For example, a full node contains three ordered CIDR blocks: 192.168.100.0/25, 192.168.100.128/25, and 192.168.100.192/26. When putting 192.168.100.0/25 into the \emph{left set}, if the \emph{left set's} number of metadata objects is between $40\%$ and $60\%$ of the original node's number of metadata objects, MetaFlow puts the rest of CIDR blocks, which are 192.168.100.128/25 and 192.168.100.192/26, into the \emph{right set}.
\item[\textbullet]  {The number of metadata objects in the \emph{left set} is more than $60\%$ of that in the full node.} In this case, MetaFlow removes the most recently considered CIDR block from the \emph{left set}. This CIDR block will be evenly split into two sub-blocks to replace the original one. For example, the CIDR block 192.168.100.0/24 will be split into two ordered sub-blocks: 192.168.100.0/25 and 192.168.100.128/25. MetaFlow uses these two sub-blocks to replace the original CIDR block 192.168.100.0/24, and continues the traversal operation. Normally, MetaFlow will next deal with the CIDR block 192.168.100.0/25 using the same operation in Step 2.
\end{enumerate}
\emph{Step3.}  After the traversal operation, MetaFlow transfers the CIDR blocks in the \emph{right set} to the newly activated node, and updates the appropriate flow entries.

It is essential to reduce the number of flow entries generated by the node split operation, since most SDN-enabled switches can only support a few thousands of flow entries. If the \emph{right set} contains exactly $50\%$ of the full node's  metadata objects, our experiments show that the node split operation usually ends after a few tens of iterations in Step 2, where each iteration generates a new flow entry for its parent switch. In real-word systems, the growing size of flow table significantly limits the system performance and scalability. When we use a value between $40\%$ to $60\%$, we find that Step 2 can end after just several iterations. Compared to the value of $50\%$, a value between $40\%$ to $60\%$ can reduce the number of new flow entries by a factor of up to $10$, although in this case MetaFlow  cannot split the full node evenly. As a trade-off between storage balance and  flow table size, MetaFlow uses a value of $40\%$ to $60\%$ for the node split operation.

\section{Numerical Results and Analysis}

In this section, we evaluate MetaFlow performance using both a large scale simulation and a testbed. In particular, we investigate MetaFlow's performance in terms of throughput and latency using extensive experiments with realistic metadata workload models. 

\subsection{Experiments' Parameters and Configurations}

\subsubsection{Performance Measures}
We evaluate the performance of a MetaFlow-based metadata management system, and compare it to existing DHT-based approaches like Chord and One-Hop.
(i) \emph{{Throughput. }}We measure the system throughput with increasing cluster size in both the simulator and the testbed. In the experiments, we define system throughput as the maximum number of metadata operations that a metadata cluster can deal with.
(ii) \emph{{Latency. }}We measure the system latency with increasing cluster size in both the simulator and the testbed. In the experiments, we define system latency as the average time used to complete a metadata operation.
(iii) \emph{{SDN Overhead. }}We evaluate the SDN-enabled switch's overhead.  The flow table size of the SDN-enabled switch can be a potential performance bottleneck for MetaFlow,  since most SDN-enabled switches can only support a few thousands of flow entries.
(iv) \emph{{NAT Agent Overhead. }}We also investigate the NAT agents' overhead in the storage servers. MetaFlow uses NAT agents to replace the destination and source IP addresses. This operation consumes CPU cycles and is the main source of the overhead in MetaFlow's current implementation.

\setlength{\minipagewidth}{0.46\textwidth}
\setlength{\figurewidthFour}{\minipagewidth}
\begin{figure}
    \centering
    \begin{minipage}[t]{\minipagewidth}
    \begin{center}
    \includegraphics[width=\figurewidthFour]{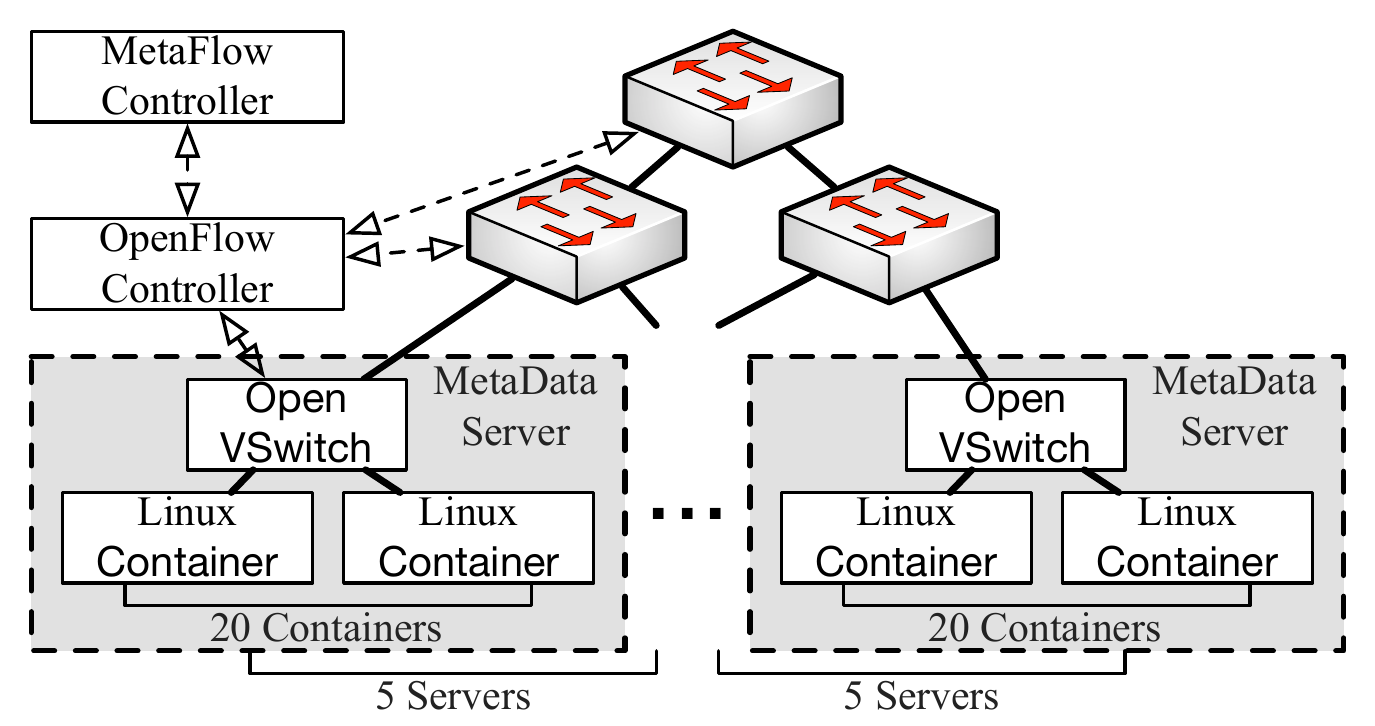}
    \end{center}
    \end{minipage}
    \centering
    \caption{Testbed architecture. }
\label{Fig: experiement_arch}
\end{figure}

\subsubsection{System Configurations}

We implement a MetaFlow-based metadata management system for distributed file systems in both a simulator and a testbed. 
(i) \emph{{Simulation Setup.}}  The simulator has up to $2000$ servers, forming a storage cluster based on the fat tree topology. In the cluster, each switch has $32$ ports. Thus, $16$ edge layer switches, $16$ aggregation layer switches, and $256$ servers form a \emph{pod} in the fat tree. There are $32$ core layer switches in total. All the network connections between switches and servers are $10$-Gbps links. The simulator uses $500$ clients to form an application cluster for generating metadata workload.
(ii) \emph{{Testbed Setup.}} The testbed has up to $200$ Linux containers, forming a storage cluster based on a three-tier tree topology as shown in Figure \ref{Fig: experiement_arch}. More specifically, in the testbed, we use a \emph{Extreme Summit x670c} switch (which has OpenFlow 1.0.0 support) as the core layer switch, use two \emph{Extreme Summit x670c} switches as the aggregation layer switches, and use OpenVSwitch \cite{pfaff2015design} (which have OpenFlow 1.0.0 support) as the edge layer switch to connect Linux containers. There are $5$ Dell R720 servers, each of which contains up to $20$ Linux containers.  Each Linux container is allocated with a $2$ GHz CPU core and $4$ GB memory. All the network connections between switches and servers are $10$-Gbps links. The testbed uses $50$ Linux containers as an application cluster for generating metadata workload. Two Linux containers are set up to manage the switches' flow tables. One is the \emph{OpenFlow Controller}, which manages flow tables for switches using OpenFlow protocols. The other one is the \emph{MetaFlow Controller}, which generates and maintains flow tables for proper lookup operations using B-tree.

\subsubsection{Workload Model}

In the experiments, we use a metadata workload in which $20\%$ are \emph{get} and $80\%$ are \emph{put} operations. This is similar to real-world metadata workloads \cite{dandong2012decentralized}. In the \emph{get} operation, a client retrieves a metadata object using the given \emph{MetaDataID}. In the \emph{put} operation, a client writes new data into a metadata object. Each metadata object for a file and directory is a key-value pair with the size of $250$ and $290$ bytes, respectively. This is similar to the metadata object size in HDFS \cite{shvachko2010hdfs}.

\subsubsection{Experiment Configurations}

We use different types of storage subsystems in the experiments to measure the system throughput and latency.
(i) \emph{{Simulator. }}We conduct several tests to find the appropriate throughput and latency parameters to be used in the simulations. In these tests, we measure the performance of a lookup subsystem and four different storage subsystems, namely Redis, LevelDB (SSD), LevelDB (HDD), and MySQL, on a single CPU core.  Based on these results, we define a throughput and a latency ratio to be used in the simulations. More specifically, the throughput ratio is obtained by dividing the throughput of the lookup subsystem to that of a storage subsystem. Similarly, the latency ratio is obtained by dividing the latency of the lookup subsystem to that of a storage subsystem. Therefore, in the simulations, we use the following throughput ratios: 1, 1.5, 2, and 100 to reflect the throughput performance of Redis, LevelDB (SSD), LevelDB (HDD), and MySQL, respectively. Similarly, the following latency ratios are used: 1, 0.7, 0.5, and 0.001. These ratios reflect the latency performance of Redis, LevelDB (SSD), LevelDB (HDD), and MySQL.
(ii) \emph{{Testbed. }}We test the MetaFlow-based system's throughput and latency using Redis, LevelDB (SSD), LevelDB (HDD), and MySQL as the storage subsystem.

\subsection{MetaFlow: Throughput}

\setlength{\minipagewidth}{0.235\textwidth}
\setlength{\figurewidthFour}{\minipagewidth}
\begin{figure}[htb]
    \centering
    \begin{minipage}[t]{\minipagewidth}
    \begin{center}
    \includegraphics[width=\figurewidthFour]{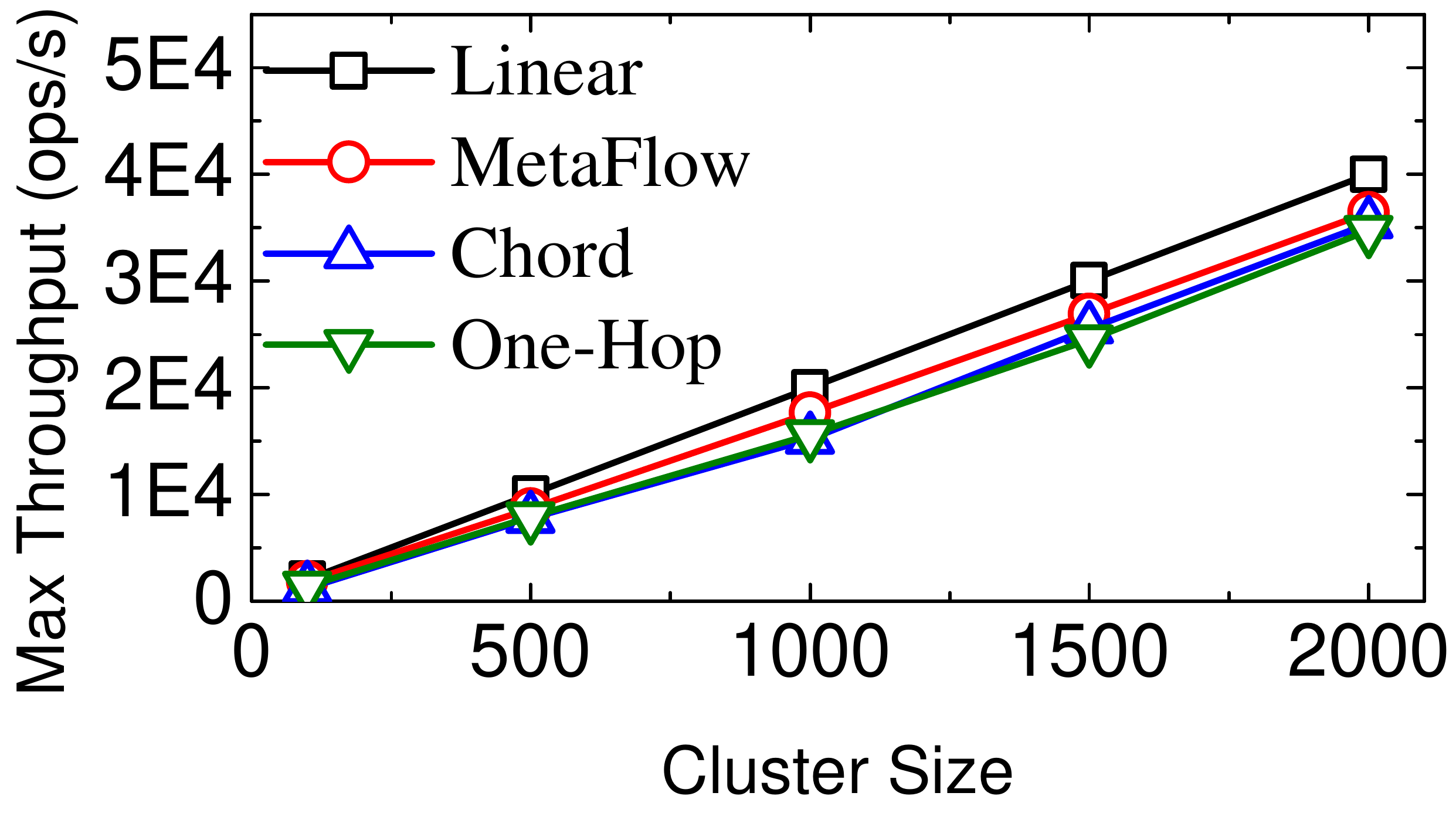}
    \subcaption{(a) 100}
    \end{center}
    \end{minipage}
    \centering
    \begin{minipage}[t]{\minipagewidth}
    \begin{center}
    \includegraphics[width=\figurewidthFour]{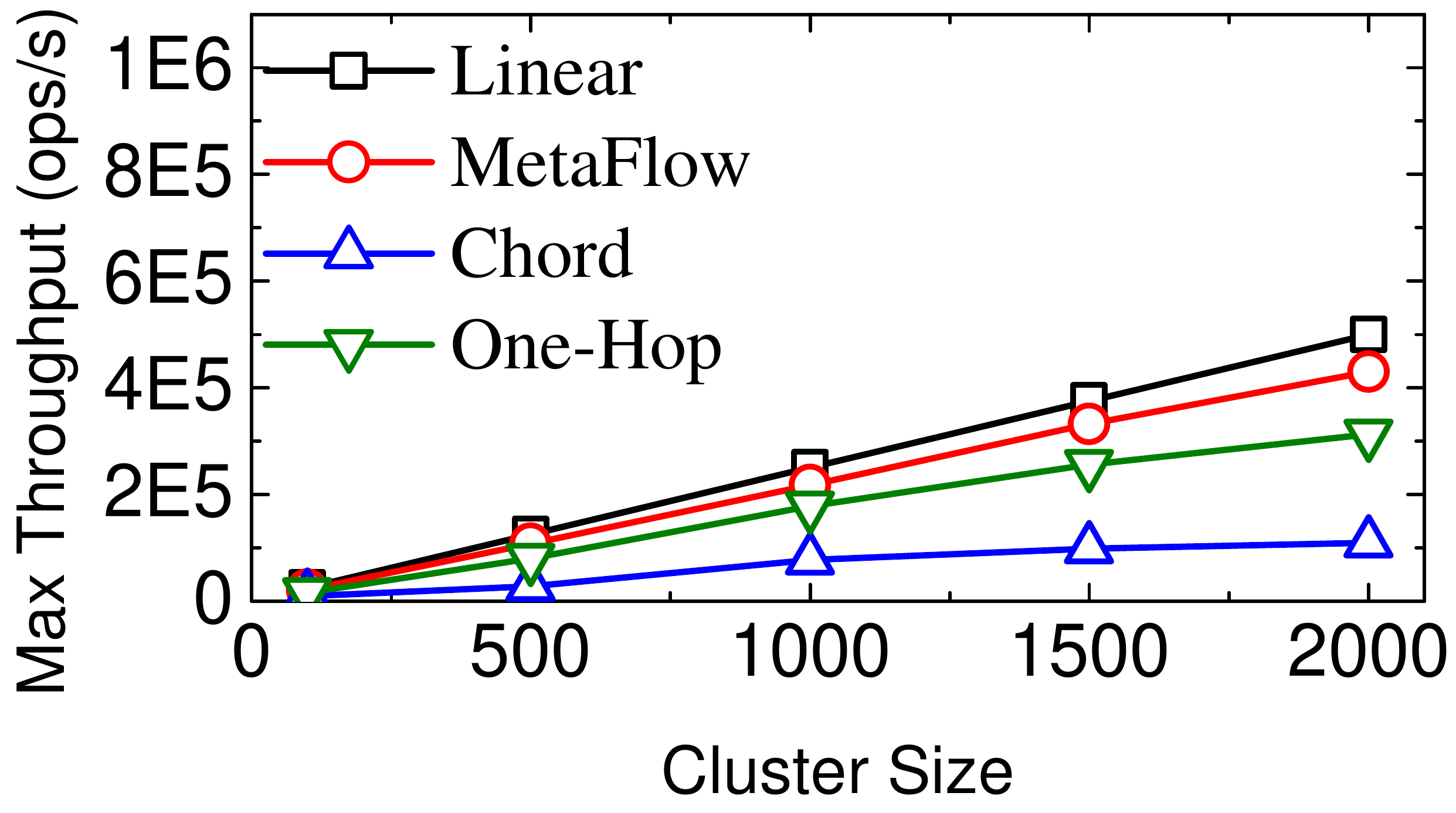}
    \subcaption{(b) 2}
    \end{center}
    \end{minipage}
    \centering
    \begin{minipage}[t]{\minipagewidth}
    \begin{center}
    \includegraphics[width=\figurewidthFour]{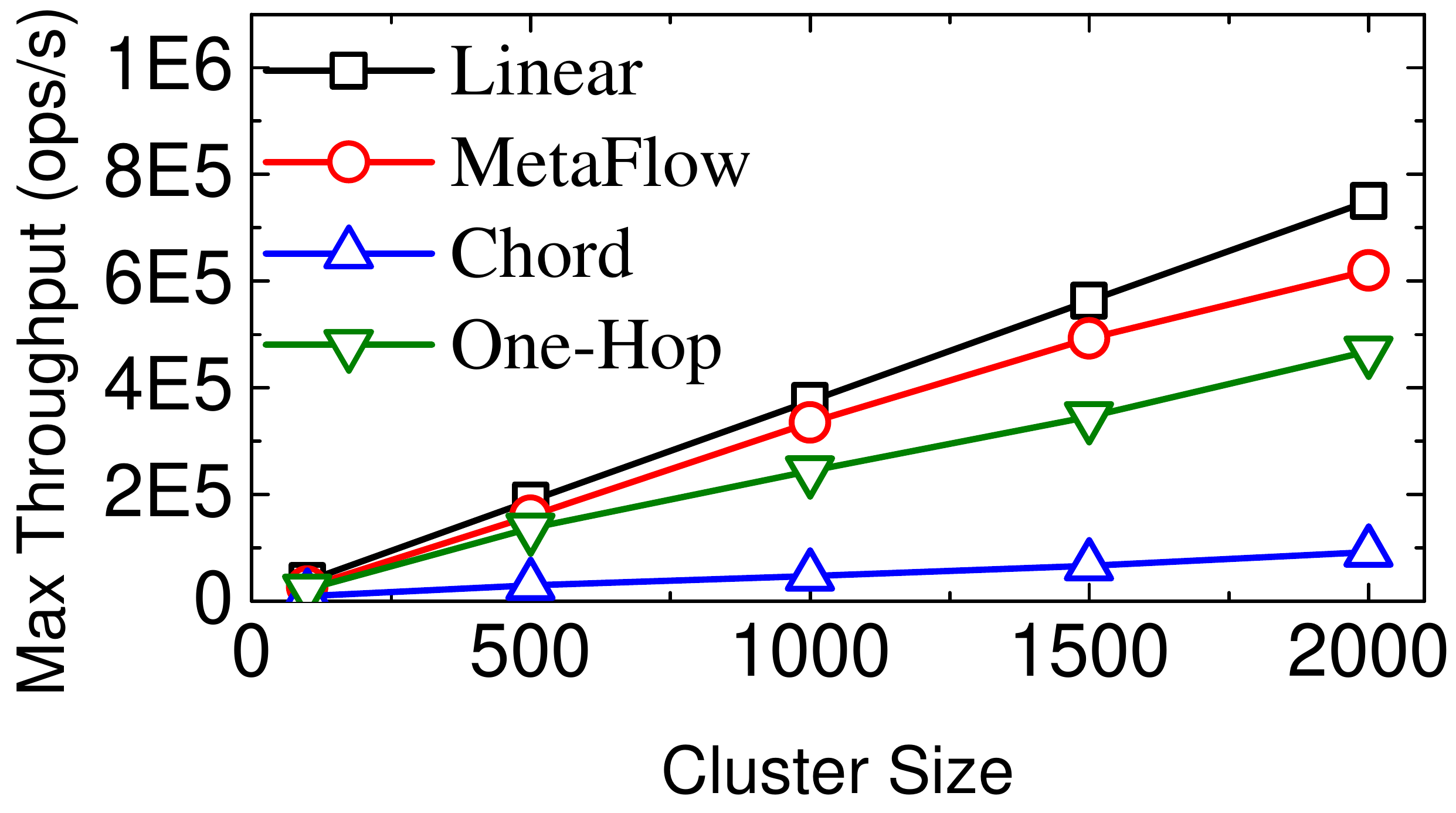}
    \subcaption{(c) 1.5}
    \end{center}
    \end{minipage}
    \centering
    \begin{minipage}[t]{\minipagewidth}
    \begin{center}
    \includegraphics[width=\figurewidthFour]{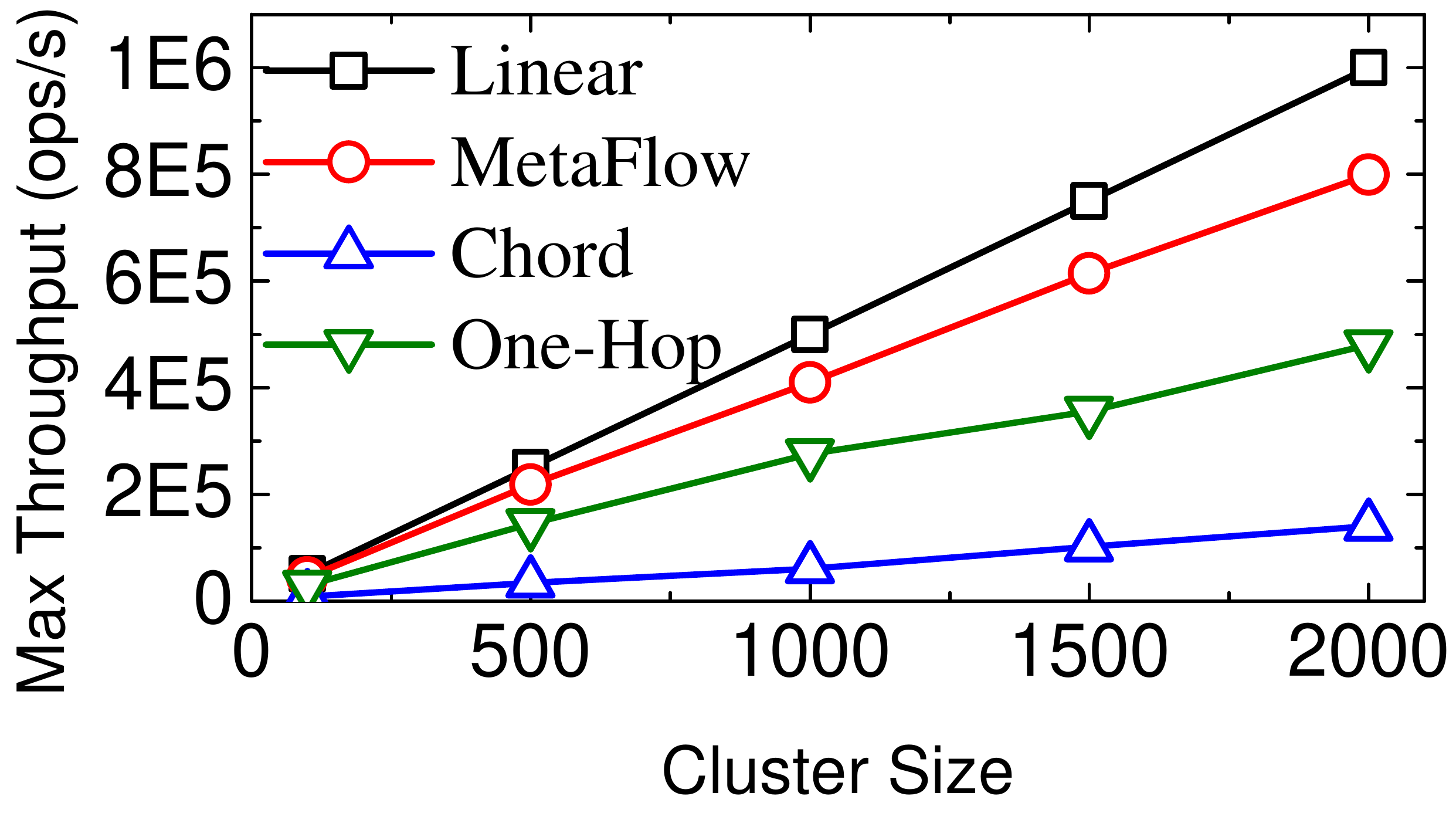}
    \subcaption{(d) 1}
    \end{center}
    \end{minipage}
    \centering
    \vspace{-0.1 in}
    \caption{Throughput comparison between the MetaFlow-based system and two DHT-based systems (Chord and One-Hop) using the simulation with 4 lookup/storage throughput ratios: 100, 2, 1.5, and 1. }
\label{Fig: Throughput_MetaFlow_simulation}
\end{figure}

\setlength{\minipagewidth}{0.235\textwidth}
\setlength{\figurewidthFour}{\minipagewidth}
\begin{figure}
    \centering
    \begin{minipage}[t]{\minipagewidth}
    \begin{center}
    \includegraphics[width=\figurewidthFour]{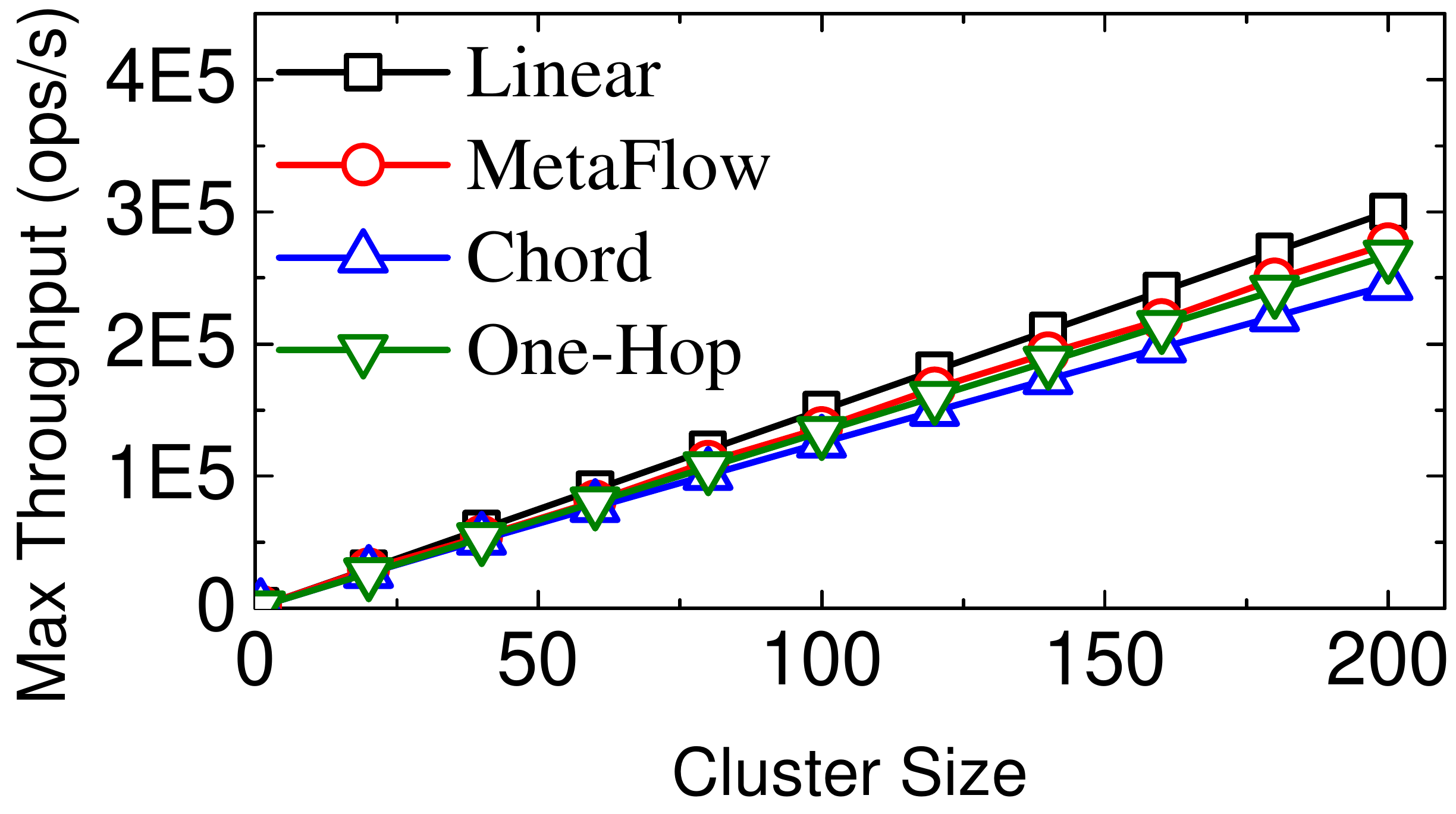}
    \subcaption{(a) MySQL}
    \end{center}
    \end{minipage}
    \centering
    \begin{minipage}[t]{\minipagewidth}
    \begin{center}
    \includegraphics[width=\figurewidthFour]{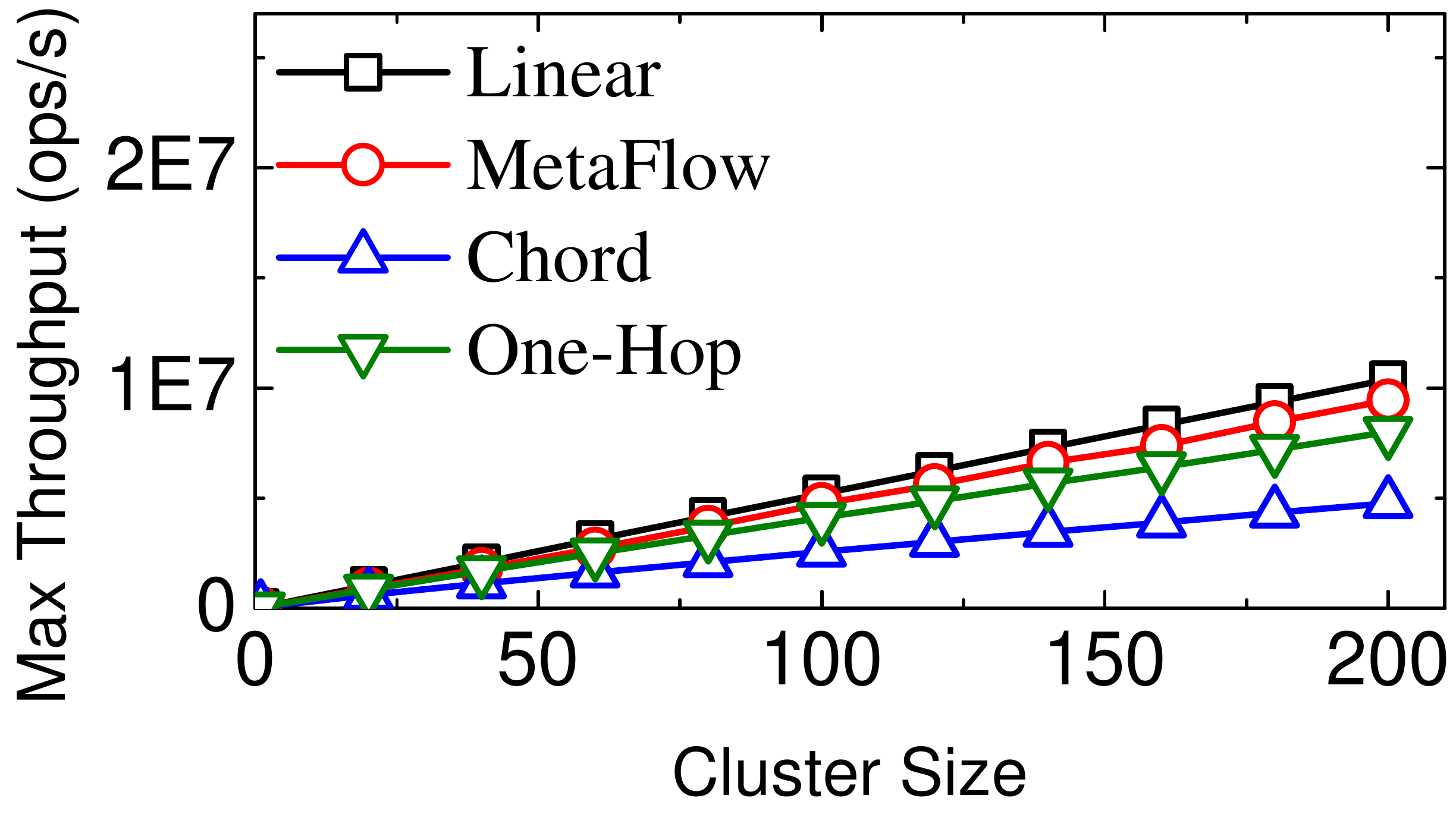}
    \subcaption{(b) LevelDB (HDD)}
    \end{center}
    \end{minipage}
    \centering
    \begin{minipage}[t]{\minipagewidth}
    \begin{center}
    \includegraphics[width=\figurewidthFour]{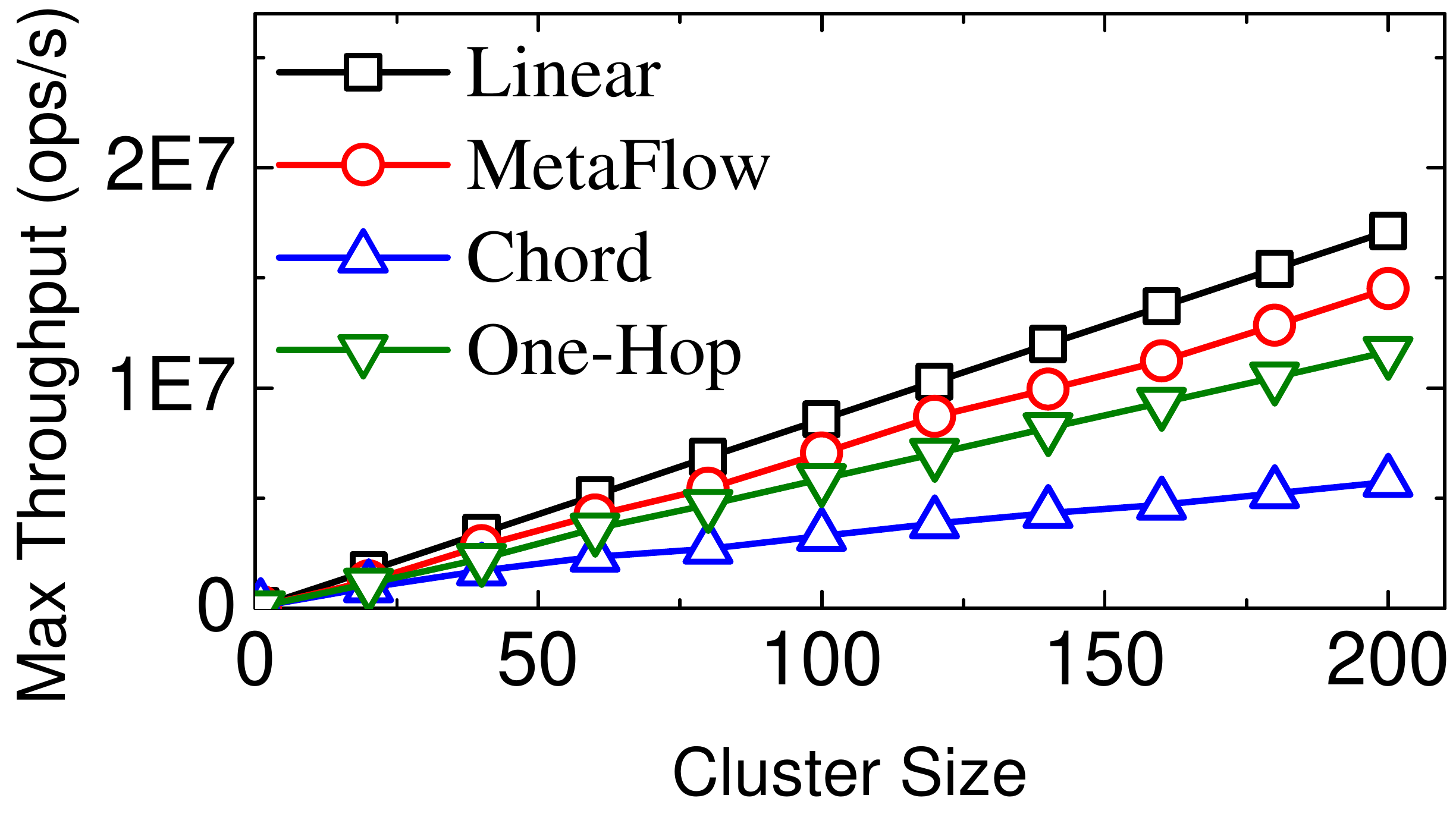}
    \subcaption{(c) LevelDB (SSD)}
    \end{center}
    \end{minipage}
    \centering
    \begin{minipage}[t]{\minipagewidth}
    \begin{center}
    \includegraphics[width=\figurewidthFour]{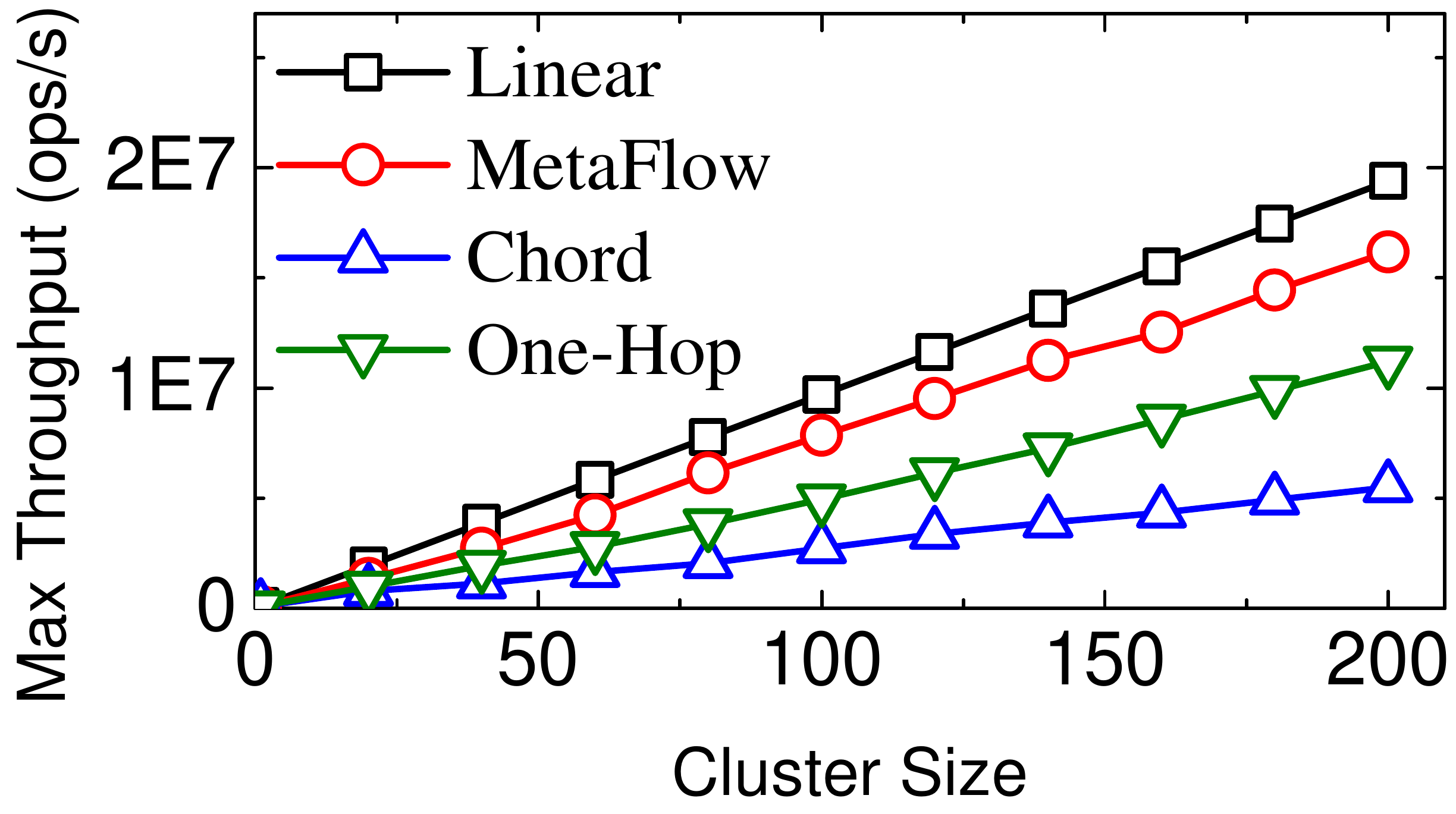}
    \subcaption{(d) Redis}
    \end{center}
    \end{minipage}
    \centering
    \vspace{-0.1 in}
    \caption{Throughput comparison between the MetaFlow-based system and two DHT-based systems (Chord and One-Hop) using the testbed with 4 types of storage subsystems.}
\label{Fig: Throughput_MetaFlow_test_bed}
\end{figure}

Figure \ref{Fig: Throughput_MetaFlow_simulation} shows the system throughputs in our simulations. In these experiments, we compare the throughputs of the MetaFlow-based system and DHT-based systems to those of an ideal system, which has linear performance (i.e., the ideal throughput increase linearly with respect to the cluster size). We observe that MetaFlow consistently performs better than Chord and One-Hop in all situations. In particular, when the throughput ratio is 1 (which means the metadata management system has similar I/O and lookup throughputs), MetaFlow has a throughput reduction of $12\%$ to $20\%$ compared to the ideal system. In contrast, Chord and One-Hop have $80\%$ to $85\%$ and $45\%$ to $50\%$ throughput reduction, respectively. In particular, when there are $2000$ servers in the cluster, MetaFlow could process $8.0 \times 10 ^{5}$ requests per second. The corresponding values for Chord and One-Hop are $2.5 \times 10 ^{5}$ and $4.0 \times 10 ^{5}$, respectively. In this case, MetaFlow could increase system throughput by a factor of $3.2$ and $2.0$ when comparing to Chord and One-Hop, respectively.

When using other throughput ratios such as $2$, MetaFlow has about 12\% to 17\% throughput reduction, as shown in Figure \ref{Fig: Throughput_MetaFlow_simulation} (b). The corresponding measures for Chord and One-Hop are 75\% to 80\% and 30\% to 36\%, respectively. Even in a low-throughput storage system such as MySQL, MetaFlow is still better, but not by much as shown in Figure \ref{Fig: Throughput_MetaFlow_simulation} (a). The reason is that the limiting factor in MySQL-based systems is actually the I/O throughput, not lookup. We should note that such low-throughput systems are not suitable for large-scale metadata management in practice. The results for MySQL provided here are mainly for highlighting the lower bound of MetaFlow's performance.

Results obtained using the real testbed, as shown in Figure \ref{Fig: Throughput_MetaFlow_test_bed}, confirm performance improvement demonstrated in the simulations. In Figure \ref{Fig: Throughput_MetaFlow_test_bed} (d), we observe that MetaFlow has roughly $15\%$ of throughput reduction compared to the ideal system when using $200$ Redis servers. On the contrary, Chord and One-Hop have throughput reductions of nearly $70\%$ and $45\%$, respectively. As shown in Figure \ref{Fig: Throughput_MetaFlow_test_bed} (b) and (c), MetaFlow has about $8\%$ and $15\%$ throughput reduction for LevelDB (HDD) and LevelDB (SSD), respectively. In contrast, One-Hop suffers roughly $20\%$ and $40\%$ performance reduction in the same cluster. At the same time, Chord has even more performance reduction, about $50\%$ and $65\%$.

\subsection{MetaFlow: Latency}

\setlength{\minipagewidth}{0.235\textwidth}
\setlength{\figurewidthFour}{\minipagewidth}
\begin{figure}
    \centering
    \begin{minipage}[t]{\minipagewidth}
    \begin{center}
    \includegraphics[width=\figurewidthFour]{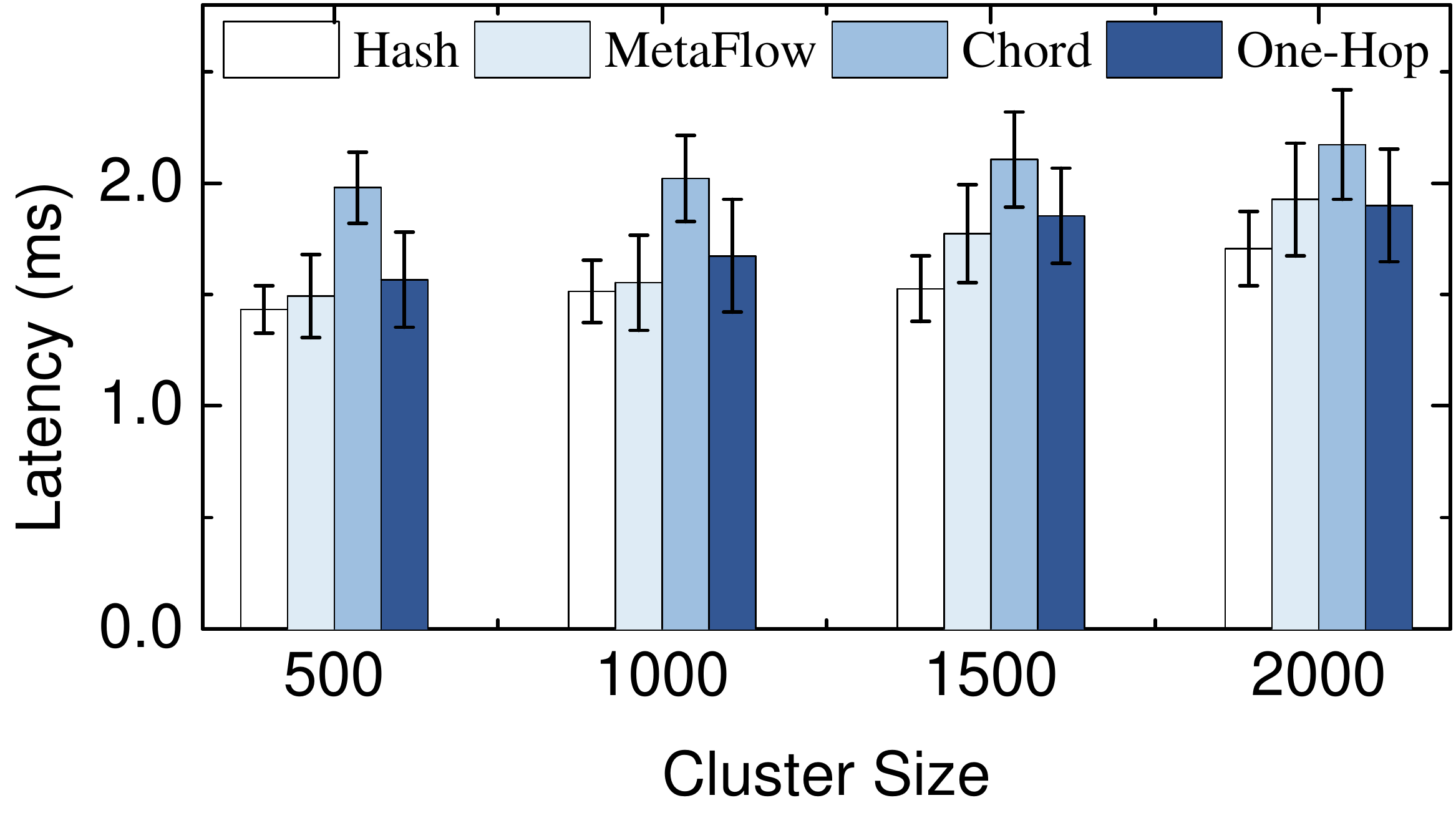}
    \subcaption{(a) 0.001}
    \end{center}
    \end{minipage}
    \centering
    \begin{minipage}[t]{\minipagewidth}
    \begin{center}
    \includegraphics[width=\figurewidthFour]{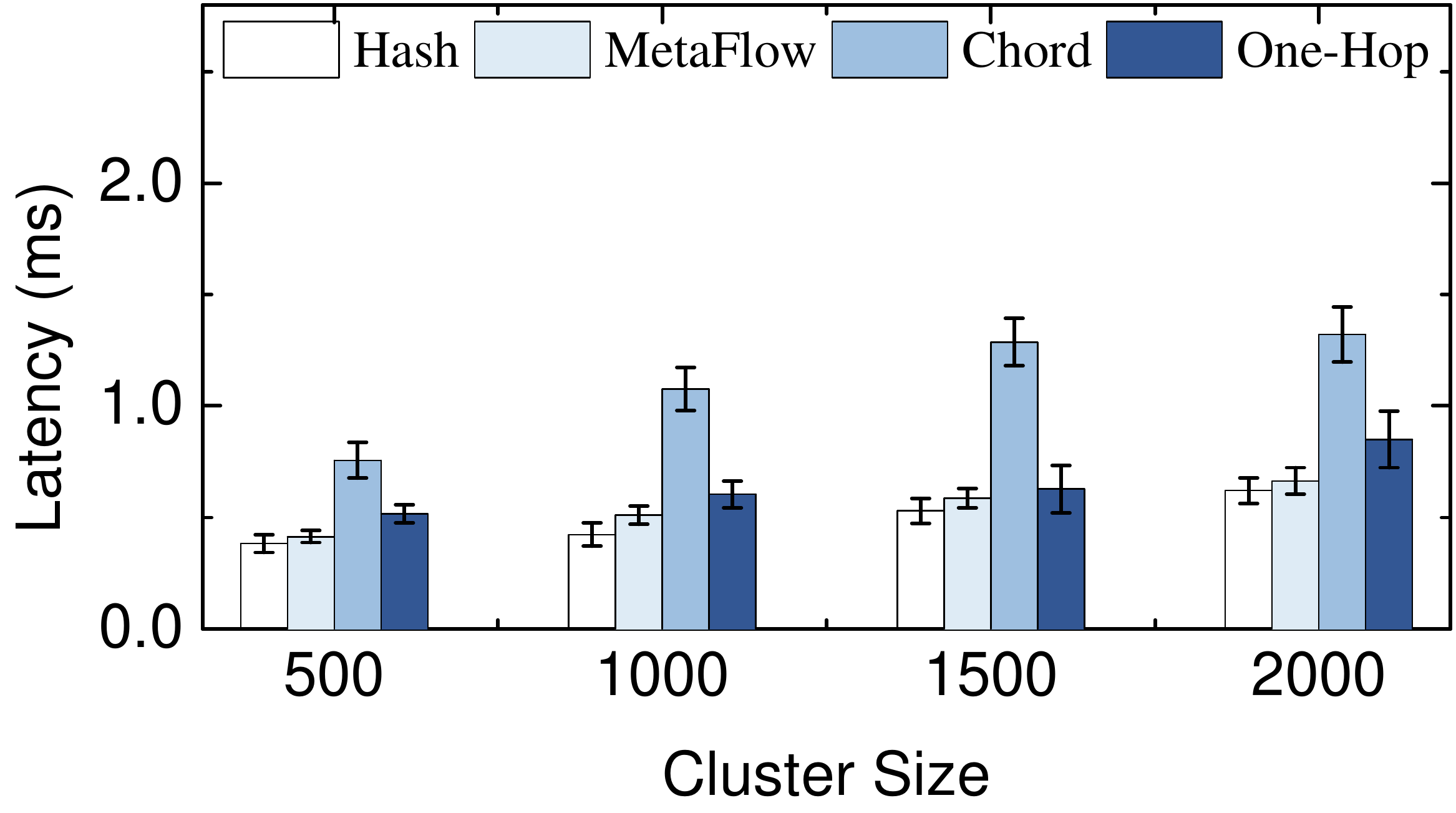}
    \subcaption{(b) 0.5}
    \end{center}
    \end{minipage}
    \centering
    \begin{minipage}[t]{\minipagewidth}
    \begin{center}
    \includegraphics[width=\figurewidthFour]{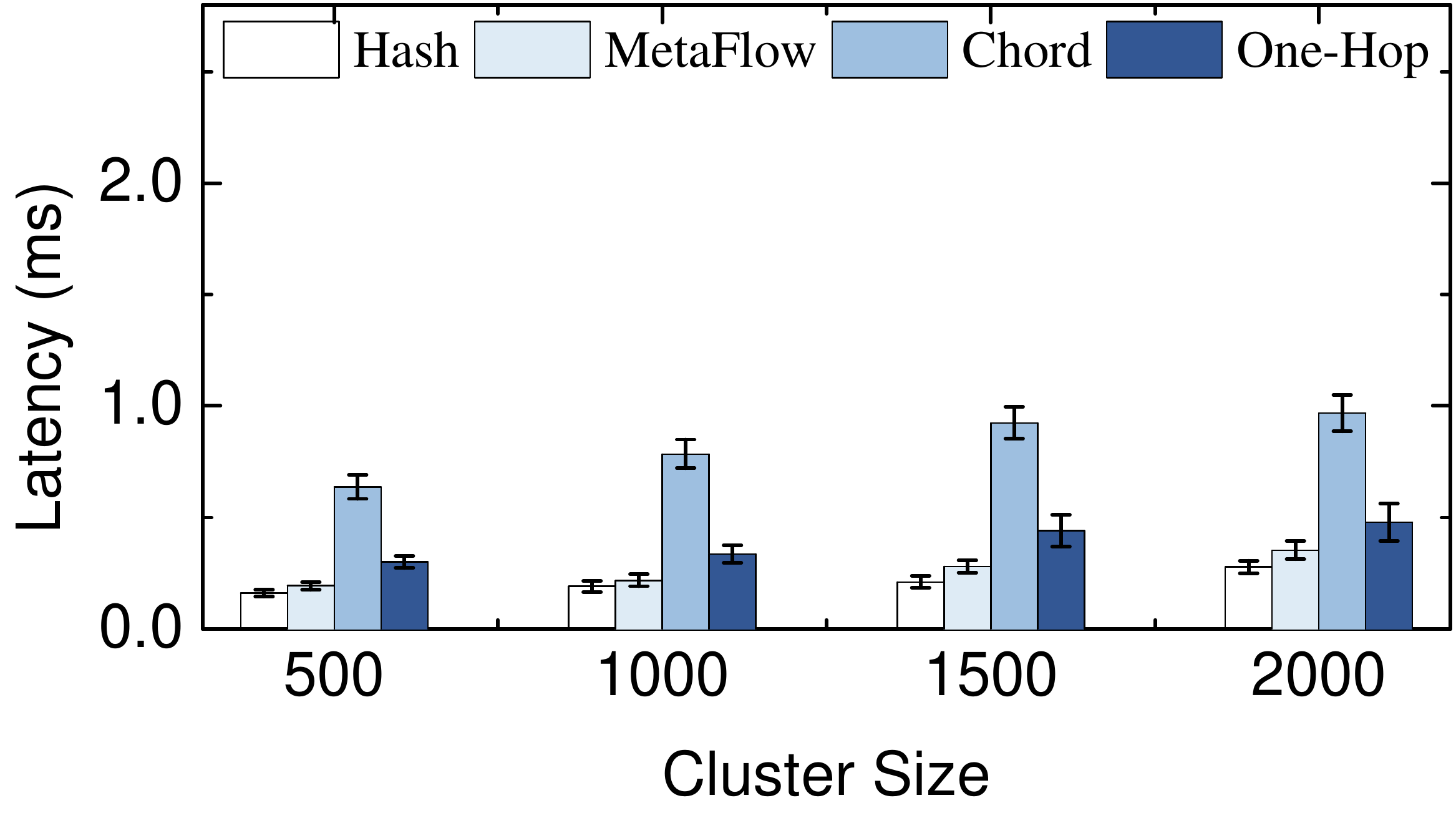}
    \subcaption{(c) 0.7}
    \end{center}
    \end{minipage}
    \centering
    \begin{minipage}[t]{\minipagewidth}
    \begin{center}
    \includegraphics[width=\figurewidthFour]{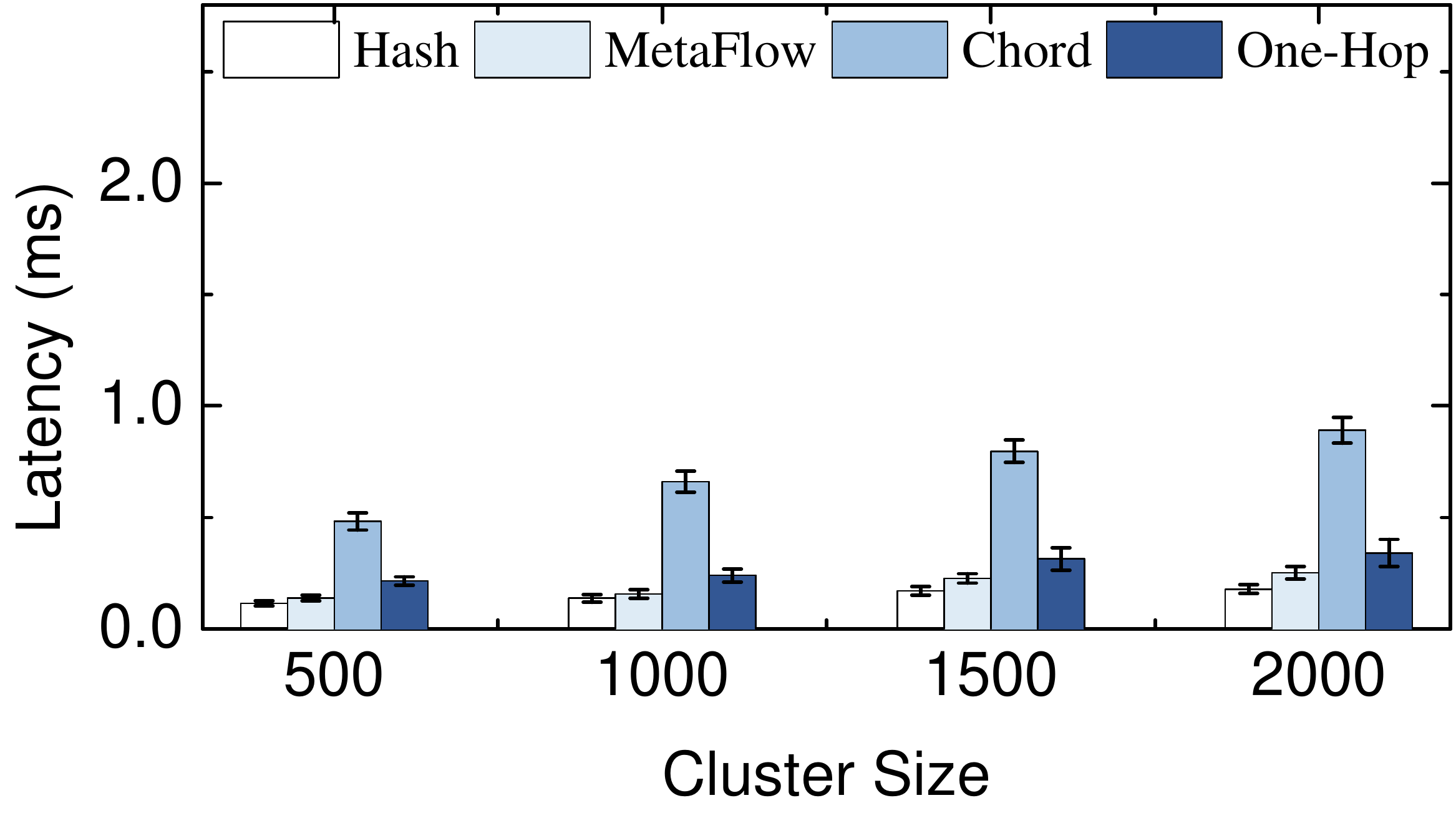}
    \subcaption{(d) 1}
    \end{center}
    \end{minipage}
    \centering
    \vspace{-0.1 in}
    \caption{Latency comparison in the simulation between the MetaFlow-based system and two DHT-based systems (Chord and One-Hop) with 4 lookup/storage latency ratios: 0.001, 0.5, 0.7, and 1.}
\label{Fig: Latency_MetaFlow_simulation}
\end{figure}

In this set of experiments, we compare the request latency of the MetaFlow-based system and other DHT-based systems with respect to a hash-based system, which has no metadata lookup latency. Figure \ref{Fig: Latency_MetaFlow_simulation} shows that the MetaFlow-based system consistently has lower latency than Chord and One-Hop in the simulations. In particular, when the latency ratio is 1, MetaFlow's latency is about up to $1.4$ times higher than the hash-based system. In contrast, Chord and One-Hop's latency are roughly $7.0$ and $2.0$ times higher than the hash-based system.  MetaFlow could reduce system latency by a factor of up to $5$ in this case.

When using other lookup/storage latency ratios like 0.7 and 0.5, MetaFlow has up to $5\%$ more latency than the hash-based system. At the same time, Chord and One-Hop have at least $50\%$ and $20\%$ more latency using the same setting, respectively as shown in Figure \ref{Fig: Latency_MetaFlow_simulation} (b) and (c). In the system with lookup/storage latency ratio of $0.001$, as shown in Figure \ref{Fig: Latency_MetaFlow_simulation} (a), MetaFlow and One-Hop have nearly the same latency performance with the hash-based system. The reason is that the high I/O latency in such systems renders lookup latency insignificant. However, we should note that such high latency systems are not suitable for metadata management in distributed file systems.

\setlength{\minipagewidth}{0.235\textwidth}
\setlength{\figurewidthFour}{\minipagewidth}
\begin{figure}
    \centering
    \begin{minipage}[t]{\minipagewidth}
    \begin{center}
    \includegraphics[width=\figurewidthFour]{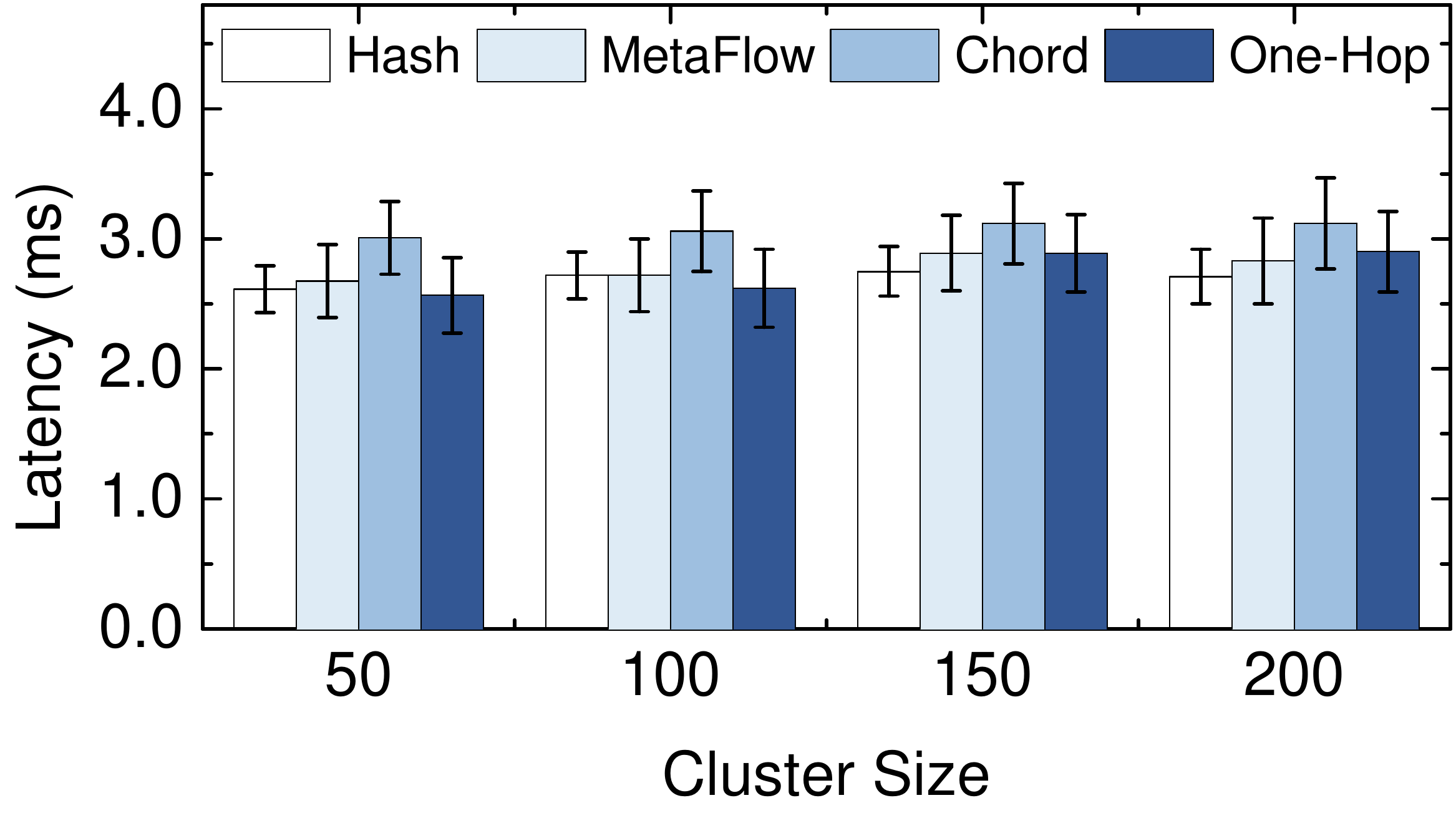}
    \subcaption{(a) MySQL}
    \end{center}
    \end{minipage}
    \centering
    \begin{minipage}[t]{\minipagewidth}
    \begin{center}
    \includegraphics[width=\figurewidthFour]{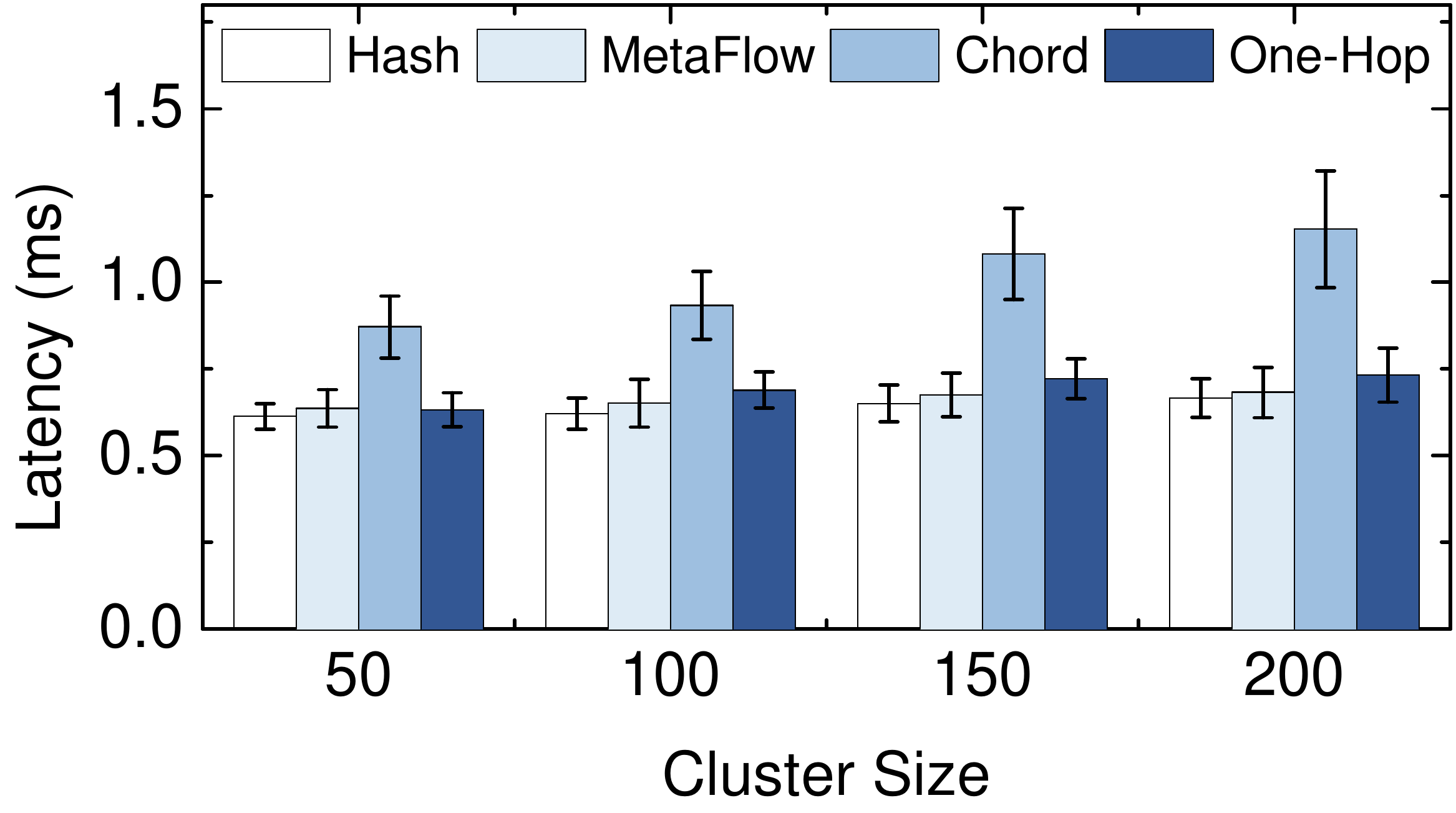}
    \subcaption{(b) LevelDB (HDD)}
    \end{center}
    \end{minipage}
    \centering
    \begin{minipage}[t]{\minipagewidth}
    \begin{center}
    \includegraphics[width=\figurewidthFour]{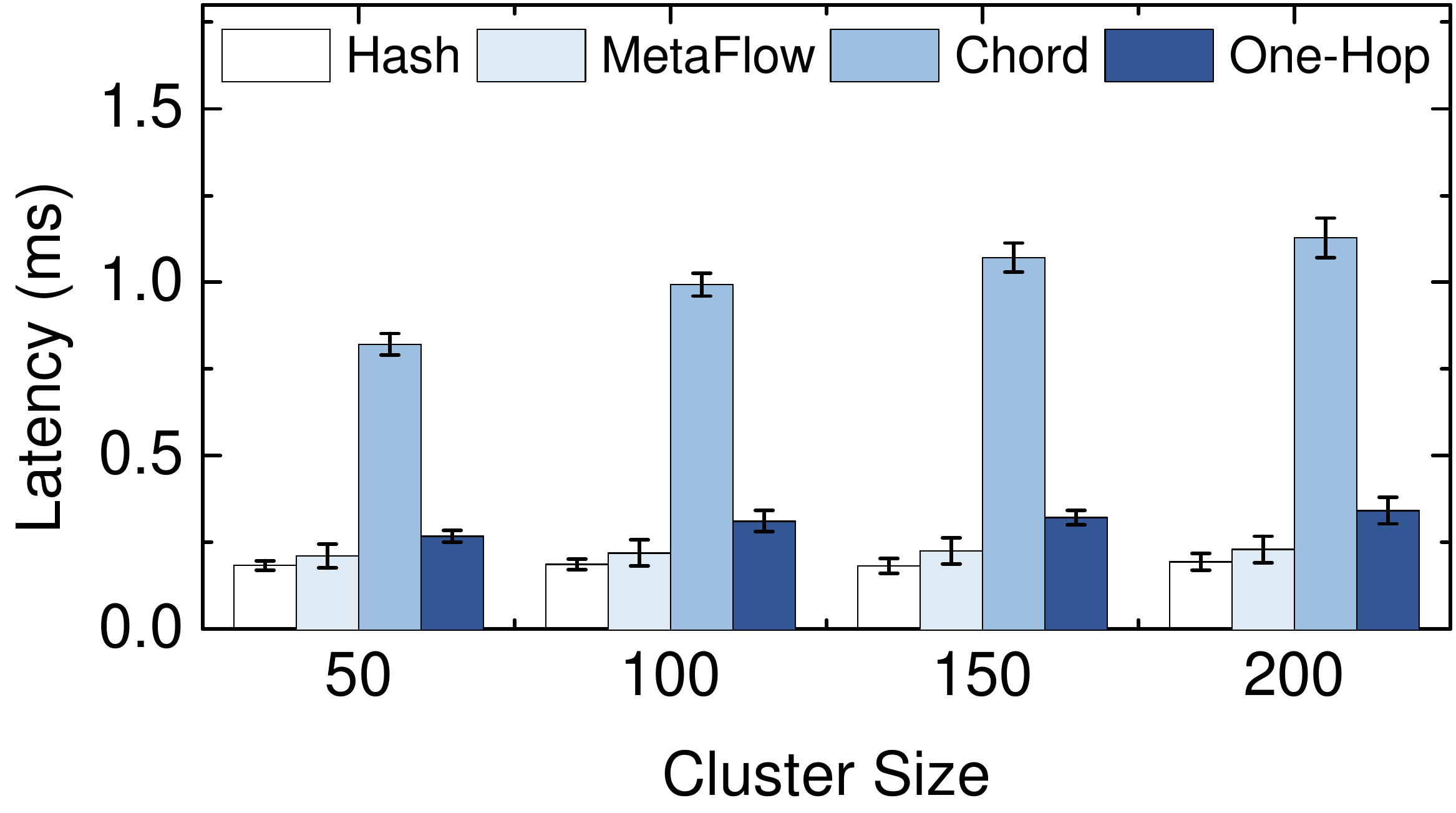}
    \subcaption{(c) LevelDB (SSD)}
    \end{center}
    \end{minipage}
    \centering
    \begin{minipage}[t]{\minipagewidth}
    \begin{center}
    \includegraphics[width=\figurewidthFour]{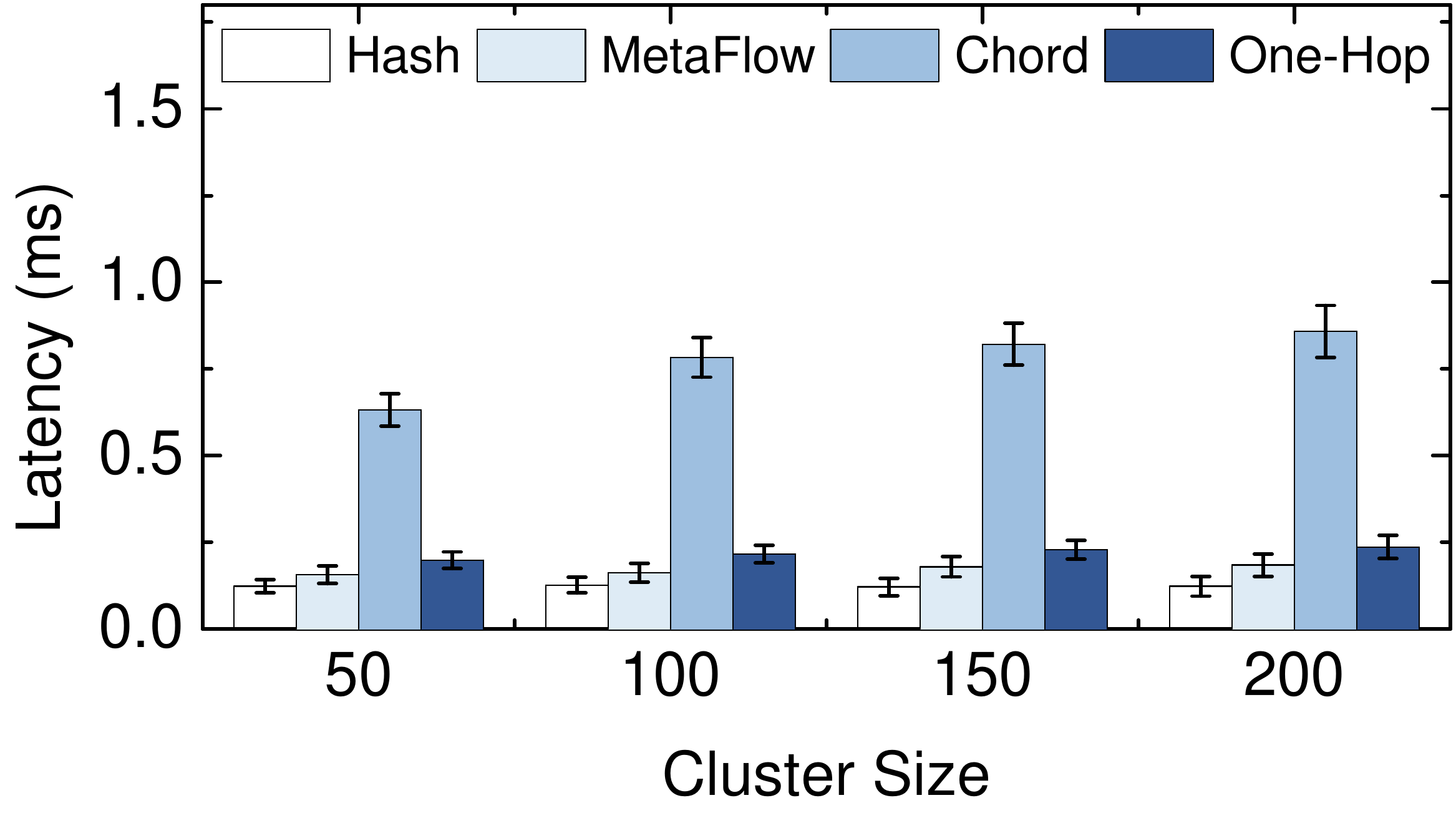}
    \subcaption{(d) Redis}
    \end{center}
    \end{minipage}
    \centering
    \vspace{-0.1 in}
    \caption{Latency comparison between the MetaFlow-based system and two DHT-based systems (Chord and One-Hop) using the testbed with 4 types of storage subsystems.}
\label{Fig: Latency_MetaFlow_test_bed}
\end{figure}

Test-bed results are similar to simulation results with regard to the system latency. Figure \ref{Fig: Latency_MetaFlow_test_bed} (d) shows that MetaFlow's latency is roughly $1.6$ times higher than the hash-based system when using $200$ Redis servers. On the contrary, Chord and One-Hop suffer up to $6.7$ and $2.1$ times more latency than the hash-based system, respectively. If the storage subsystem is LevelDB (SSD), MetaFlow has similar latency performance with the hash-based system. Meanwhile, Chord and One-Hop  have latencies that are up to $5.5$ and $1.6$ times higher than the hash-based system. When the metadata management system is deployed over a HDD-based storage system like LevelDB(HDD) and MySQL,  MetaFlow and One-Hop have similar latency with the hash-based system as shown in Figure \ref{Fig: Latency_MetaFlow_test_bed} (a) and (b). This is mainly because I/O operations on HDD constitute a large part of the system latency.

\subsection{SDN Overhead}

\setlength{\minipagewidth}{0.235\textwidth}
\setlength{\figurewidthFour}{\minipagewidth}
\begin{figure}
    \centering
    \begin{minipage}[t]{\minipagewidth}
    \begin{center}
    \includegraphics[width=\figurewidthFour]{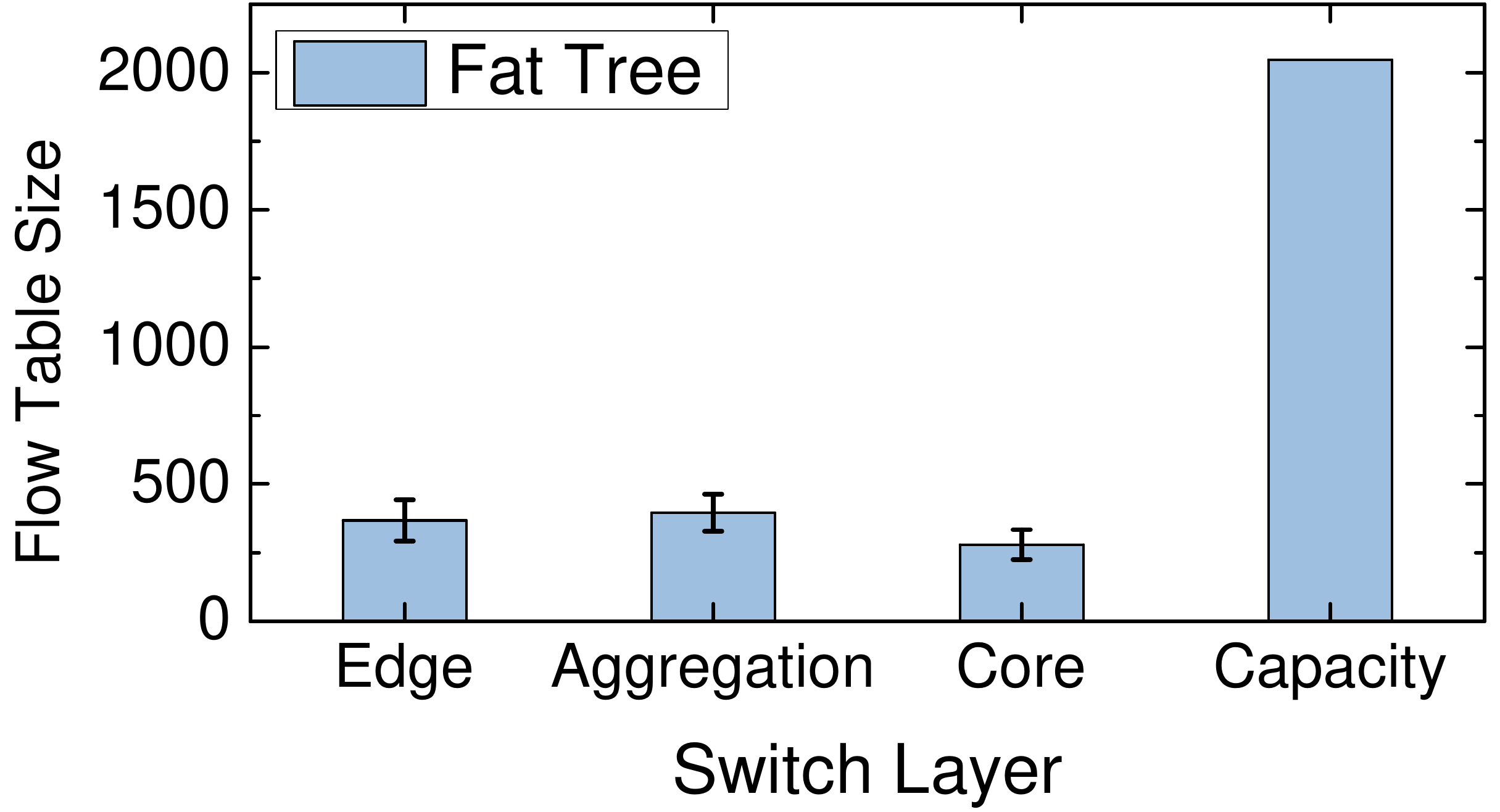}
    \subcaption{(a) Fat Tree (Simulation)}
    \end{center}
    \end{minipage}
    \centering
    \begin{minipage}[t]{\minipagewidth}
    \begin{center}
    \includegraphics[width=\figurewidthFour]{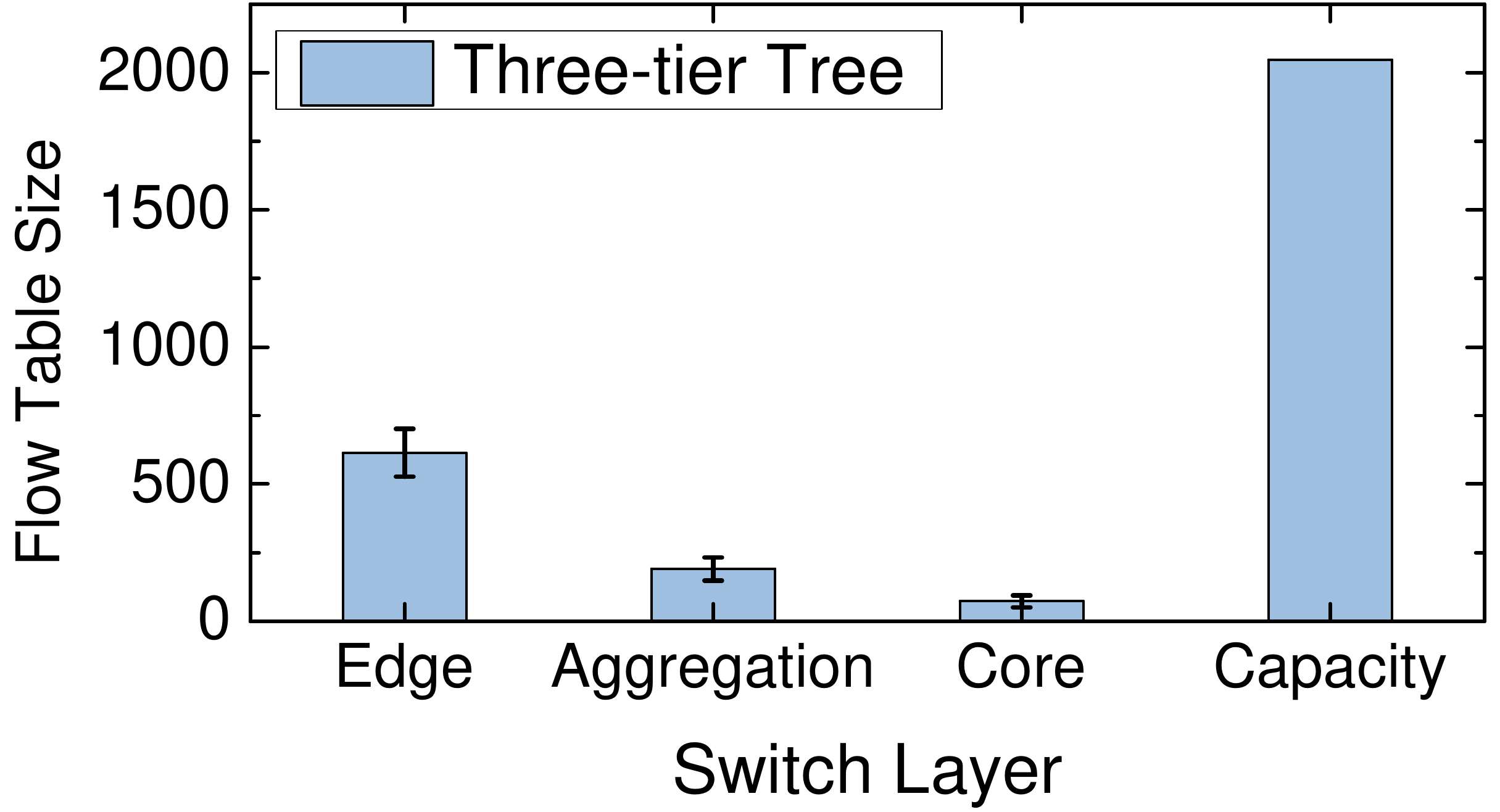}
    \subcaption{(b) Three-tier Tree (Testbed)}
    \end{center}
    \end{minipage}
    \centering
    \vspace{-0.1 in}
    \caption{SDN's flow table overhead. We measure the switches' flow table size in different layers using both a simulation and a testbed. The simulator has $2000$ servers in a fat tree topology. The testbed contains $200$ Linux containers constructed with a three-tier tree topology.}
\label{Fig: SDN_table_size}
\end{figure}

Figure \ref{Fig: SDN_table_size} shows the flow table size of the SDN-enabled switches in MetaFlow. In these experiments, we measure the flow table size  of the SDN-enabled switches in edge layer, aggregation layer, and core layer using the simulation with a fat tree topology, and the testbed with a three-tier tree topology. We observe that the flow table size is hard to limit the system performance and scalability. As shown in Figure \ref{Fig: SDN_table_size}, each SDN-enabled switch can maintain up to $2048$ flow entries. In the fat tree network, each edge layer switch  maintains roughly $360$ flow entries on average for the connected $16$ servers. The corresponding measure for the aggregation layer switch and the core layer switch are $395$ and $278$, respectively.  In the three-tier tree network, each edge layer switch (which is an OpenVSwitch daemon on the server) maintains roughly $615$ flow entries for the connected $20$ Linux containers. The corresponding measure for the aggregation layer switch and the core layer switch in the testbed are $190$ and $72$, respectively. Compared to the flow table capacity (which is $2048$ in our experiments), MetaFlow only needs a few hundreds of flow entries for each SDN-enabled switch. Therefore, the flow table size will not be the performance bottleneck in MetaFlow. The reason is that our node split algorithm uses a value between $40\%$ to $60\%$ instead of $50\%$ to split a full node, reducing the flow table size.

\setlength{\minipagewidth}{0.235\textwidth}
\setlength{\figurewidthFour}{\minipagewidth}
\begin{figure}
    \centering
    \begin{minipage}[t]{\minipagewidth}
    \begin{center}
    \includegraphics[width=\figurewidthFour]{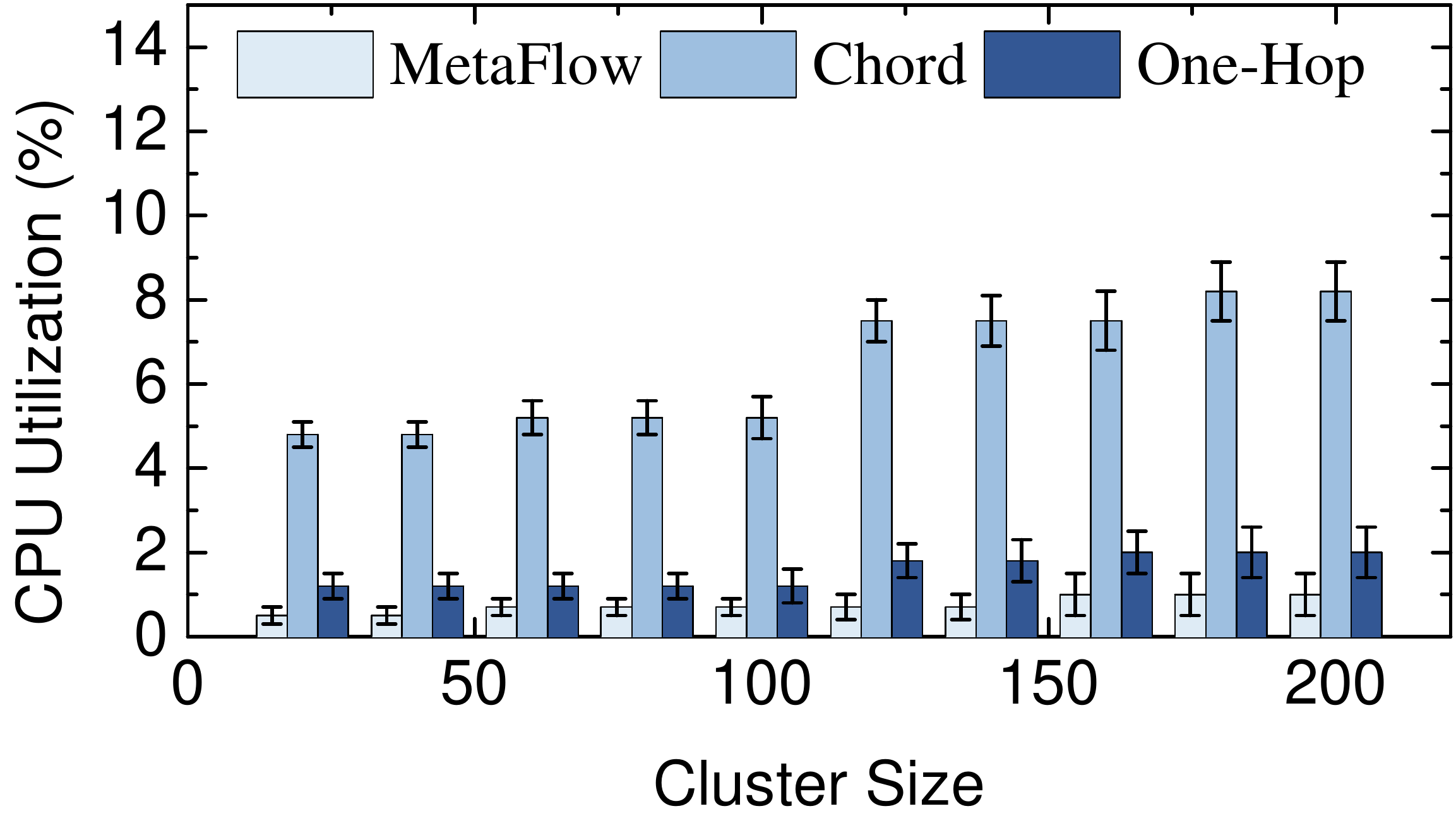}
    \subcaption{(a) MySQL}
    \end{center}
    \end{minipage}
    \centering
    \begin{minipage}[t]{\minipagewidth}
    \begin{center}
    \includegraphics[width=\figurewidthFour]{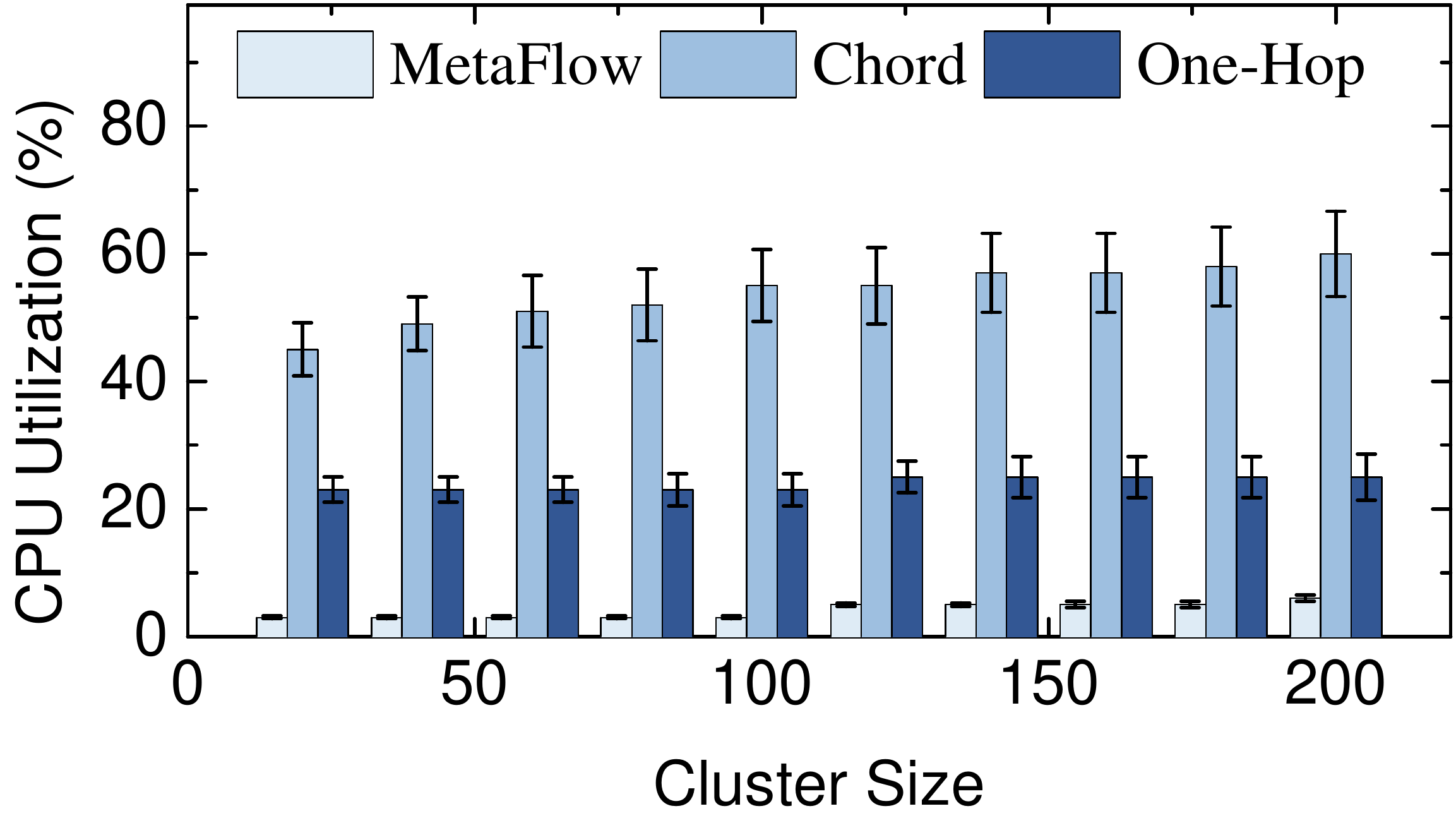}
    \subcaption{(b) LevelDB (HDD)}
    \end{center}
    \end{minipage}
    \centering
    \begin{minipage}[t]{\minipagewidth}
    \begin{center}
    \includegraphics[width=\figurewidthFour]{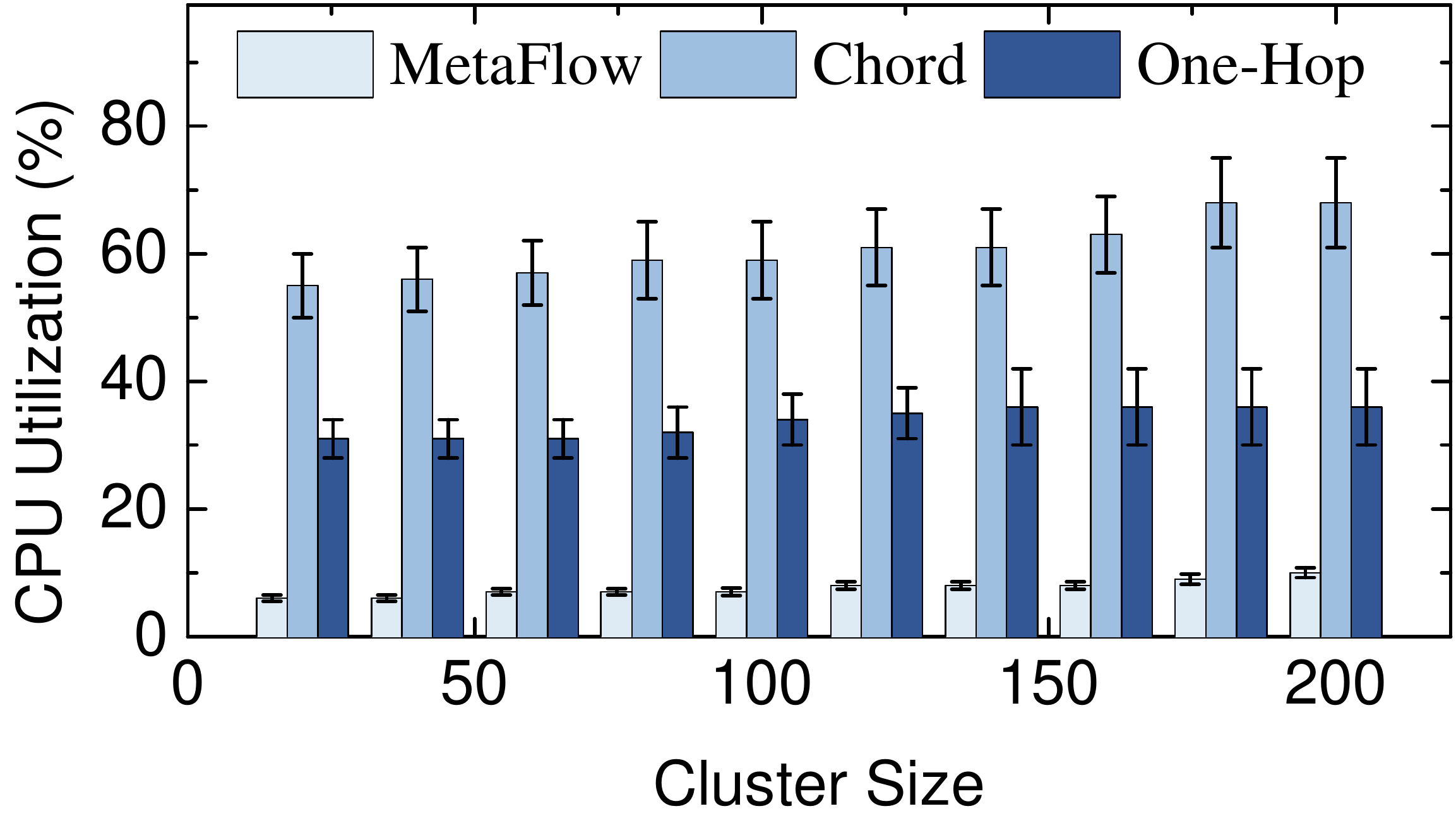}
    \subcaption{(c) LevelDB (SSD)}
    \end{center}
    \end{minipage}
    \centering
    \begin{minipage}[t]{\minipagewidth}
    \begin{center}
    \includegraphics[width=\figurewidthFour]{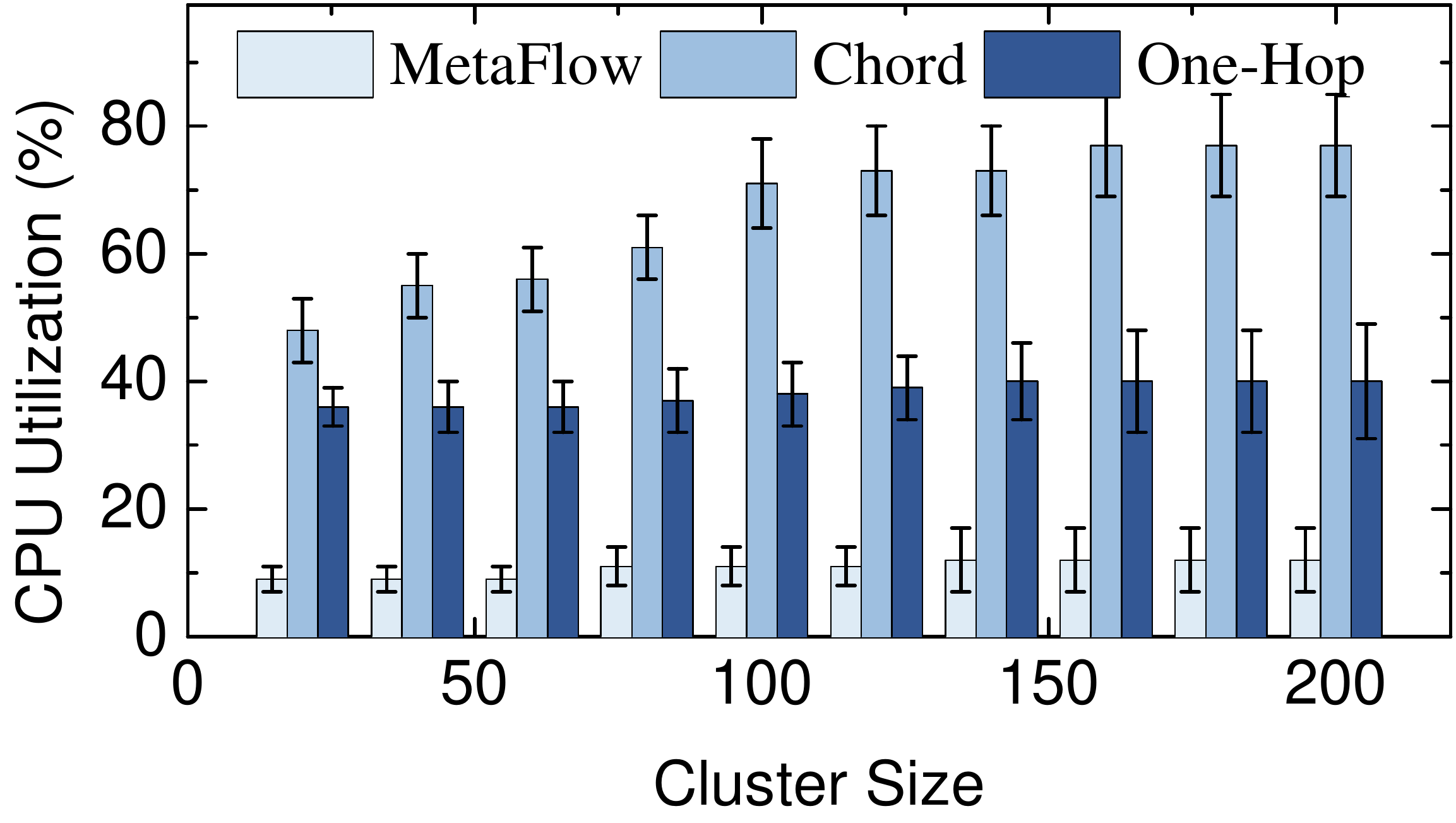}
    \subcaption{(d) Redis}
    \end{center}
    \end{minipage}
    \centering
    \vspace{-0.1 in}
    \caption{CPU overhead on the storage server. MetaFlow-based system's  NAT agent consumes CPU cycles. In Chord and One-Hop-based systems, their lookup subsystems consume CPU cycles.}
\label{Fig: MetaFlow_NAT_cpu}
\end{figure}

\subsection{NAT Agent Overhead}

\subsubsection{CPU Overhead}

We observe that the MetaFlow-based system consumes much lower CPU cycles on storage servers than DHT-based systems like Chord and One-Hop as shown in Figure \ref{Fig: MetaFlow_NAT_cpu}. This is because  MetaFlow places the lookup workload on network components. However, it still needs to set up NAT agents on  storage servers to replace source and destination IP addresses for MetaFlow  packets. These NAT agents are the main source of performance overhead in MetaFlow. In particular, when the storage subsystem is Redis, the NAT agent has up to $15\%$ of CPU utilization. This is still reasonable, considering that in the same setting Chord and One-Hop take up to $80\%$ and $40\%$ of CPU utilization, respectively. When using LevelDB (HDD) and LevelDB (SSD) as the storage subsystems, MetaFlow consumes less than $10\%$ of CPU cycles. The corresponding measures for the One-Hop-based approach are  $20\%$ and $30\%$. MetaFlow also  consumes less CPU cycles than DHT when using MySQL as the storage subsystem. Such low CPU overhead is the main reason for the higher throughput in MetaFlow.

\subsubsection{Latency Overhead}

The NAT agents on storage servers take time to translate source/destination IP addresses for MetaFlow packets. However, Figure \ref{Fig: MetaFlow_NAT_latency} shows that MetaFlow consumes much less time for the lookup service than both Chord and One-Hop.  In particular, when the storage subsystem is Redis or LevelDB (SSD), MetaFlow is responsible for less than $20\%$ of total system latency. In contrast, Chord and One-Hop take up to $60\%$ and $30\%$ of total system latency. When using LevelDB (HDD) and MySQL, MetaFlow still uses less time for the lookup service compared to One-Hop. These low overheads are the main reason for better latency performance in MetaFlow.

\setlength{\minipagewidth}{0.235\textwidth}
\setlength{\figurewidthFour}{\minipagewidth}
\begin{figure}
    \centering
    \begin{minipage}[t]{\minipagewidth}
    \begin{center}
    \includegraphics[width=\figurewidthFour]{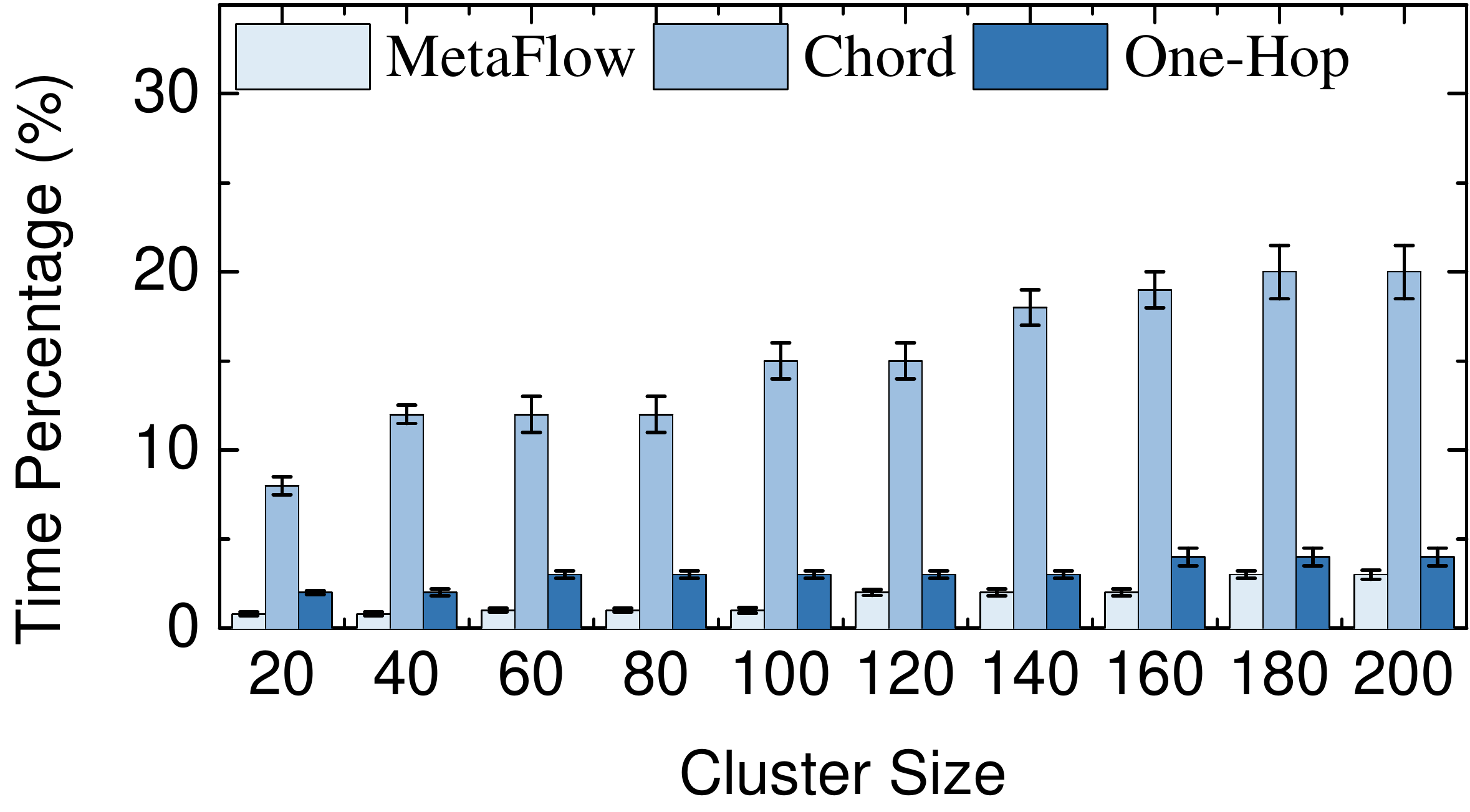}
    \subcaption{(a) Mysql}
    \end{center}
    \end{minipage}
    \centering
    \begin{minipage}[t]{\minipagewidth}
    \begin{center}
    \includegraphics[width=\figurewidthFour]{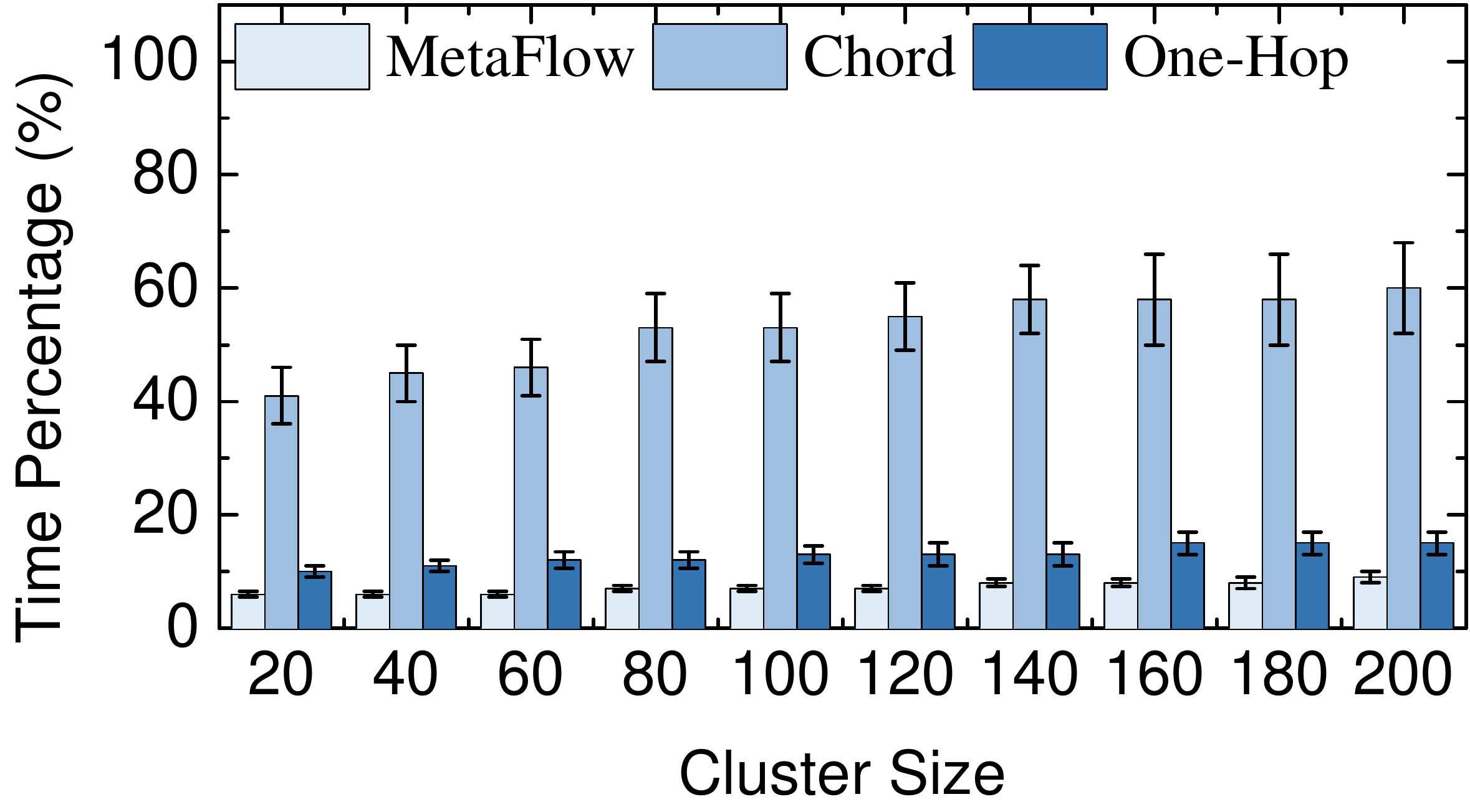}
    \subcaption{(b) LevelDB (HDD)}
    \end{center}
    \end{minipage}
    \centering
    \begin{minipage}[t]{\minipagewidth}
    \begin{center}
    \includegraphics[width=\figurewidthFour]{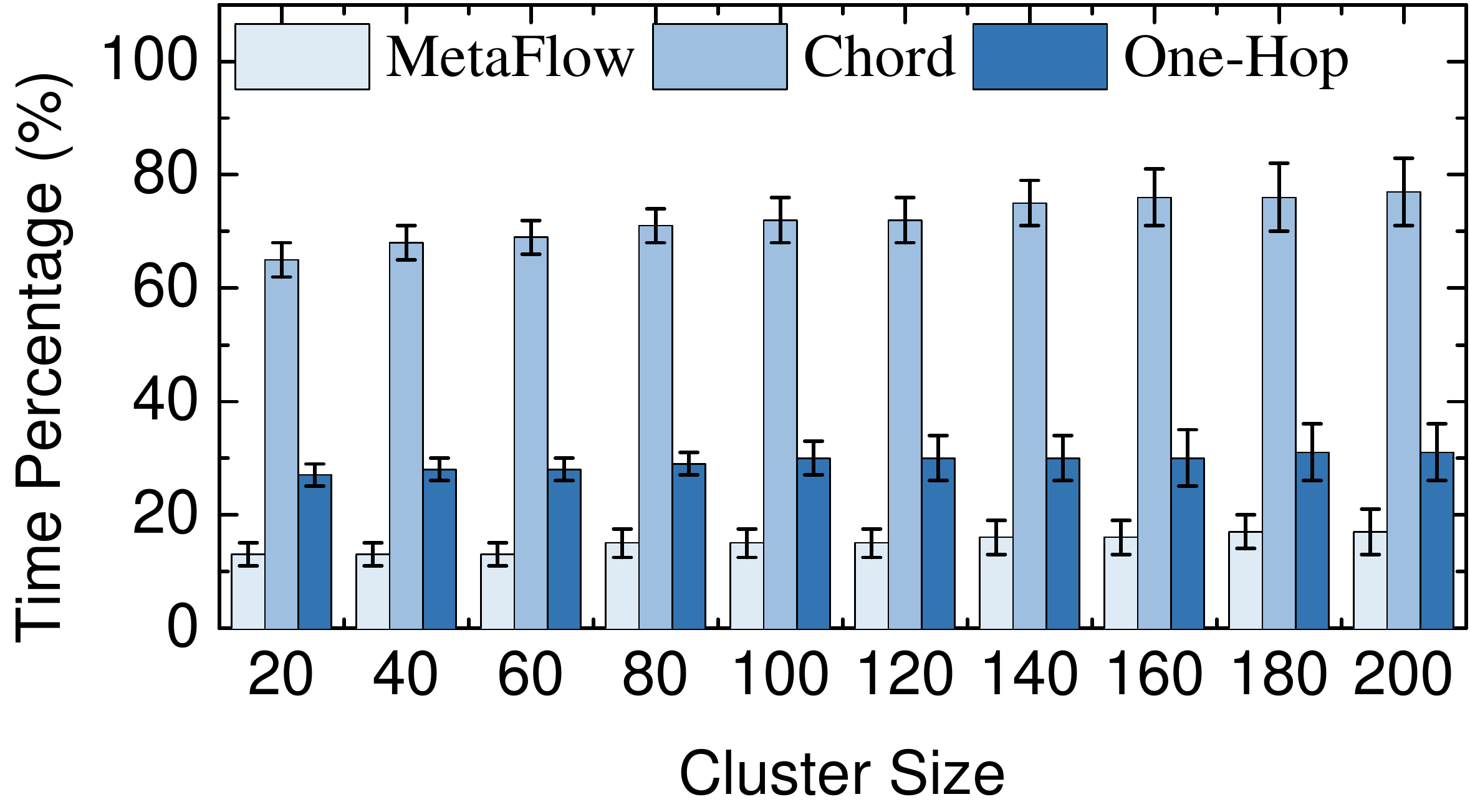}
    \subcaption{(c) LevelDB (SSD)}
    \end{center}
    \end{minipage}
    \centering
    \begin{minipage}[t]{\minipagewidth}
    \begin{center}
    \includegraphics[width=\figurewidthFour]{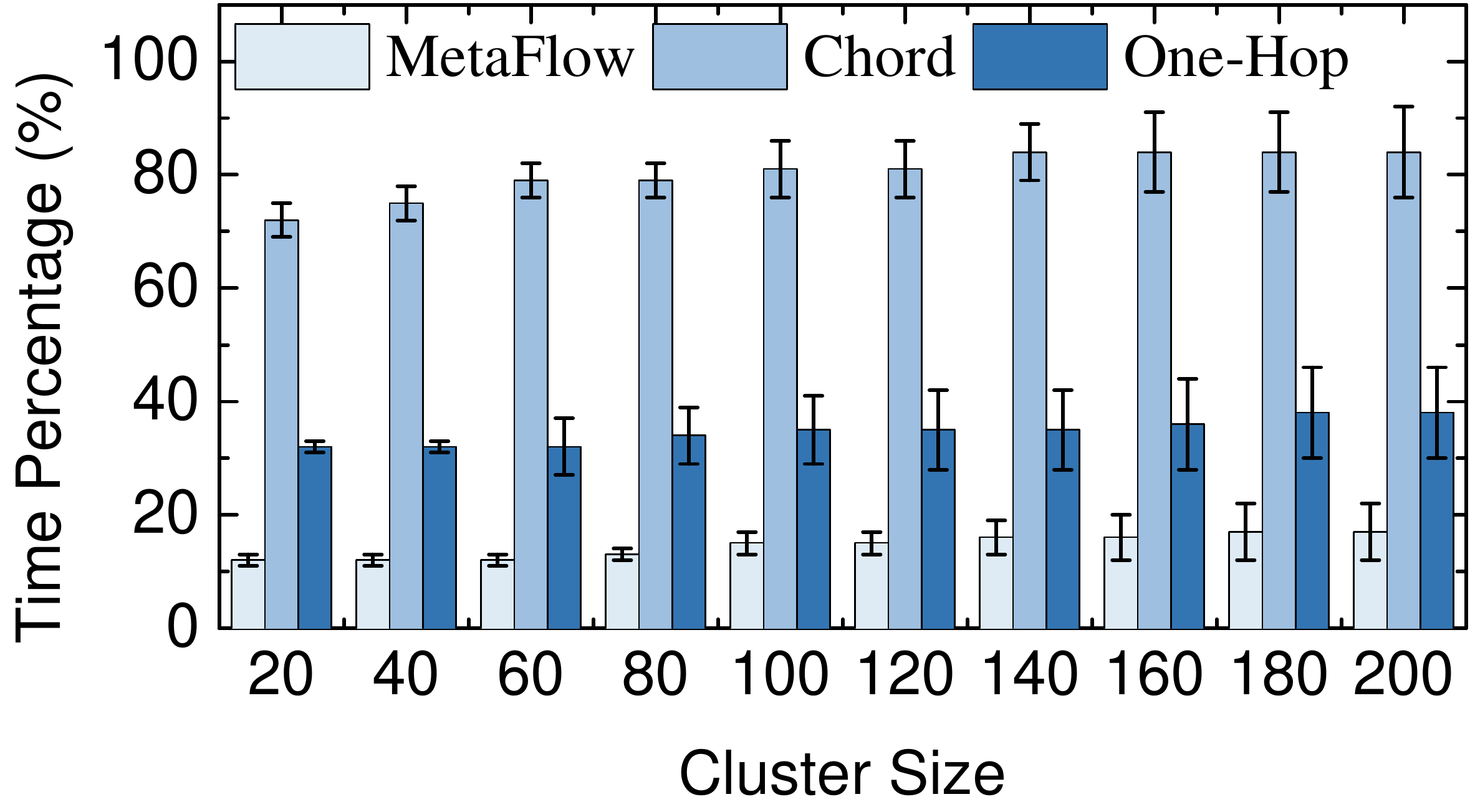}
    \subcaption{(d) Redis}
    \end{center}
    \end{minipage}
    \centering
    \vspace{-0.1 in}
    \caption{Latency overhead on the storage server. MetaFlow-based system's  NAT agent takes time to translate IP addresses for MetaFlow packets. In Chord and One-Hop-based systems, their lookup subsystems take time to locate metadata objects.}
\label{Fig: MetaFlow_NAT_latency}
\end{figure}

\subsection{Real-world Application: Distributed File System}

We investigate the performance of a distributed file system, which uses MetaFlow for managing its metadata. The testbed contains $100$ storage servers, and $10$ metadata servers. $50$ clients are set up to generate background metadata workloads in which $20\%$ are \emph{get} and $80\%$ are \emph{put} operations. We measure the completion time when writing $100$ GB of files. To investigate the impact of file sizes, we run the experiment with $4$ different file sizes,  which are $64$KB, $256$KB, $16$MB, and $64$MB.

\setlength{\minipagewidth}{0.235\textwidth}
\setlength{\figurewidthFour}{\minipagewidth}
\begin{figure}
    \centering
    \begin{minipage}[t]{\minipagewidth}
    \begin{center}
    \includegraphics[width=\figurewidthFour]{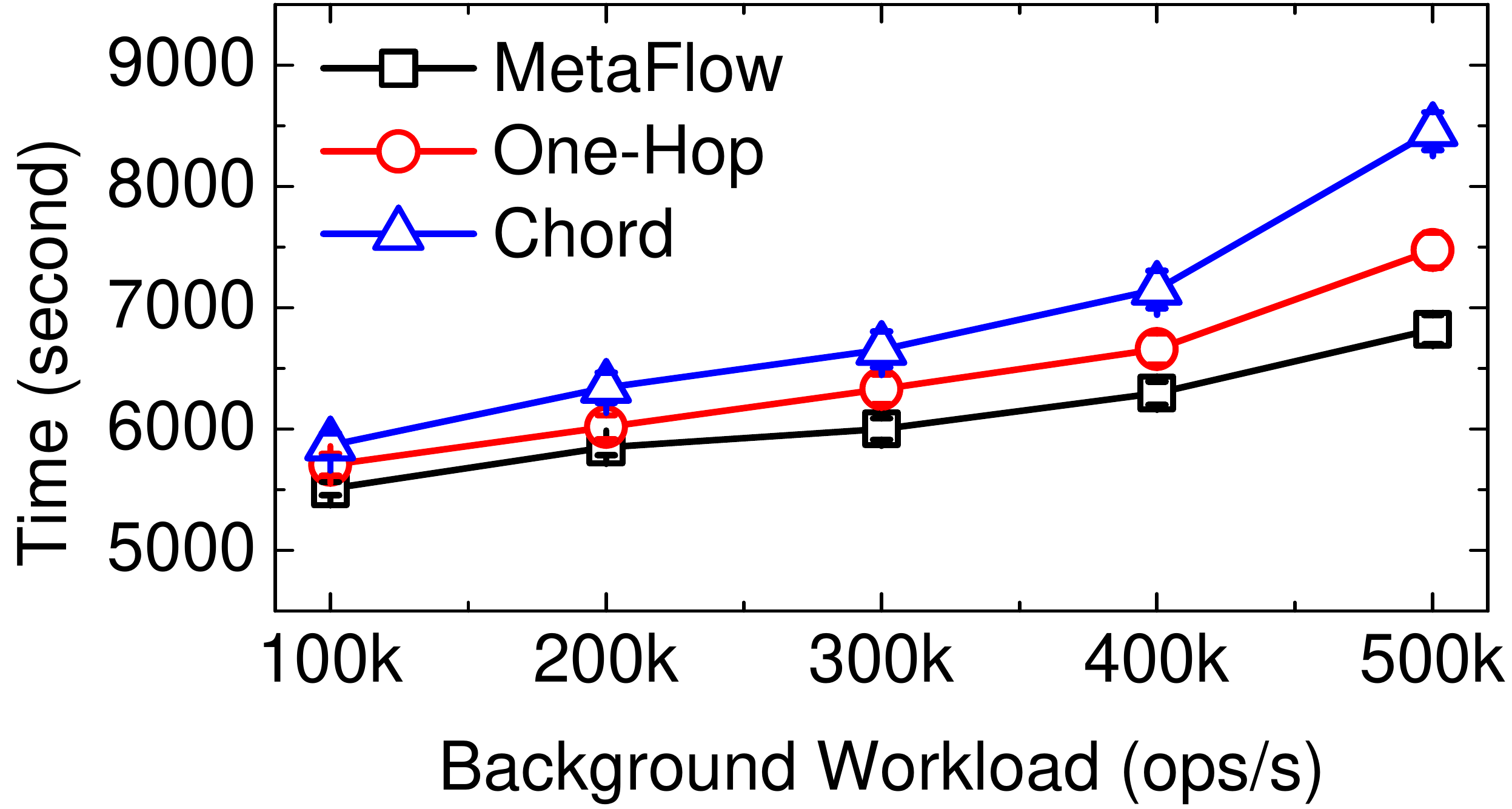}
    \subcaption{(a) 64 KB}
    \end{center}
    \end{minipage}
    \centering
    \begin{minipage}[t]{\minipagewidth}
    \begin{center}
    \includegraphics[width=\figurewidthFour]{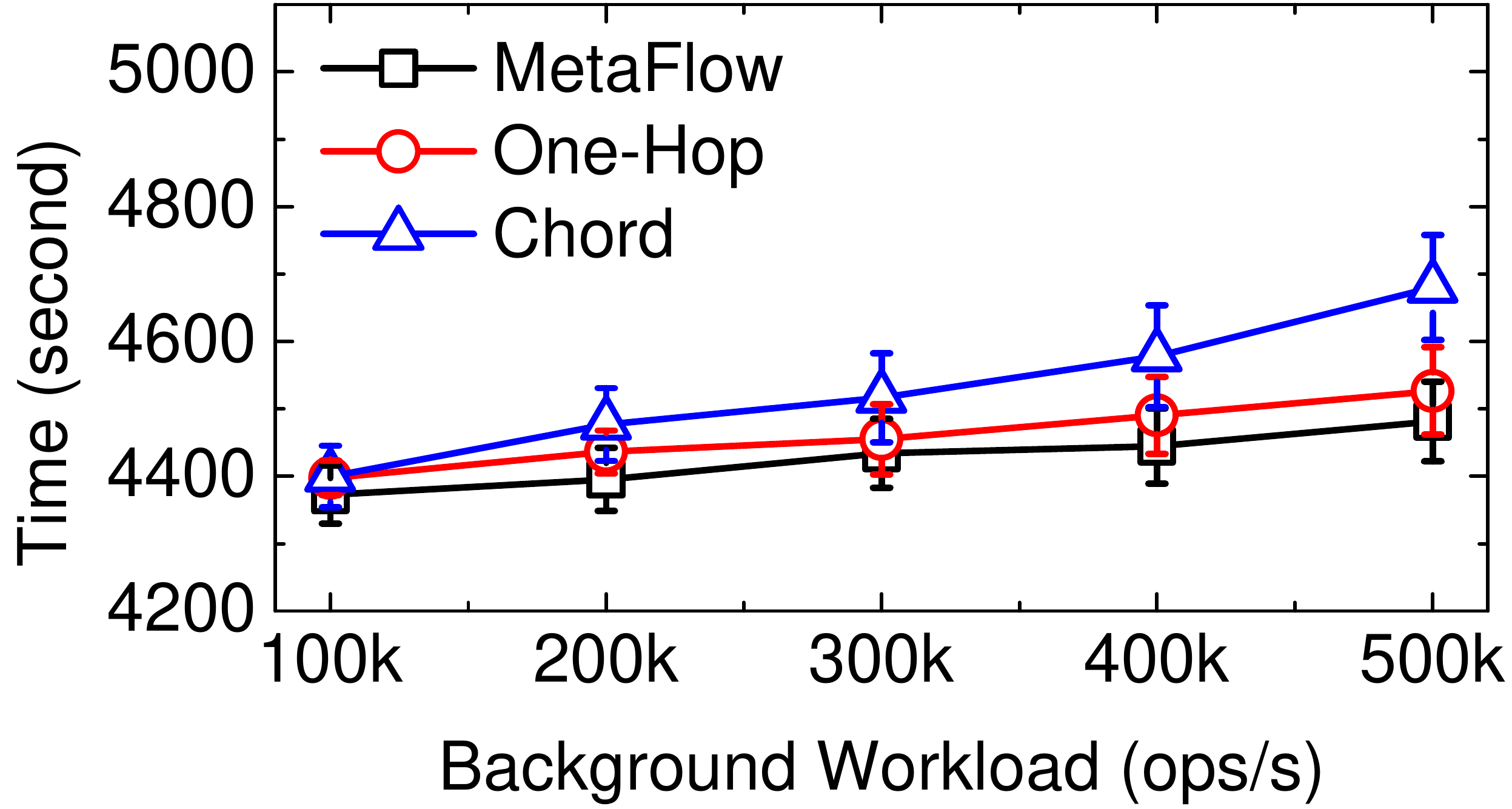}
    \subcaption{(b) 256 KB}
    \end{center}
    \end{minipage}
    \centering
    \begin{minipage}[t]{\minipagewidth}
    \begin{center}
    \includegraphics[width=\figurewidthFour]{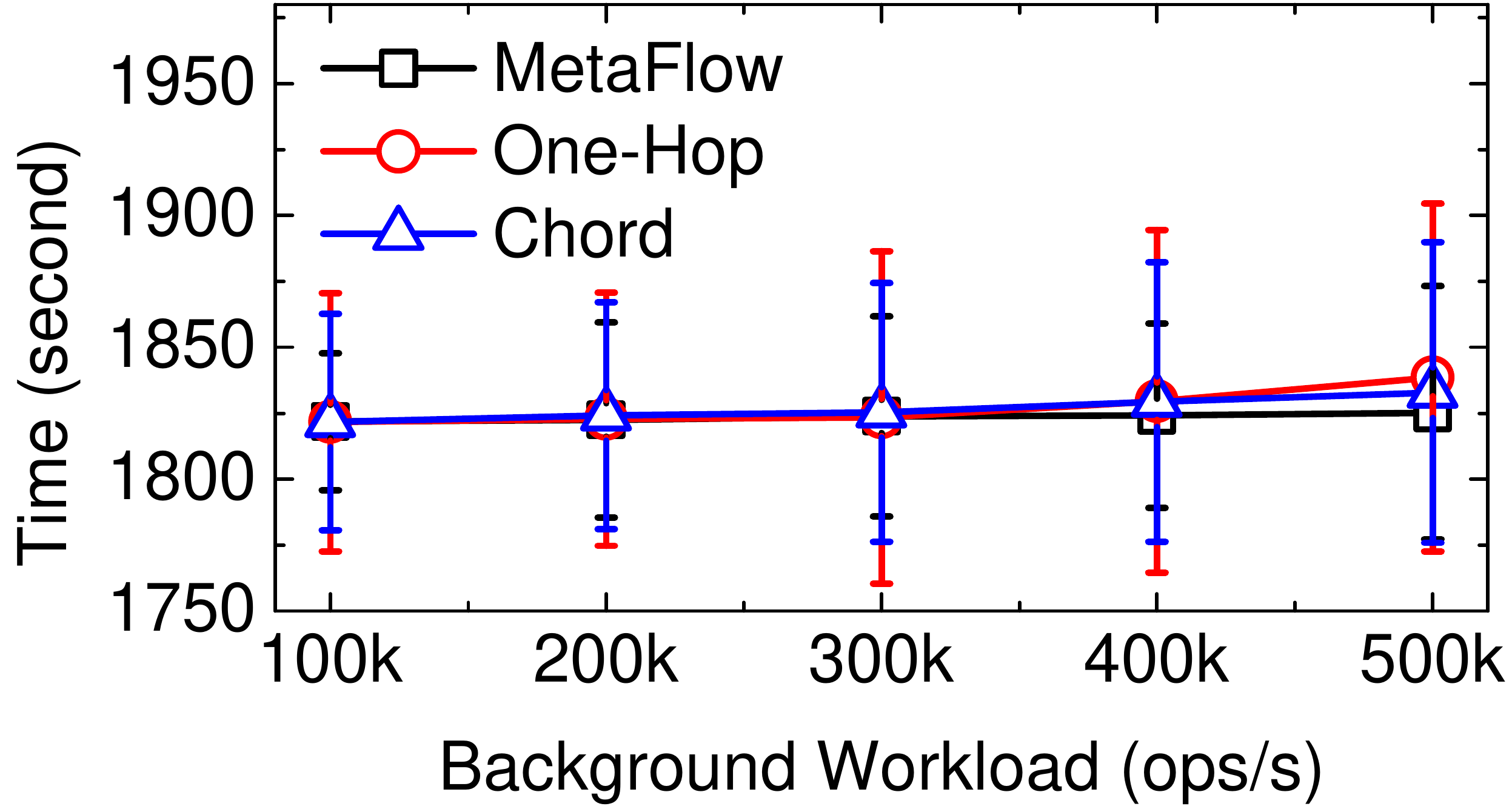}
    \subcaption{(c) 16 MB}
    \end{center}
    \end{minipage}
    \centering
    \begin{minipage}[t]{\minipagewidth}
    \begin{center}
    \includegraphics[width=\figurewidthFour]{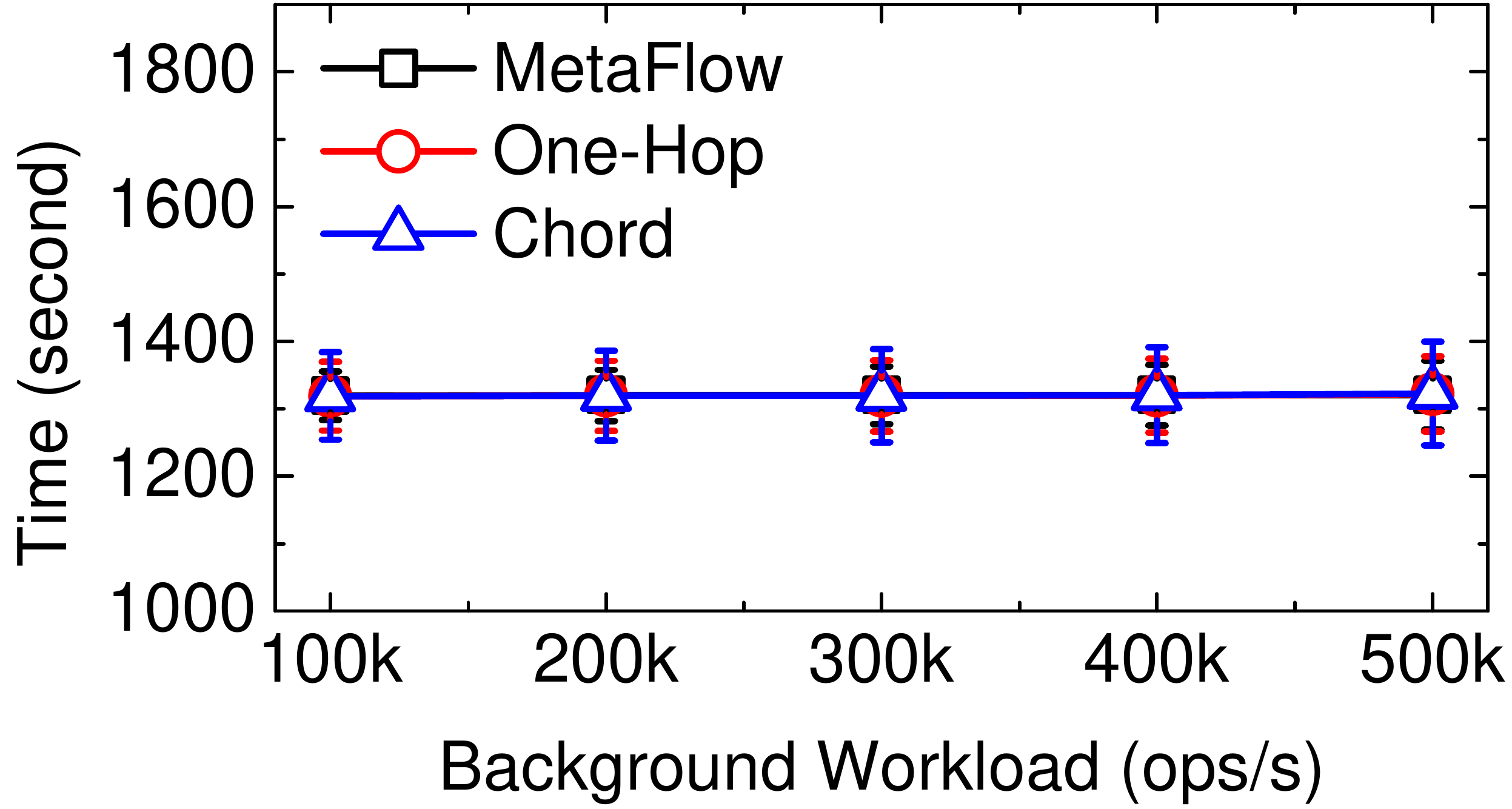}
    \subcaption{(d) 64 MB}
    \end{center}
    \end{minipage}
    \centering
    \vspace{-0.1 in}
    \caption{Distributed file system's performance comparison between MetaFlow and DHT-based approaches. We measure the time for a client to write $100$ GB of files into the distributed file system with different file sizes: 64 KB, 256 KB, 16 MB, and 64 MB.}
\label{Fig: DFS_test}
\end{figure}

The results in Figure \ref{Fig: DFS_test} show that applications using a lot of small files (e.g., the Convolutional Neural Network (CNN) \cite{krizhevsky2012imagenet} used for image classification with a large set of small images) in the distributed file system could benefit greatly from MetaFlow. In our experiment, if the file size is $64$ KB, the MetaFlow-based distributed file system consistently takes the least time to write $100$ GB of  files, regardless of the background metadata workload. Specifically, if we generate $500$ thousands of  metadata-related requests every second,  the MetaFlow-based distributed file system uses roughly $6,800$ seconds to write all $64$ KB files. In contrast, Chord and One-Hop take $8,500$ and $7,500$ seconds, which are roughly $25\%$ and $10\%$  longer than the MetaFlow-based file system. The reason for the performance improvement is that MetaFlow reduces the metadata operation time, which constitutes a large part of the total file operation time for small files.

However, if the file size is  large such as $16$ MB and $64$ MB, there are not much difference between MetaFlow and other approaches. As shown in  Figure \ref{Fig: DFS_test} (c) and (d), MetaFlow takes roughly $1820$ and $1320$ seconds to complete the write operation, which is similar to Chord and One-Hop. The reason is that the data writing operation takes much more time than metadata operation for large files.

\section{Related Work}\label{sec:related_work}

The emerging information centric networking (ICN) research also aims to eliminate separate lookup operations for network-based applications.  Named Data Networking (NDN)  \cite{jacobson2009networking} \cite{zhang2010named} is one of the pioneering fully-fledged  ICN architectures. NDN enables named-based forwarding using two types of packets, which are \emph{Interest} and \emph{Data}, to replace current IP packets. More specifically, the \emph{Interest} packet is the request packet sent by a client; and the \emph{Data} packet is the returned packet containing the requested content. Both of these two packets are identified by a resource name. Names in NDN are hierarchical and may be similar to URLs. For example, a NDN name could be \emph{/ntu.edu.sg/scse/cap}. However, compared to a normal URL, its first part (i.e., ``\emph{ntu.edu.sg}'' in this example) is not an IP address, or a DNS name, which can be parsed to an IP address. The NDN-enabled router and switch can forward an \emph{Interest} packet with the resource name to a node, which contain the target resource. Therefore, it is not needed to do a DNS lookup.

However, it is very complex to deploy NDN in real systems for two reasons. First, most existing network components work with  IP packets rather than the \emph{Interest} or the \emph{Data} packets in NDN. Second, current network-based applications such as Redis are implemented using IP rather than the NDN-based protocol. Therefore, to deploy an application using NDN, we have to  redesign both  the hardware such as switches, and the software such as  key-value storage systems.

Compared to NDN, MetaFlow is another solution for in-network lookup, which has been designed to use conventional IP-based networking. It is straightforward to deploy applications using MetaFlow for two reasons. First,  SDN capabilities provide enough  hardware support to forward MetaFlow packets using just the \emph{MetaDataID}. Second, existing applications can be easily modified to support MetaFlow: using \emph{MetaDataID} instead of destination IP address to create network connections. In this paper, we only focus on using MetaFlow to optimize distributed metadata lookup.


\section{Summary}\label{sec:conclusion}

In this paper, we propose a lookup service for metadata management. Popular DHT-based systems place the lookup subsystem and storage subsystem in the same server. These two subsystems may compete for CPU resources, which leads to reduced throughput and high latency. MetaFlow solves this problem by transferring the lookup service to the network layer. MetaFlow implements this approach by utilizing the SDN technique to forward network packets based on their \emph{MetaDataIDs} instead of conventional MAC/IP addresses. To generate and update the flow tables for the SDN-enabled switches, MetaFlow maps a data center's physical topology to a logical B-tree, and manages the flow tables using B-tree's properties. Compared to existing DHT-based approaches, MetaFlow has three key features: \emph{In-Network Lookup}, \emph{Compatibility}, and \emph{Zero-Hop}. 
Experiments show that MetaFlow could increase the system throughput by a factor of up to 3.2, and reduce the system latency by a factor of up to 5 for the metadata management compared to  existing DHT-based approaches.  
We believe that MetaFlow will be a valuable component in many distributed metadata management systems. In the future, we plan to use it in more real-word applications, such as  distributed parameter management for machine learning applications.

\bibliographystyle{IEEEtran}
\bibliography{main}

\begin{IEEEbiography}[{\includegraphics[width=1in,height=1.25in,clip,keepaspectratio]{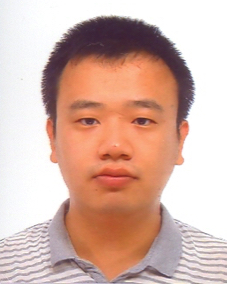}}]{Peng Sun}
received his BEng degree in automation engineering from Shandong University (SDU) in 2012. He is currently a Ph.D. student in the Energy Research Institute, Interdisciplinary Graduate School at Nanyang Technological University (NTU) in Singapore. His research interests include cloud computing, and big data processing systems.
\end{IEEEbiography}

\begin{IEEEbiography}[{\includegraphics[width=1in,height=1.25in,clip,keepaspectratio]{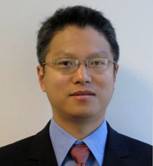}}]{Yonggang Wen}
  (S'99-M'08-SM'14) ) 
received his PhD degree in Electrical Engineering and Computer Science (minor in Western Literature) from Massachusetts Institute of Technology (MIT), Cambridge, USA. Currently, Dr. Wen is an associate professor with school of computer engineering at Nanyang Technological University, Singapore. 
Previously he has worked in Cisco to lead product development in content delivery network, which had a revenue impact of 3 Billion US dollars globally.
 His research interests include cloud computing, green data center, big data analytics, multimedia network and mobile computing.
\end{IEEEbiography}

\begin{IEEEbiography}[{\includegraphics[width=1in,height=1.25in,clip,keepaspectratio]{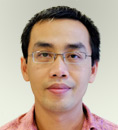}}]{Ta Nguyen Binh Duong}
 received a BEng degree from the Faculty of Information Technology, Ho Chi Minh City University of Technology, Vietnam, and a PhD degree in computer science from Nanyang Technological University, Singapore. He is currently a lecturer in the School of Computer Engineering, Nanyang Technological University. 
 Previously he worked as a scientist in the Computing Science Department, A*STAR Institute of High Performance Computing, Singapore. 
 His current research interests include distributed virtual environments, computer networking, and cloud computing. 
 \end{IEEEbiography}

\begin{IEEEbiography}[{\includegraphics[width=1in,height=1.25in,clip,keepaspectratio]{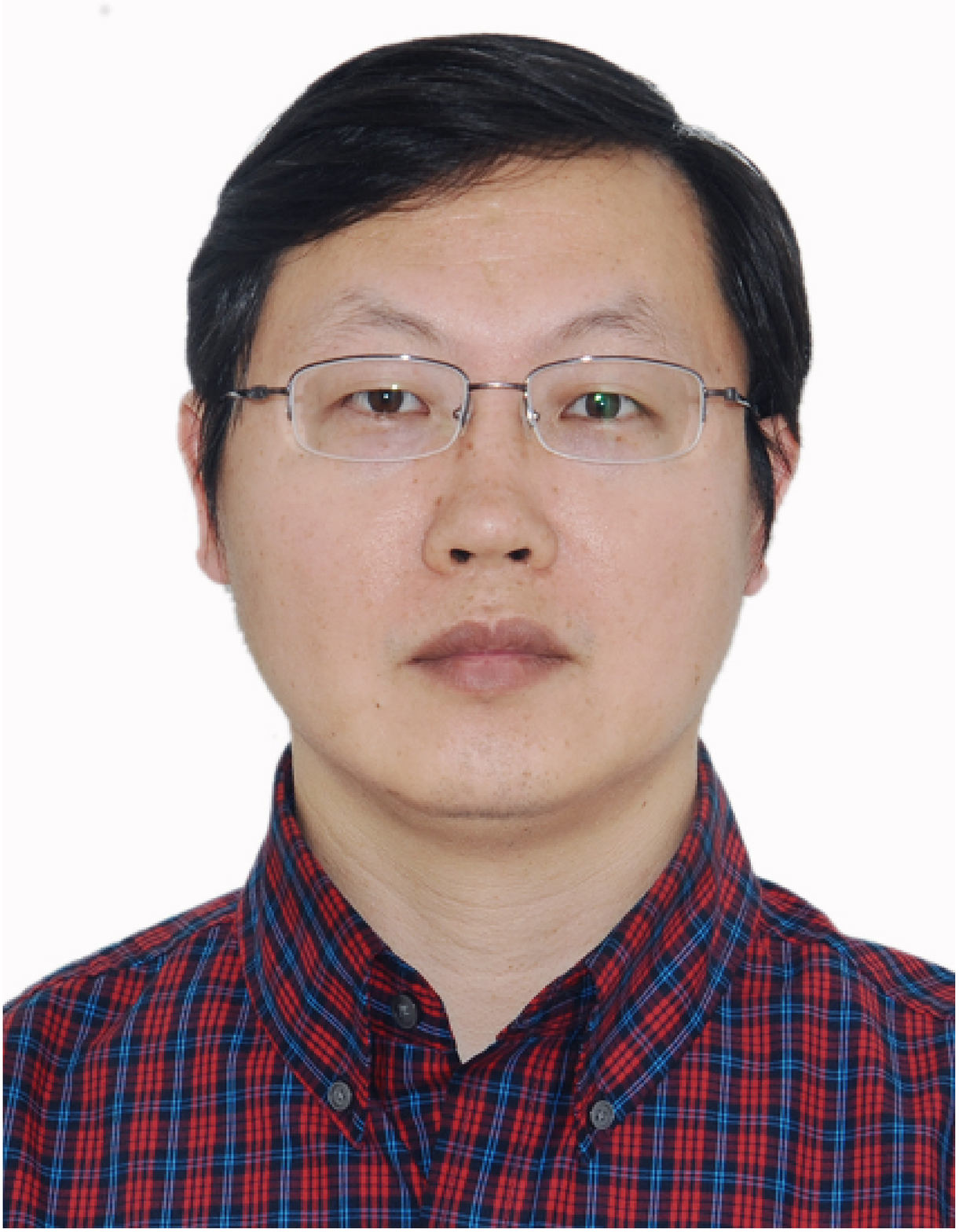}}]{Haiyong Xie} 
(S'05-M'09)  received the Ph.D. and M.S. degrees in computer science from Yale University, New Haven, CT, USA in 2005 and 2008, and the B.S. degree from the University of Science and Technology of China, Hefei, China, in 1997, respectively. He is currently the Executive Director of the CAEIT Cyberspace and Data Science Laboratory and a Professor with the School of Computer Science and Technology, University of Science and Technology of China (USTC). His research interest includes contentcentric networking, software-defined networking, future Internet architecture, and network traffic engineering.
\end{IEEEbiography}

\end{document}